\documentclass[10pt,twocolumn,showpacs,preprintnumbers,amsmath,amssymb,aps,prx,longbibliography,superscriptaddress]{revtex4-1}
\usepackage{mathrsfs}
\usepackage{graphicx}
\usepackage{dcolumn}
\usepackage{bm}
\usepackage{amsmath}
\usepackage{amsfonts}
\usepackage{color}
\usepackage{hyperref}

\begin{document}

\title{The Landau Level of Fragile Topology}

\author{Biao Lian}
\affiliation{Princeton Center for Theoretical Science, Princeton University, Princeton, New Jersey 08544, USA}
\author{Fang Xie}
\affiliation{Department of Physics, Princeton University, Princeton, New Jersey 08544, USA}
\author{B. Andrei Bernevig}
\affiliation{Department of Physics, Princeton University, Princeton, New Jersey 08544, USA}
\affiliation{Dahlem Center for Complex Quantum Systems and Fachbereich Physik, Freie Universitat Berlin, Arnimallee 14, 14195 Berlin, Germany}
\affiliation{Max Planck Institute of Microstructure Physics, 06120 Halle, Germany}

\begin{abstract}
We study the Hofstadter butterfly and Landau levels of the twisted bilayer graphene (TBG). We show that the nontrivial fragile topology of the lowest two bands near the charge neutral point makes their Hofstadter butterfly generically connected with higher bands, closing the gap between the first and second conduction (valence) bands at a certain magnetic flux per unit cell. We also develop a momentum space method for calculating the TBG Hofstadter butterfly, from which we identify three phases where the Hofstadter butterflies of the lowest two bands and the higher bands are connected in different ways. We show this leads to a crossing between the $\nu=4$ Landau fan from the charge neutral point and the zero field band gap at one flux per Moir\'e unit cell, which corresponds to a magnetic field $25\theta^2$T (twist angle $\theta$ in degrees). This provides an experimentally testable feature of the fragile topology. 
In general, we expect it to be a generic feature that the Hofstadter butterfly of topological bands are connected with the Hofstadter spectra of other bands. 
We further show the TBG band theory with Zeeman splitting being the most sizable splitting could result in Landau fans at the charge neutral point and half fillings near the magic angle, and we predict their variations under an in-plane magnetic field.
\end{abstract}

\date{\today}

\maketitle

Twisted bilayer graphene (TBG) has low energy flat electron bands at twist angles near the so-called magic angle, and exhibits superconductivity and correlated insulating phases at low temperatures
\cite{morell2010,bistritzer2011,cao2018b,cao2018,yankowitz2018,cao2016,huangs2018,
yuan2018,po2018,xu2018,roy2018,volovik2018,padhi2018,dodaro2018,baskaran2018,wu2018, isobe2018,huang2018,you2018,wux2018,zhangy2018,kang2018,koshino2018,kennes2018,zhangl2018,
pizarro2018,guinea2018,thomson2018,ochi2018,xux2018,peltonen2018,fidrysiak2018,
zou2018,gonz2018,su2018,guo2018,songz2018,po2018b,ahn2018,sherkunov2018,sboychakov2018,lingam2018,hejazi2018,
laksono2018,tarnopolsky2018,venderbos2018,chenl2018,stauber2018,fuy2018,tang2018,choi2018,angeli2018,
padhi2018b,jian2018a,kozii2018,kang2018b,wux2018b,wuf2018b,spurrier2018,liuy2018}. It was shown that the lowest two bands of TBG carry a nontrivial fragile topology \cite{zou2018,songz2018,po2018b,ahn2018,po2018f}, in sharp contrast to the lowest two bands of monolayer graphene which are as a whole trivial. However, what observable effects the TBG nontrivial band topology brings is yet unknown. Moir\'e edge states are predicted in TBG \cite{fleischmann2018}, while whether they are related to the fragile topology is still unclear. In addition, the lowest two bands of TBG near the magic angle exhibit unconventional Landau fans in the magnetic fields \cite{cao2018,yankowitz2018}, lacking a theoretical understanding.

In this paper, we study the single-particle Hofstadter butterfly and Landau level (LL) of TBG. We show that the Hofstadter butterfly of the fragile topological bands is quite generically connected with the Hofstadter butterfly of other bands. We first verify this in two tight-binding models of Refs. \cite{songz2018,po2018b} which have the fragile topology of TBG. We then develop a momentum space method to calculate the Hofstadter butterfly of the TBG continuum model \cite{bistritzer2011}. For the continuum model, we find the Hofstadter butterfly of the lowest two fragile topological bands are connected with higher bands for angles $\theta\gtrsim 1.9^\circ$ and $\theta\lesssim 1.1^\circ$ at finite magnetic field $B$ due to fragile topology, while for angles $1.1<\theta<1.9^\circ$ the Hofstadter butterfly are connected at $B=\infty$. 
We then show the large field Landau fans have fingerprints from the fragile topology, which are experimentally measurable. 
We expect connected Hofstadter butterfly to be a generic feature of topological bands. 
At low energies, we show that the band theory with Zeeman splitting can give rise to Landau fans at the charge neutral point (CNP) and half fillings near magic angle as observed in experiments \cite{cao2018,yankowitz2018}. Given that the experimental fan splittings \cite{yankowitz2018} are not spin splitting \cite{priv}, other effects may overwhelm the Zeeman effect \cite{zhangy2019}.


The band structure of TBG can be derived using the continuum model \cite{bistritzer2011,santos2007}. Each graphene layer contains two flavors of Dirac electrons at valleys $K$ and $K'$ of the graphene Brillouin zone (GBZ), described by Hamiltonians with opposite helicities $h^K(\mathbf{k})=\hbar v(k_x\sigma_x-k_y\sigma_y)=\hbar v\bm{\sigma}^*\cdot\mathbf{k}$ and $h^{K'}(\mathbf{k})=-\hbar v\bm{\sigma}\cdot\mathbf{k}$, respectively. Here $v$ is the graphene Fermi velocity, $\sigma_{x,y,z}$ are Pauli matrices for sublattice indices, and $\mathbf{k}=(k_x,k_y)$ is the momentum measured from $K$ or $K'$ point. The interlayer hopping couples electrons at the same graphene valley, which leads to two decoupled sets of bands at $K$ and $K'$, respectively.
As a result, all the bands are 4-fold degenerate (graphene valley $K,K'$ and spin $\uparrow$, $\downarrow$).
For each spin and valley, there are $2$ Dirac points at $K_M$ and $K_M'$ points (subindex $M$ for distinguishing from GBZ) of the Moir\'e Brillouin zone (MBZ) around the CNP, described by Hamiltonian
\begin{equation}\label{HD}
H_{\text{eff}}^{\eta}(\mathbf{k})=\hbar v_*(\eta k_x\sigma_x-k_y\sigma_y)\ ,
\end{equation}
where the helicity $\eta=\pm1$ for graphene valleys $K$ and $K'$, respectively. The momentum $\mathbf{k}$ is measured from $K_M$ or $K_M'$. The Fermi velocity $v_*$ depends on $\theta$ and the corrugation (relaxation) $u_0$. The corrugation $u_0\in[0,1]$ is defined as the ratio of the effective AA and AB interlayer hoppings (\cite{suppl} Sec. S1). For $u_0=1$ (no corrugation), $v_*=0$ for $\theta=\theta_m\approx1.0^\circ$ in our parameters, which is defined as the magic angle. The lowest two bands become extremely flat near $\theta_m$ \cite{bistritzer2011}.

For each graphene valley and spin, the two Dirac points at $K_M$ and $K_M'$ between the lowest two TBG bands have the same helicity \cite{bistritzer2011,po2018b} (see Eq. (\ref{HD})), which indicates the lowest two bands by themselves are non-Wannierizable and have a nontrivial fragile topology protected by $C_{2z}T$ symmetry \cite{songz2018,po2018b}. This fragile topology of the lowest two bands yields a Wilson loop winding number $1$ in the MBZ \cite{songz2018,ahn2018}.

In general, if a set of bands is topologically trivial, its LLs or Hofstadter butterfly \cite{hofstadter1976} will be roughly bounded by the energy span of the set of bands, isolated from other bands. As the magnetic field $B$ (out-of-plane) increases, the LLs carrying Chern number $+1$ from the band maxima and minima will approach the van Hove singularities, and annihilate with the negative Chern number Hofstadter bands therein, leading to an energetically bounded butterfly. In contrast, we expect that the Hofstadter butterfly of a set of topologically nontrivial bands are generically unbounded, until it connects with the butterfly of another set of bands which trivializes the band topology (see \cite{suppl} Sec. S3D for the example of Chern bands and Sec. S3E for general argument).


In the case of two bands connected by two Dirac points (such as monolayer graphene or TBG), each Dirac point contributes a zero mode Landau level at small $B$.
If the two Dirac points have opposite helicities so that the two bands are topologically trivial (e.g., in monolayer graphene), the two zero modes will repel each other in energy when $B$ grows large (when the inverse magnetic length $\ell_B^{-1}=\sqrt{eB/\hbar}$ exceeds the distance between two Dirac points), and eventually merge into the van Hove singularities of the two bands, respectively, resulting an energetically bounded Hofstadter butterfly. More explicitly, consider a toy model Hamiltonian
\begin{equation}\label{Hnq}
H(\mathbf{k})=A[(k_+-k_D)(k_-+k_D)\sigma_++h.c.]\ ,
\end{equation}
which has two Dirac points of opposite helicities at $(k_x,k_y)=(\pm k_D,0)$, where $k_\pm=k_x\pm ik_y$ and $\sigma_\pm=(\sigma_x\pm i\sigma_y)/2$. One can show that the two Dirac zero mode LLs split into energies $E_{0,\pm}\approx\pm A \ell_B^{-2}e^{-k_D^2\ell_B^2}$ (\cite{suppl} Sec. S2), where $\ell_B=\sqrt{\hbar/eB}$ is the magnetic length. When $k_D\ell_B\lesssim1$, $E_{0,\pm}$ become large and comparable to the van Hove energies around $Ak_D^2$.

\begin{figure}[tbp]
\begin{center}
\includegraphics[width=3.4in]{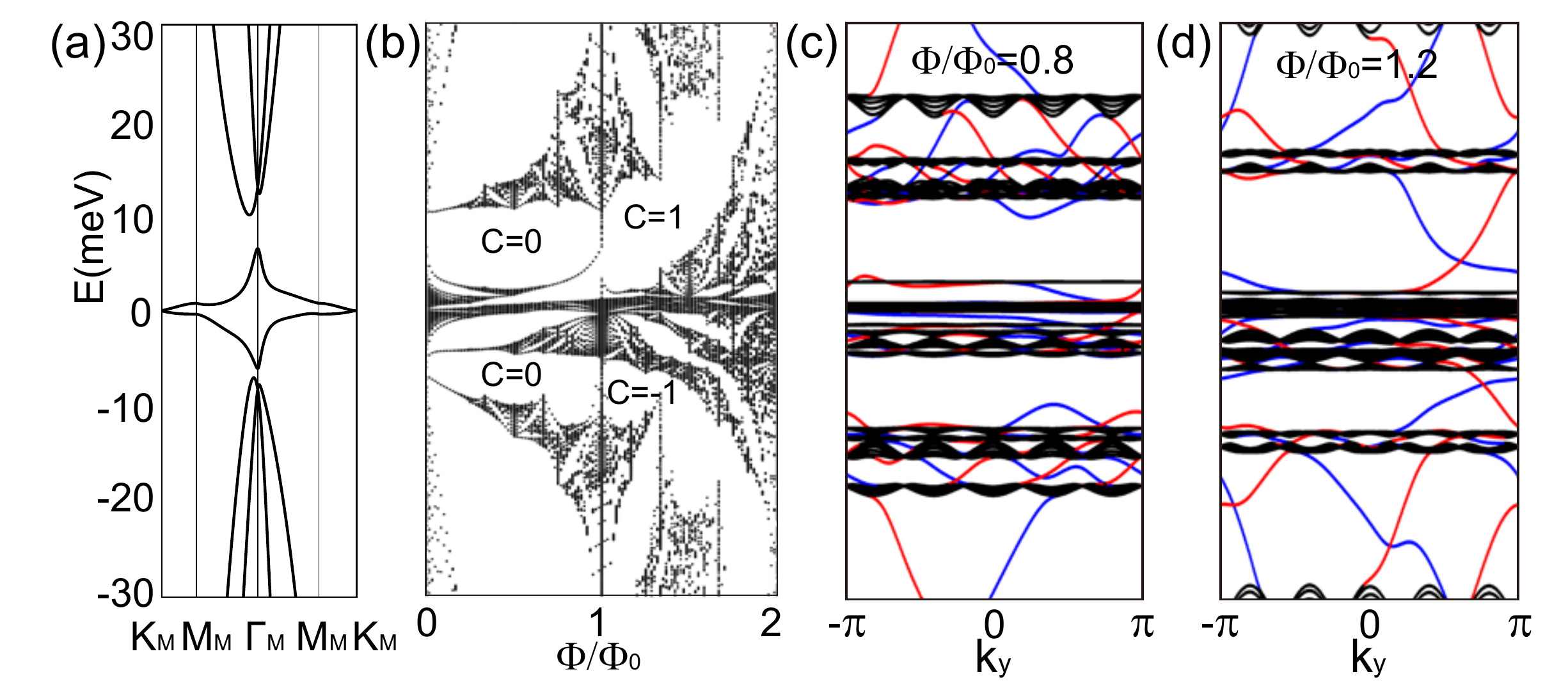}
\end{center}
\caption{(a) The band structure and (b) the Hofstadter butterfly of the ten-band model in Ref. \cite{po2018b}, where $\frac{\Phi}{\Phi_0}$ is the magnetic flux per supercell. The open boundary spectrum at $\frac{\Phi}{\Phi_0}=0.8$ and $1.2$ are given in (c) and (d), respectively, where red (blue) lines are the edge states on the left (right) edge, and black lines are the bulk states.
}
\label{Ten}
\end{figure}

On the contrary, if the two Dirac points have the same helicity, as is the case of the topologically nontrivial TBG flat bands, at large $B$ one expects them to behave together as a quadratic band touching, with the two zero mode LLs staying robust. To see this, consider a toy Hamiltonian
\begin{equation}\label{Hq}
H'(\mathbf{k})=A[(k_+^2-k_D^2)\sigma_++(k_-^2-k_D^2)\sigma_-]\ ,
\end{equation}
which has two Dirac points of the same helicity at $(k_x,k_y)=(\pm k_D,0)$. The LLs can be obtained by substituting $(k_+,k_-)\rightarrow\sqrt{2}\ell_B^{-1}(a,a^\dag)$ with the LL lowering and raising operators $a,a^\dag$ (\cite{suppl} Sec. S4), from which one can obtain two exactly zero energy LLs $\psi_\pm=(0,|\Omega_\pm\rangle)^T$ carrying Chern number $C=1$ disregarding the value of $k_D\ell_B$, where $|\Omega_\pm\rangle=e^{\pm\frac{k_D\ell_B}{\sqrt{2}}a^\dag}|0\rangle$.

If a perturbation $H_{p}(\mathbf{k})$ is added to the Hamiltonian $H'(\mathbf{k})$ in Eq. (\ref{Hq}), it may contribute nonzero energies to the two zero mode LLs $\psi_\pm$. However, as long as the perturbation $H_p(\mathbf{k})$ is smaller than $H'(\mathbf{k})$, one expects the two zero model LLs $\psi_\pm$ to be isolated from higher LLs (\cite{suppl} Sec. S2). This will therefore lead to a $C=1$ ($C=-1$) gap above (below) $\psi_\pm$ extending to higher bands, forcing the Hofstadter butterfly of the lowest two bands to be connected with higher bands, until the higher bands trivializes the fragile topology. In TBG, the ``stable" index proposed in \cite{songz2018,ahn2018} cannot be trivialized, the effect of which on the Hofstadter butterfly is yet unknown.

To verify our claim of the unboundness and connectivity of the Hofstadter spectrum for topological bands, we first study the ten-band tight-binding model for TBG in Ref. \cite{po2018b} (\cite{suppl} Sec. S3C), where the lowest two bands near zero in Fig. \ref{Ten}(a) have the fragile topology of TBG. As shown in Fig. \ref{Ten}(b), the Hofstadter butterfly of the lowest two bands is connected with the higher band butterfly at flux per unit cell $\Phi/\Phi_0=1$, leading to two Chern number $\pm1$ gaps emerging from the lowest two bands and extending to higher bands as expected. The in-gap Chern numbers are determined from the number of chiral edge states in open boundary calculations as shown in Fig. \ref{Ten}(c),(d).

\begin{figure}[tbp]
\begin{center}
\includegraphics[width=3.4in]{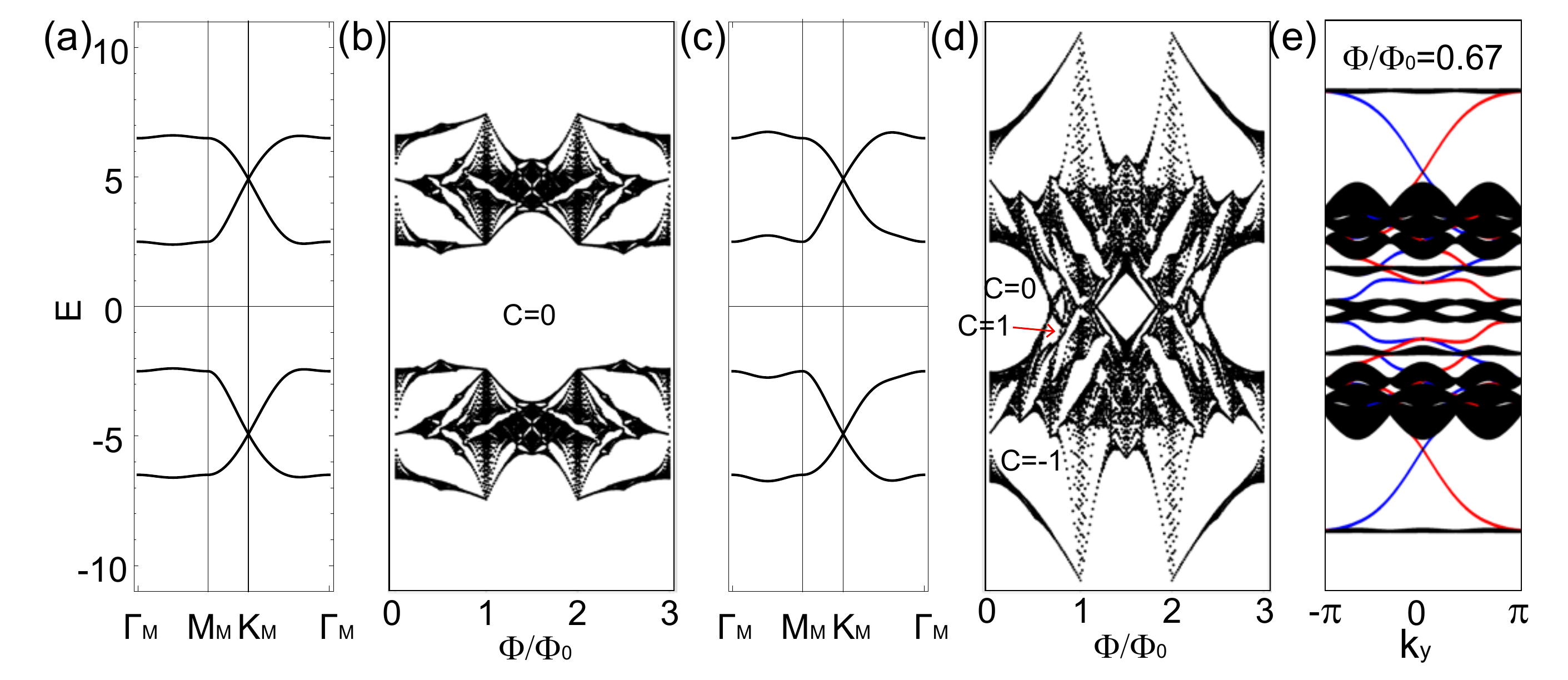}
\end{center}
\caption{Band structure and Hofstadter butterfly of the TB4-1V model in Ref. \cite{songz2018}, where $\frac{\Phi}{\Phi_0}$ is the magnetic flux per supercell. (a) and (b): the trivial phase. (c) and (d): the topological phase (the lower two bands reproduce the topology of the TBG flat bands). (e) The open boundary spectrum of the topological phase in (c) at $\frac{\Phi}{\Phi_0}=0.67$, where red (blue) lines are the edge states on the left (right) edge.
}
\label{Butterfly}
\end{figure}

To gain some further insight, we also study the Hofstadter butterfly of the TB4-1V model for TBG in Ref. \cite{songz2018}. This model has a topological phase where the lower two bands (Fig. \ref{Butterfly}(c)) faithfully reproduce the topology of the two TBG flat bands, and also a trivial phase with similar band dispersions (Fig. \ref{Butterfly}(a)). To faithfully reproduce TBG, we use Wannier functions at AB (BA) stackings with charge densities concentrated at AA stackings, as anticipated before \cite{po2018,kang2018b}. 
The Peierls substitution gauge phases accumulate along paths from AB (BA) to AA and then to another AB (BA) positions (see \cite{suppl} Sec. S3A).
As shown in Fig. \ref{Butterfly}(b), the Hofstadter butterflies of the trivial phase are bounded within the energy range of the lower (higher) two trivial bands. By contrast, in the topological phase, the butterflies of the lower and higher two bands are heavily connected, closing the gap between them.


\begin{figure}[tbp]
\begin{center}
\includegraphics[width=3.4in]{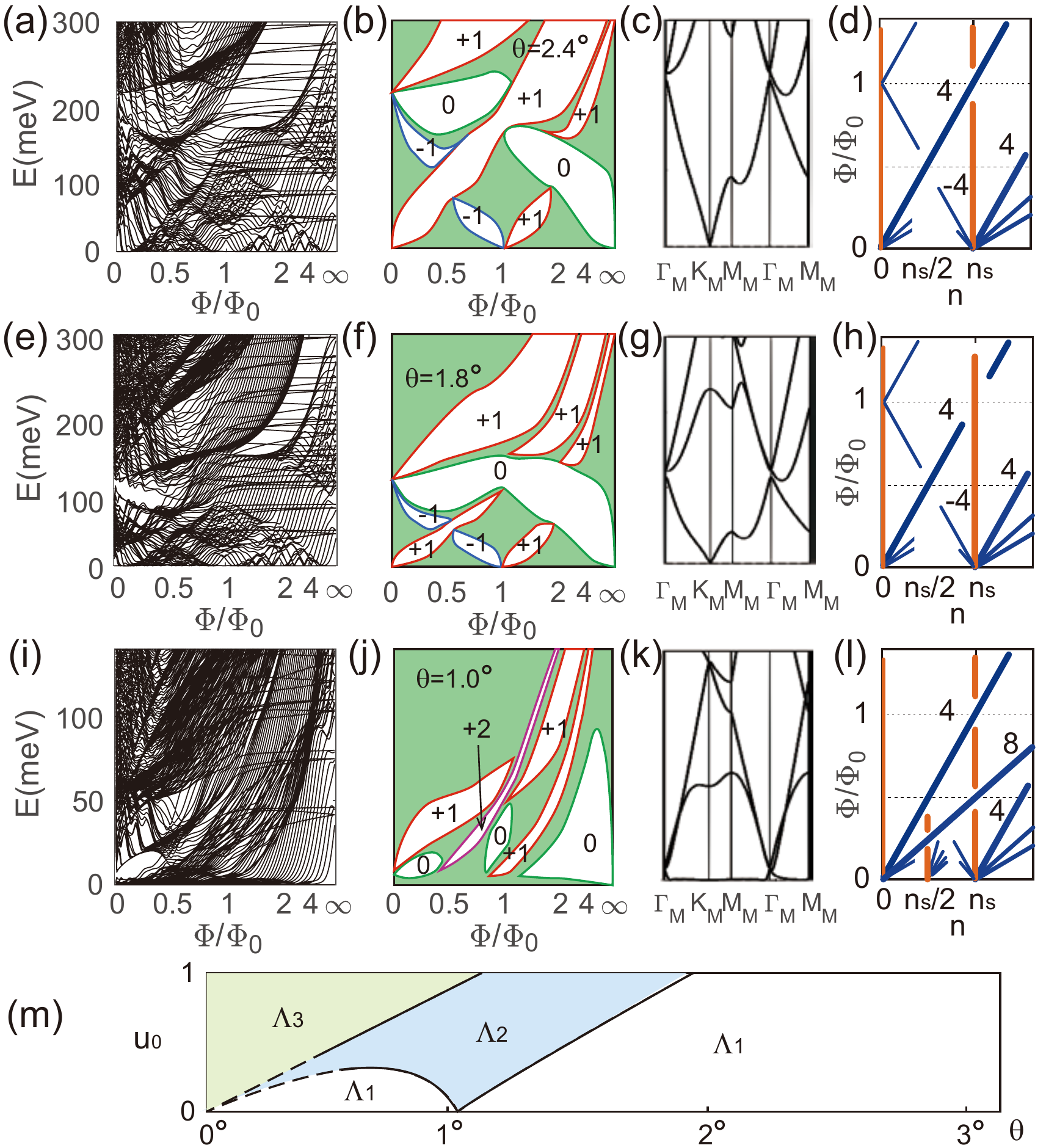}
\end{center}
\caption{The numerical Hofstadter spectrum, sketched Hofstadter butterfly, energy spectrum and large $B$ Landau fan for $\theta=2.4^\circ$ ((a)-(d)), $1.8^\circ$ ((e)-(h)) and $1.0^\circ$ ((i)-(l)), where the corrugation $u_0=1$. (m) The Hofstadter spectrum of these three angles are in $\Lambda_1$, $\Lambda_2$ and $\Lambda_3$ phases of phase diagram, respectively (dashed lines for phase boundaries extrapolated but not explored numerically). }
\label{TopLL}
\end{figure}

We then develop an open momentum space method to calculate the Hofstadter butterfly of the continuum model \cite{bistritzer2011}. We first truncate the continuum model at an open momentum shell enclosing an area of $N_{BZ}$ MBZs, and then substitute $k_\pm$ with $\sqrt{2}\ell_B^{-1}(a,a^\dag)$, after which we diagonalize the Hamiltonian with a LL number cutoff $|a^\dag a|\le N$. Here we take $N_{BZ}=36$ and $N=60$. Fig. \ref{TopLL} (a), (e) and (i) show such spectrum at $\theta=2.4^\circ$, $1.8^\circ$ and $1.0^\circ$ for corrugation $u_0=1$, respectively, where the $x$ axis is the flux per Moir\'e unit cell $\Phi/\Phi_0=\sqrt{3}a_0^2eB/16\pi\hbar\sin^2(\theta/2)$, with $a_0=0.246$nm being the graphene lattice constant. We take the particle-hole symmetric approximation, and only the positive energy spectrum is shown. $\Phi/\Phi_0$ is plotted linearly within $[0,1]$, while is mapped to $2-\Phi_0/\Phi$ when $\Phi/\Phi_0>1$, so that infinite $B$ is mapped to a finite value. One can see outline of the Hofstadter butterfly, and also the so-called in-gap spectral flows \cite{streda1982,wannier1978,asboth2017} due to edge states of the open momentum space area. The spectral flow allows us to determine an in-gap dual Chern number in momentum space $C_K=\frac{1}{N_{BZ}}\frac{d N_{occ}}{d(\Phi_0/\Phi)}$, where $\frac{d N_{occ}}{d(\Phi_0/\Phi)}$ is the number of flowing levels per inverse flux in the Hofstadter gap. The conventionally defined Chern number $C$ can be shown to be related to $C_K$ by $C=\frac{N_{occ}}{N_{BZ}}-\frac{\Phi_0}{\Phi}C_K$, where $N_{occ}$ is the number of levels between CNP and the Hofstadter gap in the calculation (\cite{suppl} Sec. S5). Details of this method will be derived in a separate paper \cite{LLmethod}. Compared to the periodic Hofstadter method for TBG \cite{bistritzer2011a,moon2012}, this method is much faster and converges easily (the Hamiltonian matrix is sparse), and gives accurate enough Hofstadter butterflies as well as spectral flows which tells us the in-gap Chern numbers.

We first set $u_0=1$. At large angles $\theta\gtrsim 1.9^\circ$, the Hofstadter spectrum has a $C=1$ gap from the lowest conduction band extending to higher bands as we expected from Eq. (\ref{Hq}), yielding a connection of Hofstadter butterflies between the lowest band and the second band at $\Phi/\Phi_0=1$ (Fig. \ref{TopLL}(b)). We denote such a spectrum as $\Lambda_1$ phase. This yields a Landau fan at large $B$ in Fig. \ref{TopLL}(d), where the LL filling $\nu=4$ fan ($4$ due to 4-fold degeneracy, $\nu$ corresponds to electron density $n=\nu eB/2\pi\hbar$) passes through the density $n_s$ of fully occupying the lowest conduction band (4 electrons per Moir\'e unit cell) at $\Phi/\Phi_0=1$. This is a unique feature of the fragile topology.

When $1.1^\circ<\theta<1.9^\circ$, the Hofstadter spectrum enters a different phase $\Lambda_2$: the extended $C=1$ gap breaks into two at $\Phi/\Phi_0=1$, giving way to a nonvanishing $C=0$ band gap within $0\le\Phi/\Phi_0<\infty$. 
Our argument of the persistent $C=1$ gap in Eq. (\ref{Hq}) breaks down, because 
$H_p(\mathbf{k})$ is no longer perturbative relative to $H'(\mathbf{k})$. The Hofstadter butterfly in this case is disconnected between the lowest two bands and the second bands at any finite $\Phi/\Phi_0$, ``contradictory" to our expectation from the fragile topology. However, since the Hofstadter spectrum of continuum model is not periodic, we need include the infinite flux point $\Phi/\Phi_0=\infty$ by identifying $+\infty$ with $-\infty$ to define the full ``periodic" Hofstadter butterfly. We then find the Hofstadter butterflies of the lowest two bands and the higher bands in phase $\Lambda_2$ still connected at $\Phi/\Phi_0=\infty$ at zero energy. This is numerically seen in Fig. \ref{TopLL}(e) and (f), and can be analytically proved (\cite{suppl} Sec. S5). The Landau fan changes into Fig. \ref{TopLL}(h): the $\nu=4$ fan is broken at density $n_s$ and arises again afterwards. Note that the crossing of $\nu=4$ fan and density $n_s$ is analogous to that in Fig. \ref{TopLL}(d), as a remaining feature of the fragile topology.

When $\theta<1.1^\circ$, the Hofstadter spectrum reconnects the lowest two bands with higher bands at finite $\Phi/\Phi_0$ (Fig. \ref{TopLL}(j)), which we denote as $\Lambda_3$ phase. The spectrum has both a $C=2$ gap and a $C=1$ gap extending to higher bands from the lowest two bands, leading to a Landau fan as shown in Fig. \ref{TopLL}(l), where both $\nu=4$ and $8$ fans pass through density $n_s$.

For different $\theta$ and corrugation $0\le u_0\le1$, we obtain a phase diagram Fig. \ref{TopLL}(m)(\cite{suppl} Sec. S5). The Hofstadter butterfly of phases $\Lambda_1$ and $\Lambda_3$ is connected at finite $B$, while that of phase $\Lambda_2$ is connected at $B=\infty$.

\begin{figure}[tbp]
\begin{center}
\includegraphics[width=3.4in]{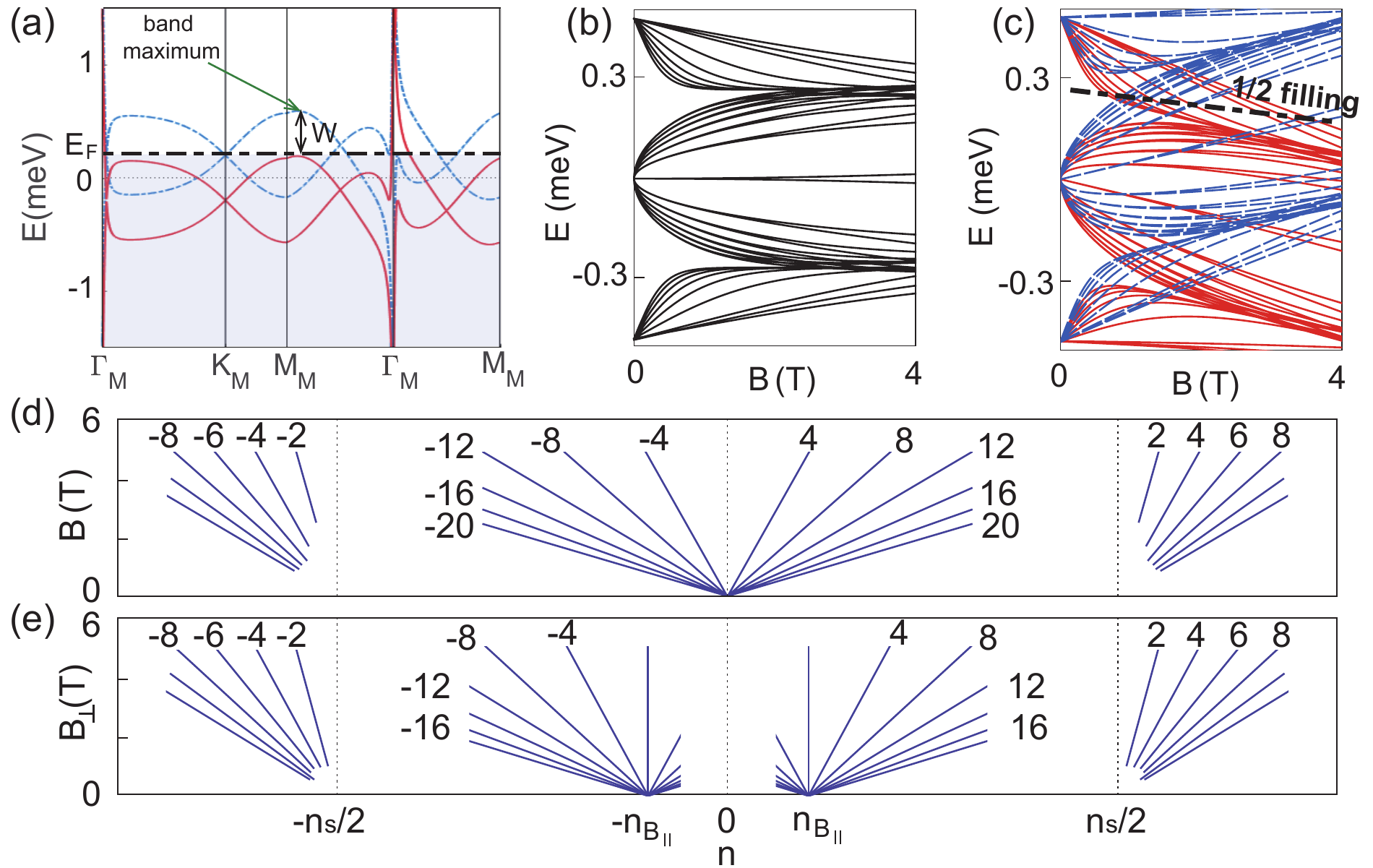}
\end{center}
\caption{(a) The flat bands of $\theta=1.04^\circ$ with a $B\approx 4$T Zeeman energy. The red solid (blue dashed) line represents spin $\uparrow$ ($\downarrow$). (b) The main sets of LLs of the lowest bands without Zeeman energy. (c) LLs with Zeeman energy, where the red solid (blue dashed) line represents spin $\uparrow$ ($\downarrow$) LLs. (d) The Landau fans in an out-of-plane magnetic field $B$. (e) The Landau fans in an out-of-plane magnetic field $B_\perp$ at a fixed in-plane field $B_\parallel>0$.}
\label{halfLL}
\end{figure}

We now study the small $B$ Landau fans for $\theta$ near $\theta_m$ (assuming $v_*\neq0$) with the Zeeman energy $E_Z^{\pm}(B)=\pm\mu_BB$ (for spin $\downarrow,\uparrow$) taken into account.
Fig. \ref{halfLL}(b) shows the main sets of TBG LLs at $\theta=1.04^\circ$ without Zeeman energy based on both quantum and semiclassical calculations (\cite{suppl} Secs. S6, S9). At small $B$, the Dirac LLs around zero energy are 8-fold degenerate.  As $B$ increases ($B\gtrsim3$T), these LLs become 4-fold degenerate due to the mutual hopping between $K_M$ and $K_M'$ (\cite{suppl} Sec. S7A). The LLs at the top (bottom) of Fig. \ref{halfLL}(b) are contributed by the conduction (valence) band maximums (minimums), which are 4-fold degenerate for $B\gtrsim1$T (\cite{suppl} Sec. S6B).


With Zeeman energy, the spin degeneracy is strongly broken (Fig. \ref{halfLL}(c), red solid and blue dashed lines for spin $\uparrow,\downarrow$, respectively), and all the LL degeneracies are further reduced by a factor of $2$.
This yields a dominant Landau fan at $\nu_j=4j$ at small $B$ at the CNP $n=0$ (Fig. \ref{halfLL}(d)), which splits into $\nu_{j'}'=2j'$ at large $B$ ($j,j'\in\mathbb{Z}$).


The strong Zeeman splitting can also give rise to Landau fans at half fillings $n=\pm n_s/2$.
This is because when the Zeeman splitting $E_Z(B)=E_Z^+(B)-E_Z^-(B)=2\mu_BB$ exceeds the conduction (valence) flat band width $W$, the spin $\downarrow$ and $\uparrow$ Dirac points at $K_M,K_M'$ will shift to electron densities $n_0\approx\pm n_s/2$, respectively. For $\theta=1.04^\circ$, this requires a Zeeman field $B\gtrsim4$T (Fig. \ref{halfLL}(a), without orbital effect of $B$ field yet). Therefore, when $B>W/2\mu_B$, one expects to see the spin $\uparrow$ and $\downarrow$ Dirac Landau fans shifted to $n_0\approx\pm n_s/2$, respectively (Fig. \ref{halfLL}(d), see also \cite{suppl} Sec. S7B).

When a tilted magnetic field is added ($B_\perp$ and $B_\parallel$ for out-of-plane and in-plane), the Zeeman energy becomes $E_Z^\pm(B)=\pm\mu_B(B_\perp^2+B_\parallel^2)^{1/2}$, while only $B_\perp$ has orbital effects. For fixed $B_\parallel>0$, at $B_\perp=0$, the opposite spin Dirac points are shifted to energies $\pm\mu_BB_\parallel$, respectively. Therefore, the Landau fan at $n_0=0$ at small $B_\perp$ will split into two fans at $n=\pm n_{B_\parallel}$ as shown in Fig. \ref{halfLL}(e).
When $B_\parallel>W/2\mu_B$, one will have $n_{B_\parallel}\approx n_s/2$, and the Landau fans at $n_0=\pm n_{B_\parallel}$ will merge with those at $n_0=\pm n_s/2$, respectively.

For $\theta$ near $\theta_m$, additional non-high symmetric Dirac points arise between two flat bands \cite{songz2018,hejazi2018}, and our calculation shows their LLs are $2$-fold degenerate for $B\gtrsim0.5$T under Zeeman energy (\cite{suppl} Sec. S6B), and may cross with the approximately $4$-fold Dirac LLs from $K_M,K_M'$. Such crossings will induce a shift of the Landau fan at $n_0=0$ from $\nu_j=4j$ to $\nu_j=4j-2\text{sgn}(j)$ (\cite{suppl} Sec. S7C), Such phenomena is observed recently \cite{yankowitz2018}.




\emph{Discussion.}
We have shown that the Hofstadter butterfly of a set of bands with fragile topology is generically unbounded and connected with other bands. Such a feature is closely related to the non-Wannierizability of topological bands (\cite{suppl} Sec. S3E). In TBG, we identified three phases where the Hofstadter butterflies of the lowest two bands and higher bands are connected differently (Fig. \ref{TopLL}(m)). 
This leads to experimentally testable Landau fans around $\Phi/\Phi_0=1$ (Fig. \ref{TopLL}), which corresponds a magnetic field around $25\theta^2$T ($\theta$ in degrees).
Further, we show the band theory of TBG with Zeeman energy near the magic angle can give Landau fans at both $n_0=0$ and $n_0=\pm n_s/2$, like that observed in experiments \cite{cao2018,yankowitz2018}.
However, we do note that the observed Landau fans at $n_0=\pm n_s/4$ and $\pm 3n_s/4$ \cite{yankowitz2018} cannot be explained by the single-particle picture, and require interactions.

\begin{acknowledgments}
\emph{Acknowledgments.}
We acknowledge helpful conversation with Andrea Young and Steven Kivelson. We also thank Kasra Hejazi, Chunxiao Liu \cite{hejazi2019} and Ya-Hui Zhang for pointing out a misidentified Chern number in the initial version of our work. BL is supported by Princeton Center for Theoretical Science at Princeton University. BB is supported by the Department of Energy Grant No. DE-SC0016239, the National Science Foundation EAGER Grant No. DMR 1643312, Simons Investigator Grants No. 404513, ONR N00014-14-1-0330, and No. NSF-MRSEC DMR-142051, the Packard Foundation, the Schmidt Fund for Innovative Research.

\emph{Note added}. During the preparation of our paper, we noticed the very recently posted Ref. \cite{liuj2018}, from whose model of the topological bands the claim of Chern number of the flat bands at large magnetic field can also be derived in agreement with our results, using a different argument.
\end{acknowledgments}

\bibliography{TBG_ref}

\newpage

\begin{widetext}

\section*{Supplementary Material for ``The Landau Level of Fragile Topology"}

The supplementary material sections are organized as follows:

\begin{itemize}
\item Sec. \ref{SecI}: Review of the continuum model of TBG.
\item Sec. \ref{SecZM}: Discussion of the behavior of the zero mode LLs for two Dirac points with the same/opposite helicity.
\item Sec. \ref{SecTB}: Hofstadter butterfly calculations of the TB41V-model in Ref. \cite{songz2018}, the 10 band tight-binding model in Ref. \cite{po2018b}, and the Chern insulator.
\item Sec. \ref{SecII}: The open momentum space numerical method we used for Hofstadter butterfly and Landau levels of the continuum model.
\item Sec. \ref{SecFlow}: Determination of the in-gap (Hofstadter gap) Chern numbers from the spectral flows in the open momentum space method, and discussion of the Hofstadter spectrum up to infinite magnetic field.
\item Sec. \ref{SecNLL}: Numerical results of small magnetic field low energy LLs for various twist angles around the magic angle.
\item Sec. \ref{SecShift}: Discussion of Landau fans slightly away from the magic angle.
\item Sec. \ref{SecQ}: Discussion of Landau fans exactly at the magic angle.
\item Sec. \ref{SecSemi}: Semiclassical LL calculations for TBG.
\end{itemize}

\section{The continuum model Hamiltonian for TBG}\label{SecI}
The TBG band structure can be calculated using the continuum model constructed in Ref. \cite{bistritzer2011}. The low energy band structure of a graphene monolayer near the graphene BZ $K$ (or $K'$) point is a Dirac fermion with Hamiltonian $h^K(\mathbf{k})=v(k_x\sigma_x-k_y\sigma_y)=\hbar v\bm{\sigma}^*\cdot\mathbf{k}$ (or $h^{K'}(\mathbf{k})=-\hbar v\bm{\sigma}\cdot\mathbf{k}$), where $v$ is the fermi velocity, and $\sigma_{x,y,z}$ are the Pauli matrices for sublattice indices, and momentum $\mathbf{k}$ is measured from the graphene BZ $K$ (or $K'$) point. In a TBG, the Hamiltonian consists of Dirac fermions from the $K$ (or $K'$) point of both layers, and the interlayer hopping between them. We consider the Dirac fermions at the $K$ point of both layers. To the lowest approximation, an electron state at momentum $\mathbf{k}$ in layer $1$ can hop to a state at momentum $\mathbf{p}'$ in layer $2$ if $\mathbf{k}-\mathbf{p}'=\mathbf{q}_{j}$, where
\begin{equation}
\mathbf{q}_1=k_\theta(0,-1)^T\ ,\qquad \mathbf{q}_2=k_\theta\left(\frac{\sqrt{3}}{2},\frac{1}{2}\right)^T\ ,\qquad \mathbf{q}_3=k_\theta\left(-\frac{\sqrt{3}}{2},\frac{1}{2}\right)^T\ ,
\end{equation}
and $k_\theta=|\mathbf{q}_j|=(8\pi/3a_0)\sin(\theta/2)$. Here $a_0=0.246$ nm is the graphene lattice constant, and $\theta$ is the twist angle of TBG. The two reciprocal vectors of the Moir\'e superlattice are given by $\bm{g}_{1}=\mathbf{q}_2-\mathbf{q}_3=k_\theta\left(\sqrt{3},0\right)^T$ and $\bm{g}_{2}=\mathbf{q}_3-\mathbf{q}_1=k_\theta\left(-\frac{\sqrt{3}}{2},\frac{3}{2}\right)^T$. To distinguish Moir\'e high symmetry points from the high symmetry points ($K$, $K'$, $M$, $\Gamma$) of the graphene BZ (GBZ), we denote the high symmetry points in the Moir\'e BZ (MBZ) as $K_M$, $K_M'$, $M_M$, $\Gamma_M$, respectively. The TBG Hamiltonian takes the form \cite{bistritzer2011}
\begin{equation}\label{SeqH}
H^{K}(\mathbf{k})=\left(\begin{array}{ccccc}
h^{K}_{\theta/2}(\mathbf{k})& wT_1& wT_2&wT_3&0\\
wT_1^\dag&h^{K}_{-\theta/2}(\mathbf{k}-\mathbf{q}_1)&0&0&\cdots\\
wT_2^\dag&0&h^{K}_{-\theta/2}(\mathbf{k}-\mathbf{q}_2)&0&\cdots\\
wT_3^\dag&0&0&h^{K}_{-\theta/2}(\mathbf{k}-\mathbf{q}_3)&\cdots\\
0&\vdots&\vdots&\vdots&\ddots\\
\end{array}\right)\ ,
\end{equation}
where the momentum $\mathbf{k}$ is measured from $K_M$ point of the MBZ, $h^{K}_{\pm\theta/2}(\mathbf{k})$ is the Dirac Hamiltonian rotated by angle $\pm\theta/2$ ($\pm$ for layer $1$ and layer $2$, respectively):
\begin{equation}\label{hk}
h^{K}_{\pm\theta/2}(\mathbf{k})=\hbar v\left(\begin{array}{cc}0&e^{\mp i\theta/2}(k_x+ik_y) \\ e^{\pm i\theta/2}(k_x-ik_y)&0\end{array}\right)\ ,
\end{equation}
$w=110$ meV is the interlayer hopping amplitude, and $T_j$ ($1\le j\le3$) are given by
\[T_1=u_0\mathbf{1}_2+\sigma_x\ ,\qquad T_2=u_0\mathbf{1}_2-\frac{1}{2}\sigma_x-\frac{\sqrt{3}}{2}\sigma_y\ ,\qquad T_3=u_0\mathbf{1}_2-\frac{1}{2}\sigma_x+\frac{\sqrt{3}}{2}\sigma_y\ .\]
Here $\mathbf{1}_2$ is the $2\times2$ identity matrix, and $u_0\in[0,1]$ characterizes the corrugation and relaxation of the Carbon atoms in the Moir\'e lattice. Roughly speaking, $u_0$ is the ratio between the effective interlayer hopping at $AA$ stackings and that at $AB$ ($BA$) stackings. $u_0=1$ corresponds to no corrugation/relaxation, which is adopted in the original model of \cite{bistritzer2011}. In realistic devices, $u_0$ is generically slightly smaller than $1$.

This model can be understood as a "tight-binding model" in a momentum space lattice at positions $\mathbf{k}+\mathbf{Q}_m$ ($m\in\mathbb{Z}$), where $\mathbf{Q}_m$ are the honeycomb momentum lattice sites (with hexagon edge length $k_\theta$) as shown in Fig. \ref{Mhop}, with blue solid and red hollow sites from layer $1$ and layer $2$, respectively. The matrix $T_j$ is the (interlayer) hopping between two sites $\mathbf{Q}_m$ and $\mathbf{Q}_{m'}$ (differing by momentum $\mathbf{q}_j$), while $h^{K}_{\pm\theta/2}(\mathbf{k}-\mathbf{Q}_j)$ ($\pm$ for layer $1$, $2$) serve as the on-site energies of this momentum space "tight-binding model". Each honeycomb plaquette is exactly a Moir\'e BZ. We calculate the band structures of TBG by truncating the momentum space honeycomb lattice at a certain momentum away from $\mathbf{k}=0$, which encloses a momentum space area $\mathcal{A}_k=N_{BZ}\Omega_{BZ}$ as shown in Fig. \ref{Mhop}. Here $\Omega_{BZ}$ is the area of the Moir\'e BZ, and $N_{BZ}$ is the number of plaquettes we keep. Then we diagonalize the Hamiltonian restricted in the momentum space area $\mathcal{A}_k$. As long as $N_{BZ}$ is large enough, we can obtain accurate low energy band structures.

\begin{figure}[htbp]
\begin{center}
\includegraphics[width=3in]{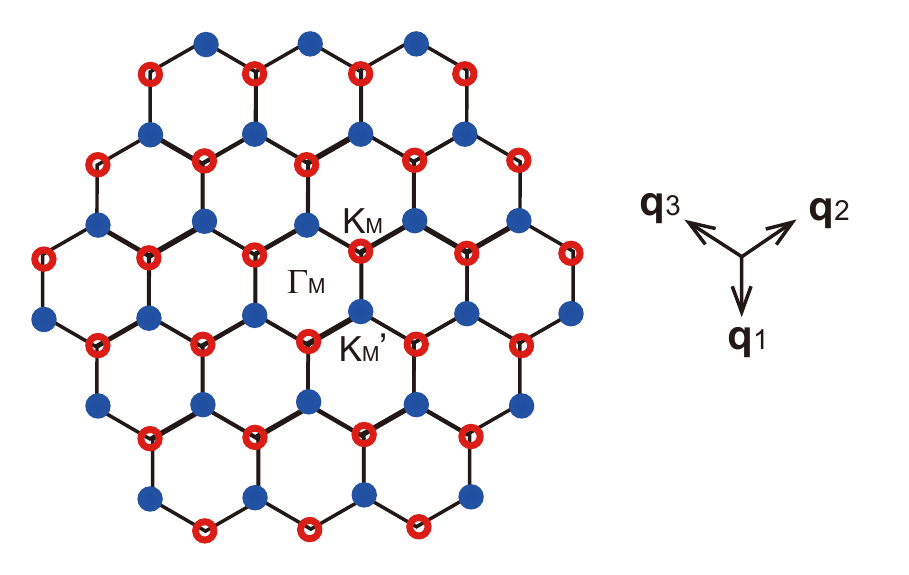}
\end{center}
\caption{The continuum model can be viewed as a ``tight binding model" on the momentum space honeycomb lattice $\mathbf{k}-\mathbf{Q}_m$ shown here. The red hollow and blue solid sites originate from layer $1$ and $2$, respectively. Our numerical calculation is done by cutting off the momentum space lattice at a finite momentum away from $\mathbf{k}=0$, so the area of the lattice we keep is finite, with an area $\mathcal{A}_{k}=N_{BZ}\Omega_{BZ}$.}
\label{Mhop}
\end{figure}

The low energy bands at $K_M$ and $K_M'$ points of MBZ near zero energy are still Dirac fermions.
Taking into account the spin and graphene valley degeneracy, there are $4$ Dirac fermions at $K_M$ and $4$ Dirac fermions at $K_M'$ in total in the MBZ, each of which has an effective Hamiltonian
\begin{equation}
H_{\text{eff}}^{\eta}(\mathbf{k})=\hbar v_*(\eta k_x\sigma_x-k_y\sigma_y)\ ,
\end{equation}
where $\eta=\pm1$ for graphene valleys $K$ and $K'$, respectively, and the Fermi velocity $v_*$ is a function of twist angle $\theta$.

\begin{figure}[htbp]
\begin{center}
\includegraphics[width=7in]{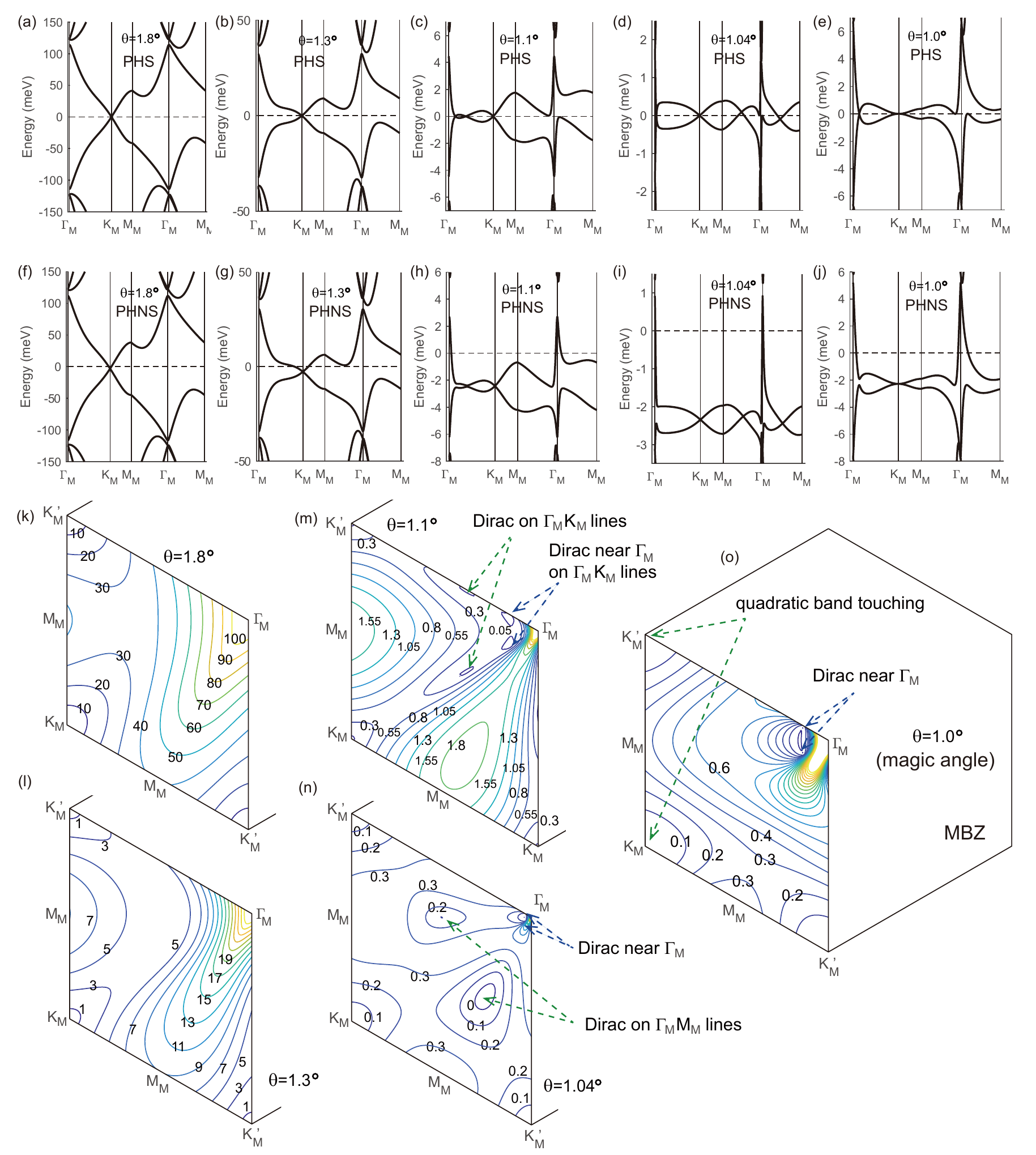}
\end{center}
\caption{(a)-(j) The high symmetry line plot of lowest two Moir\'e bands for various twist angles. The bands with the zero twist angle approximation (thus particle-hole symmetric (PHS)) are shown in (a) $\theta=1.8^\circ$ ($\alpha=0.336$), (b) $\theta=1.3^\circ$ ($\alpha=0.466$), (c) $\theta=1.1^\circ$ ($\alpha=0.551$), (d) $\theta=1.04^\circ$ ($\alpha=0.583$) and (e) magic angle $\theta=\theta_m=1.0^\circ$ ($\alpha=0.605$). The bands without the zero twist angle approximation (thus particle-hole non-symmetric (PHNS)) are shown in (f) $\theta=1.8^\circ$, (g) $\theta=1.3^\circ$, (h) $\theta=1.1^\circ$, (i) $\theta=1.04^\circ$ and (j) magic angle $\theta=\theta_m=1.0^\circ$. In (e) and (j) at the magic angle, one can see the Dirac point at $K_M$ becomes quadratic. (k)-(o) The energy contour plot of the lowest conduction band (PHS) for (k) $\theta=1.8^\circ$, (l) $\theta=1.3^\circ$, (m) $\theta=1.1^\circ$, (n) $\theta=1.04^\circ$ and (o) magic angle $\theta=\theta_m=1.0^\circ$. The numbers on the curves are energies in units of meV. The additional Dirac points not at $K_M$, $K_M'$ are labeled, which exist in (m)-(o).}
\label{band}
\end{figure}


In the $u_0=1$ case, the lowest perturbation theory shows the Dirac velocity $v_*=\frac{1-3\alpha^2}{1+6\alpha^2}v$ at $K_M$ and $K_M'$ points vanishes at $\alpha=1/\sqrt{3}$, where $\alpha=w/\hbar vk_\theta$, and the corresponding twist angle $\theta_M$ is defined as the first magic angle. More rigorous numerical calculation shows $v_*=0$ is achieved at $\alpha=0.605$. Here we adopt the graphene parameters $\hbar v=611$meV$\cdot$nm and $w=110$meV, under which $\theta_m=1.0^\circ$ ($\alpha=0.605$).

The Moir\'e band spectrum of TBG is approximately particle-hole symmetric (PHS) for small twist angles $\theta$, largely due to the particle-hole symmetry of the graphene Dirac Hamiltonian $h^K(\mathbf{k})$. In fact, if one \cite{bistritzer2011} approximates the $\pm\theta/2$ in $h^K_{\pm\theta/2}(\mathbf{k})$ as zero, the TBG bands becomes exactly particle-hole symmetric (PHS) \cite{songz2018}. Such an approximation is legitimate when $\theta$ is small. More explicitly, under the zero angle approximation, the Hamiltonian (\ref{SeqH}) satisfies
\begin{equation}
H^{K}(\mathbf{k})=-U^\dag H^{K}(\mathbf{q}_1-\mathbf{k})U\ ,
\end{equation}
where $U=i\tau_y\otimes\mathbf{1}$, with $\tau_{x,y,z}$ the Pauli matrices in the layer ($1$ and $2$) basis. Therefore, the band energies of $H^{K}(\mathbf{k})$ at momentum $\mathbf{k}$ is exactly opposite to those at momentum $\mathbf{q}_1-\mathbf{k}$. Since momentum $\mathbf{k}$ is measured from the MBZ $K_M$ point, the two momentums $\mathbf{k}$ and $\mathbf{q}_1-\mathbf{k}$ are opposite about $\Gamma_M$ point (up to superlattice reciprocal vectors). This indicates the band structure is PHS.

Fig. \ref{band} (a)-(j) shows the high-symmetry-line plot of the lowest two Moir\'e bands for angles ranging from $1.8^\circ$ to the magic angle $1.0^\circ$ with and without the zero twist angle approximation, respectively, which are calculated by taking momentum space cutoff $N_{BZ}=36$. They are not very different (PHS/PHNS) except for a shift of the charge neutral point in energy. Fig. \ref{band} (k)-(o) show the energy contour plots of the lowest conduction Moir\'e band in the PHS case. In the PHS band structure and for $\theta=1.1^\circ$ (Fig. \ref{band} (c) and (m)) and $\theta=1.04^\circ$ (Fig. \ref{band} (d) and (n)), one can see additional Dirac points on the high symmetry lines away from $K_M$ \cite{songz2018}. When the zero twist angle approximation is not used, the Dirac points on the $\Gamma_MK_M$ line deviates from the $\Gamma_MK_M$ line (due to the loss of PHS). This can be seen in Fig. \ref{band} (h) and (i), where the $\Gamma_MK_M$ line becomes gapped (not at the $K_M$ point). For the magic angle, three of the additional Dirac points merge with the Dirac point at $K_M$, and the dispersion at $K_M$ becomes quadratic (Fig. \ref{band} (e) and (j)).

In most of our discussions in the main text, we adopt the zero twist angle approximation so that the electron bands and LLs are PHS, which qualitatively does not change the low energy physics. The particle-hole non-symmetric (PHNS) spectrum and LL results are mainly shown and discussed in Sec. \ref{SecNLL}. 

\section{Behavior of zero mode LLs of Dirac points with the same helicity}\label{SecZM}

In the main text Eq. (2) and Eq. (3), we have discussed two toy models, one with two Dirac points of opposite helicities and one with two Dirac points of the same helicity, respectively. In this section, we show the zero mode LLs in the model with two Dirac points of the same helicity (main text Eq. (3)) are more robust than those of opposite helicities.

In the first model of main text Eq. (2) with two opposite helicity Dirac points, the Hamiltonian is $H(\mathbf{k})=A[(k_+-k_D)(k_-+k_D)\sigma_++h.c.]$. To calculate LLs, we substitute $(k_+,k_-)$ by $\sqrt{2}\ell_B^{-1}(a,a^\dag)$ (Eq. (\ref{kplus})), after which the Hamiltonian becomes
\begin{equation}\label{Hpm}
H=A\left\{\left[(\sqrt{2}\ell_B^{-1}a^\dag+k_D)(\sqrt{2}\ell_B^{-1}a-k_D)+\ell_B^{-2}\right]\sigma_++ \left[(\sqrt{2}\ell_B^{-1}a^\dag-k_D)(\sqrt{2}\ell_B^{-1}a+k_D)+\ell_B^{-2}\right]\sigma_-\right\}\ .
\end{equation}
Note that $k_+k_-\rightarrow \ell_B^{-2}(2a^\dag a+1)$. At small magnetic fields, the zero mode LLs of the two Dirac points are given by $|\psi_{1}\rangle=\left(|\Omega_-\rangle,0\right)^T$ and $|\psi_{2}\rangle=\left(0,|\Omega_+\rangle\right)^T$, which satisfy $(\sqrt{2}\ell_B^{-1}a+k_D)|\psi_{1}\rangle=0$ and $(\sqrt{2}\ell_B^{-1}a-k_D)|\psi_{2}\rangle=0$, respectively, where $|\Omega_\pm\rangle=e^{\pm\frac{k_D\ell_B}{2}a^\dag}|0\rangle$. Under the Hamiltonian ($\ref{Hpm}$), there will be a hopping between $|\psi_{1}\rangle$ and $|\psi_{2}\rangle$ given by
\begin{equation}
t_{12}=\frac{\langle\psi_{1}|H|\psi_{2}\rangle}{\sqrt{\langle\psi_{1}|\psi_{1}\rangle \langle\psi_{2}|\psi_{2}\rangle}}=\frac{A\ell_B^{-2}\langle0|e^{-\frac{k_D\ell_B}{2}a} e^{\frac{k_D\ell_B}{2}a^\dag}|0\rangle}{\sqrt{\langle0|e^{-\frac{k_D\ell_B}{2}a} e^{-\frac{k_D\ell_B}{2}a^\dag}|0\rangle\langle0|e^{\frac{k_D\ell_B}{2}a} e^{\frac{k_D\ell_B}{2}a^\dag}|0\rangle}}=A\ell_B^{-2}e^{-k_D^2\ell_B^2}\ .
\end{equation}
Therefore, the two zero mode LLs in this model will split into energies $E_{0,\pm}\approx\pm A \ell_B^{-2}e^{-k_D^2\ell_B^2}$. When $k_D\ell_B\sim 1$, the energy splitting will be large, and comparable to the van Hove singularity energy which is of order $Ak_D^2$ (in graphene or TBG, $k_D$ is about the size of the Brillouin zone), so one expect them to merge with other Hofstadter bands around the van Hove singularities.

In the second model of main text Eq. (3) of Dirac points of the same helicity, the Hamiltonian $H'(\mathbf{k})=A[(k_+^2-k_D^2)\sigma_++(k_-^2-k_D^2)\sigma_-]$ under magnetic field becomes
\begin{equation}\label{Hpp}
H'(\mathbf{k})=A\left[(a-k_D)(a+k_D)\sigma_++(a^\dag-k_D)(a^\dag+k_D)\sigma_-\right]\ .
\end{equation}
One can then easily verify $\psi_\pm=\left(0,|\Omega_\pm\rangle\right)^T$ are two exact zero energy eigenstates of Hamiltonian (\ref{Hpp}), independent of $k_D\ell_B$ (where we have defined $|\Omega_\pm\rangle=e^{\pm\frac{k_D\ell_B}{2}a^\dag}|0\rangle$ as before). Therefore, the two zero mode LLs in this model are more robust.

We now consider adding a perturbation, for instance $H_{p}(\mathbf{k})=\epsilon(\mathbf{k})\sigma_z$, to the Hamiltonians of the two models, where $\epsilon(\mathbf{k})\ge0$ vanishes at the two Dirac points. We also assume the momentum $\mathbf{k}$ is bounded in the Brillouin zone (BZ), so that $\epsilon(\mathbf{k})$ is energetically bounded. We note that the two-band toy model $H'(\mathbf{k})$ and the form of perturbation $H_{p}(\mathbf{k})=\epsilon(\mathbf{k})\sigma_z$ we consider here are not meant to obey all the symmetries of the one-valley TBG continuum model ($C_{2x}$, $C_{2z}T$, $C_{3z}$, etc), since it is impossible to write down a two-band model for the lowest two TBG bands continuously defined in the entire MBZ (which would indicate a well defined real space Wannier function) due to the fragile topology (non-Wannierizable). We only take the perturbation $H_{p}(\mathbf{k})$ here as a simplest example to illustrate the robustness of the zero mode LLs of the toy model $H'(\mathbf{k})$.

One can think of $H_{p}(\mathbf{k})$ by itself as a model of two trivial bands with dispersions $\pm\epsilon(\mathbf{k})$, respectively, and the two bands touch at zero energy at the positions of the two Dirac points $(k_x,k_y)=(\pm k_D,0)$. In the small $B$ field (large $\ell_B$) limit, the states $\left(0,|\Omega_\pm\rangle\right)^T$ and $\left(|\Omega_\pm\rangle,0\right)^T$ are simply the lowest LLs from the band touching points $(k_x,k_y)=(\pm k_D,0)$ of such a model $H_{p}(\mathbf{k})$. If we consider $H_{p}(\mathbf{k})$ by itself as a 2-band model, the LLs contributed by its touching points in between the two bands will generically be obstructed by the van Hove singularities of the two bands. Namely, if a LL state $|\psi\rangle$ is from the band touching points $(k_x,k_y)=(\pm k_D,0)$, its ``energy" $\langle\psi|H_{p}(\hat{\mathbf{k}}) |\psi\rangle$ ($\hat{\mathbf{k}}$ stands for the momentum operator expressed in terms of $a$, $a^\dag$) in the model $H_{p}(\mathbf{k})$ will be bounded between $[-\epsilon_V,\epsilon_V]$, where $\epsilon_V$ is the van Hove singularity energy of the conduction band $\epsilon(\mathbf{k})$ of model $H_{p}(\mathbf{k})$. This is generically true: the LLs from the band edges (band maximum, minimum or band touching points) are obstructed by the neighbouring van Hove singularities. This is because semiclassically, the $n$-th LL from a band edge correspond to a constant energy contour in the $\mathbf{k}$ space around the band edge enclosing a quantized area $\sim 2\pi n/\ell_B^2$, while the van Hove singularity is the energy at which the area of the constant energy contour tends to infinity (see Sec. \ref{SecSemi}).

Now we turn to the actual models we study. In the second model (Dirac points with the same helicity), the full Hamiltonian with perturbation is $H'(\mathbf{k})+H_p(\mathbf{k})$. If we restrict ourselves in the subspace of the two zero modes $\psi_\pm=\left(0,|\Omega_\pm\rangle\right)^T$, the term $H'(\mathbf{k})$ becomes zero, and the energies of $\psi_\pm$ will be solely contributed by $H_p(\mathbf{k})$. Therefore, by the argument of last paragraph, the energies of $\psi_\pm$ will be bounded within the energy range $[-\epsilon_V,\epsilon_V]$. However, the energy dispersion of the full Hamiltonian $H'(\mathbf{k})+H_p(\mathbf{k})$ at zero magnetic field is $E(\mathbf{k})=\sqrt{\epsilon(\mathbf{k})^2+A^2|k_+^2-k_D^2|^2}$, so the actual van Hove singularity energy is around $\sim\sqrt{\epsilon_V^2+A^2k_D^4}$, which is greater than $\epsilon_V$. Therefore, the zero mode LLs $\psi_\pm$ in this model with two Dirac points of the same helicity cannot reach the van Hove singularities $\sim\pm\sqrt{\epsilon_V^2+A^2k_D^4}$, and we expect they have a $C=\pm1$ gap with all the other Hofstadter bands, as long as $H_p(\mathbf{k})$ can be treated as a perturbation. In TBG, $k_D$ is the size of the MBZ $k_\theta$, which is proportional to the twist angle $\theta$. Since the energy scale of $H'(\mathbf{k})\sim Ak_D^2$ is larger at larger $\theta$, we expect this perturbation argument of the existence of a persistent $C=\pm1$ gap to hold for large enough $\theta$, where the effective $H_{p}(\mathbf{k})$ is perturbative. This is indeed the case, as we have seen in the main text Fig. 3(m), the large angle phase $\Lambda_1$ has a persistent $C=\pm1$ gap extending to higher bands.

In contrast, in the first model with opposite helicity Dirac points, the two zero mode LLs already split in energy without the perturbation. In the perturbed Hamiltonian $H(\mathbf{k})+H_p(\mathbf{k})$, both terms will contribute to the energies of the two zero mode LLs. When $k_D\ell_B$ is of order $1$, the energies of the two zero mode LLs are comparable to the van Hove singularity energy, and there is no restriction which forbids the zero mode LLs to reach the van Hove singularities.

\section{Two effective tight binding models of TBG in magnetic field, and the example of Chern insulator in magnetic field}\label{SecTB}
In this section, we present the Hofstadter butterfly calculations of two tight binding models of TBG given in Refs \cite{songz2018,po2018b}, respectively, which correctly reproduce the nontrivial topology of the lowest two flat bands. In particular, we show that the Hofstadter butterfly of nontrivial fragile topology is generically connected with other bands in both models, which confirms our expectation in the main text.

\subsection{The TB4-1V model in the magnetic field}
In Ref.\cite{songz2018} the authors proposed a four-band one-valley (TB4-1V) effective tight binding model of TBG.
In this model there are $s$ and $p_z$ orbitals are situated on each honeycomb sites (or the AB and BA stacking positions of the Moir\'{e} lattice.) We use Pauli matrices $\sigma_i$ to represent the Moir\'{e} A and B sublattices, and Pauli matrices $\mu_i$ for the $s$ and $p_z$ orbitals ($i=x,y,z$). The symmetry operators $C_{2x}$, $C_{2z}T$ and $C_{3z}$ under this basis can be written as $C_{2x} = \mu_z$, $C_{2z}T = \sigma_x K$, $C_{3z} = 1$,
These symmetries give the constraints on the form of the Hamiltonian; for example, the term $\mu_x\sigma_0$ is banned by the symmetry $C_{2x}$. The Hamiltonian of this model is given by
\begin{align}\label{H-TB41V}
	H^{TB4-1V}(\mathbf{k}) =& \Delta\mu_z\sigma_0 + \mu_0\sigma_x \sum_{i=1}^3[t\cos(\bm{\delta}_i\cdot \mathbf{k})+t' \cos(-2\bm{\delta}_i\cdot \mathbf{k})] \nonumber\\
	-&\mu_0\sigma_y\sum_{i=1}^3[t \sin(\bm{\delta}_i\cdot \mathbf{k})+t' \sin(-2\bm{\delta}_i\cdot \mathbf{k})]-2\lambda \mu_y\sigma_z \sum_{i=1}^3\sin(\bm{d}_i\cdot \mathbf{k})\,,
\end{align}
As shown in Fig. \ref{TB41Vhopping}(e), the model includes a nearest neighbour (differing by displacement vectors $\bm{\delta}_i$) hopping $t$ and a 3rd nearest neighbour (differ by displacement vectors $-2\bm{\delta}_i$) hopping $t'$ between two $s$ orbitals or two $p$ orbitals, and a 2nd nearest neighbour (differ by displacement vectors $\bm{d}_i$) hopping $\lambda$ between an $s$ orbital and a $p$ orbital. We follow the definition of basis vectors
(the nearest neighbour vectors $\bm{\delta}_i$ and the second nearest neighbour vectors $\bm{d}_i$) in Ref. \cite{songz2018}.

A numerical calculation in TBG \cite{Laissardiere2010} shows that the charge density is mainly localized around the AA stacking region. However, because the irreducible representation of the lowest two Moir\'e bands cannot be written as the positive sum of Elementary Band Representations \cite{bradlyn2017}, the tight binding model is constructed by $4$ bands (two sets of topological bands) whose orbitals are located on either AB and BA stacking positions. This indicates the $s$ and $p_z$ orbital wave functions should have the shape as shown in Fig. \ref{TB41Vhopping}(a).

To study the LLs or Hofstadter butterfly of the TB4-1V model, we first review how the tight binding model is coupled to the gauge field $\bm{A}(\bm{r})$ (Peierls substitution). Since the orbitals are rather extended instead of well localized on AB or BA stacking sites, and have charge densities dominantly away from AB or BA stacking sites, we need to carefully reconsider the gauge field coupling of their hoppings. Suppose the local orbital Wannier functions are given by $W_{\alpha}(\bm{r}-\bm{R}_i)$. The hopping between $W_{\bm{R}_i}$ and $W_{\bm{R}_j}$ without a magnetic field is given by integral
\begin{equation}
 	t_{ij} = \int d^d \bm{r} W^*_{\bm{R}_i}(\bm{r})\hat{H}W_{\bm{R}_j}(\bm{r})\ ,
\end{equation}
where the operator $\hat{H}$ is the bare Hamiltonian for electron $\hat{H}=\frac{-\nabla^2}{2m}+U(\bm{r})$, the potential energy $U(\bm{r})$ is periodic with period given by the Bravais lattice, and $m$ is the electron mass. In the presence of a gauge field, the Hamiltonian becomes
$$
\widetilde{H} = \frac{(-i\nabla+e\bm{A}(\bm{r}))^2}{2m} +U(\bm{r})\ .
$$
We define a set of new local Wannier basis as
$$
\widetilde{W}_{\bm{R}}(\bm{r}) = \exp\left(-ie\int_{\bm{R}}^{\bm{r}}d\bm{r}'\cdot\bm{A}(\bm{r}')\right)W_{\bm{R}}(\bm{r})\ ,
$$
where $\bm{r}'=\bm{R}+\lambda(\bm{r}-\bm{R})$ in the integral is along the straight line between $\bm{R}$ and $\bm{r}$ ($\lambda\in[0,1]$). We can then evaluate the hopping matrix elements of the Hamiltonian in the gauge field under this new basis as $\widetilde{t}_{ij} = \langle \widetilde{W}_i | \widetilde{H}|\widetilde{W}_j\rangle$. The reason we define the hopping $\widetilde{t}_{ij}$ in this way (as opposed to, say $\widetilde{t}_{ij} = \langle W_i | \widetilde{H}|W_j\rangle$) is that, when $\bm{A}(\bm{r})=\nabla f(\mathbf{r})$ is a pure gauge transformation, $\widetilde{W}_{\bm{R}}(\bm{r})$ and $\widetilde{H}$ will simply be the gauge transformation of the original Wannier function $W_{\bm{R}}(\bm{r})$ and Hamiltonian $H$, and one will find $\widetilde{t}_{ij}=t_{ij}e^{ie[f(\bm{R}_j)-f(\bm{R}_i)]}$ as expected (In contrast, $\langle W_i | \widetilde{H}|W_j\rangle$ is not a gauge transformation of $t_{ij}$).

When the Wannier function $W_{\bm{R}}(\bm{r})$ has a size far smaller than the unit cell size and the magnetic length $\ell_B$, and is well localized at site $\bm{R}$, following the derivation given by Luttinger \cite{Luttinger1951}, one can obtain the Peierls substitution in the textbook $\widetilde{t}_{ij} = t_{ij}\exp\left({ie\int_{\bm{R}_i}^{\bm{R}_j}\bm{A}\cdot d\bm{l}}\right)$. In contrast, when the Wannier function size is not small or has a delocalized shape, we need a better approximation than the Peierls substitution. First, note that
\begin{align}
	\nabla \widetilde{W}_{\bm R}(\bm{r}) &= \nabla \left(e^{-ie\int_{\bm R}^{\bm r}\bm{A}\cdot d\bm{r}'}W_{\bm R}\right) =e^{-ie\int_{\bm R}^{\bm r}\bm{A}\cdot d\bm{r}'}\left[\nabla_{\bm{r}} -ie\left(\nabla_{\bm r}\int_{\bm R}^{\bm r}\bm{A}(\bm{r}')\cdot d\bm{r}' \right)\right]W_{\bm R}(\bm{r}) \nonumber\\
	&= e^{-ie\int_{\bm R}^{\bm r}\bm{A}\cdot d\bm{r}'} \left[\nabla_{\bm{r}} - ie\left(\nabla_{\bm r}\int_0^1d\lambda (\bm{r}-\bm{R})\cdot \bm{A}(\bm{R}+\lambda(\bm{r}-\bm{R}))\right)\right]W_{\bm{R}}(\bm{r})\label{eqn:dwtilde}\ .
\end{align}
Using the identity $\nabla(\bm{X}\cdot\bm{Y}) = (\bm{X}\cdot \nabla)\bm{Y} + (\bm{Y}\cdot \nabla)\bm{X} + \bm{X}\times(\nabla \times \bm{Y})+\bm{Y}\times(\nabla \times \bm{X})$, one can rewrite the second term in Eq. (\ref{eqn:dwtilde}) as (recall that $\bm{r}'=\bm{R}+\lambda(\bm{r}-\bm{R})$):
\begin{align}
	&\nabla_{\bm r}\int_0^1d\lambda (\bm{r}-\bm{R})\cdot \bm{A}(\bm{R}+\lambda(\bm{r}-\bm{R}))\nonumber\\
	=&\int_0^1 d\lambda \left\{((\bm{r}-\bm{R})\cdot \nabla_{\bm r})\bm{A}(\bm{r}') + (\bm{A}(\bm{r}')\cdot \nabla_{\bm r})(\bm{r}-\bm{R})+(\bm{r}-\bm{R})\times (\nabla_{\bm{r}}\times \bm{A}(\bm{r}'))+\bm{A}(\bm{r}')\times (\nabla_{\bm r}\times(\bm{r}-\bm{R}))\right\}\nonumber\\
=& \int_0^1 d\lambda \left\{\bm{A}(\bm{r}') + \lambda \left[(\bm{r}-\bm{R})\cdot \nabla_{\bm{r}'}\right]\bm{A}(\bm{r}') + \lambda (\bm{r}-\bm{R})\times (\nabla_{\bm{r}'}\times \bm{A}(\bm{r}')) \right\}\nonumber\\
=& \int_0^1 d\lambda \left\{\frac{d}{d\lambda}\left[\lambda\bm{A}(\bm{r}')\right] + \lambda (\bm{r}-\bm{R})\times (\nabla_{\bm{r}'}\times \bm{A}(\bm{r}')) \right\}
= \bm{A}(\bm{r}) + \int_0^1d\lambda\,\lambda(\bm{r}-\bm{R})\times (\nabla_{\bm{r}'}\times \bm{A}(\bm{r}'))\ ,
\label{eqn:dwtilde2}
\end{align}
where we have used the facts $\nabla_{\bm r} \times (\bm{r}-\bm{R}) = 0$, $\nabla_{\bm r}(\bm{r}-\bm{R}) =1$ and $\nabla_{\bm{r}} \bm{A}(\bm{r}') = \lambda \nabla_{\bm{r}'} \bm{A}(\bm{r}')$.
Since we are only interested in uniform magnetic fields, we have $\nabla_{\bm{r}'} \bm{A}(\bm{r}') = \bm{B}$ independent of coordinate $\bm{r}'$, so we find
\begin{equation}\label{eqn:derivativewl}
\nabla_{\bm{r}}\int_{\bm{R}}^{\bm r}d\bm{r}'\cdot \bm{A}(\bm{r}') = \bm{A}(\bm{r}) + \frac{1}{2}(\bm{r}-\bm{R})\times \bm{B}\,.
\end{equation}
With Eq. (\ref{eqn:derivativewl}), we can then evaluate the hopping matrix elements in the gauge field as:
\begin{align}
	\widetilde{t}_{ij}&=\int d^d\bm{r} \widetilde{W}_{\bm{R}_i}^*(\bm{r})\left[\frac{-(\nabla+ie\bm{A})^2}{2m}+U(\bm{r})\right] \widetilde{W}_{\bm{R}_j}(\bm{r})\nonumber\\
	&=\int d^d\bm{r}\exp\left(ie\int_{\bm{r}'\in c_{ij,\bm{r}}} d\bm{r}'\cdot \bm{A}(\bm{r}')\right) W^*_{\bm{R}_i}(\bm{r})\left\{\frac{-\left[\nabla + \frac i2 e(\bm{r}-\bm{R}_j)\times\bm{B}\right]^2}{2m}+U(\bm{r})\right\}W_{\bm{R}_j}(\bm{r})\ ,\label{eqn:ttilde}
\end{align}
where the integral in the exponent is done along contour $c_{ij,\bm{r}}$ which goes from $\bm{R}_{i}$ to $\bm{r}$ and then to $\bm{R}_j$ in straight lines. The main contributions to $\widetilde{t}_{ij}$ come from positions $\mathbf{r}$ where $W_{\bm{R}_i}(\bm{r})$ and $W_{\bm{R}_j}(\bm{r})$ have a large overlap probability.

\begin{figure}[htbp]
\begin{center}
\includegraphics[width=6in]{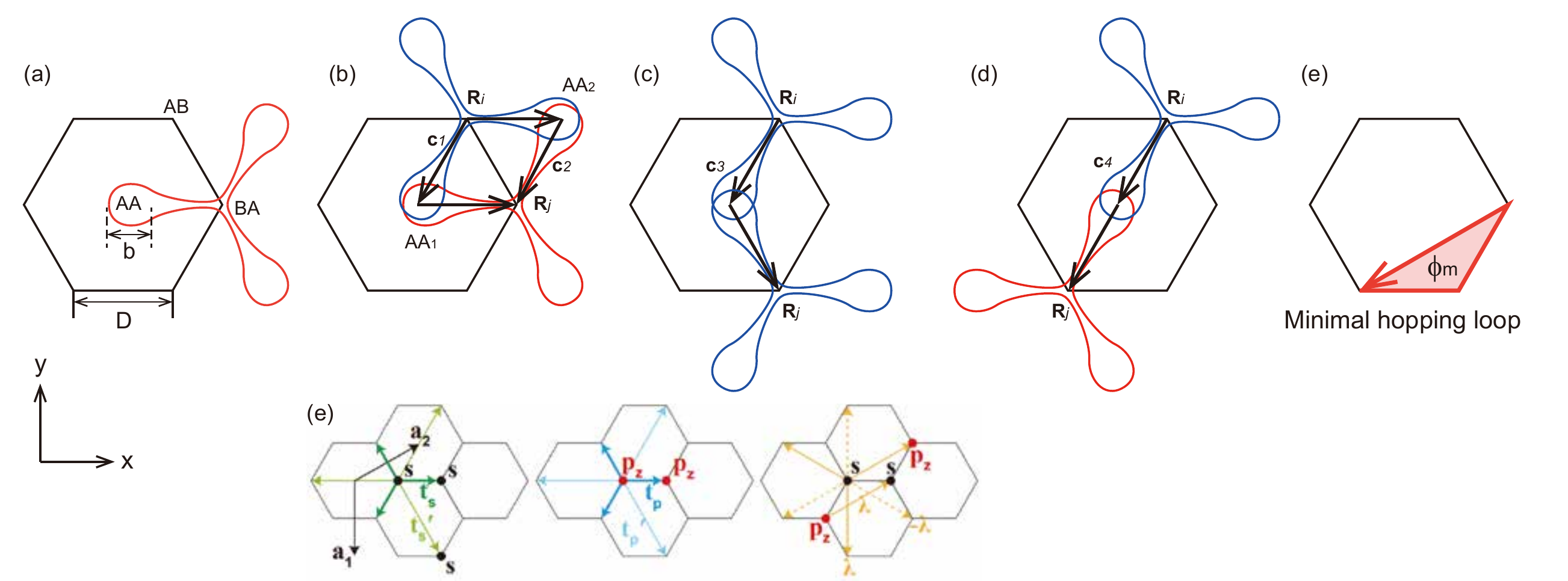}
\end{center}
\caption{(a) The shape of the local orbital Wannier wave function, of which the probability density is localized at AA positions. (b) The nearest neighbour hopping of Wannier functions. (c) The second nearest neighbour hopping. (d) The third nearest neighbour hopping. (e) Illustration of the hopping parameters in the TB4-1V model (from Ref. \cite{songz2018}).}
\label{TB41Vhopping}
\end{figure}

In the TB4-1V model, the amplitude of Wannier function $W_{\bm{R}_i}(\bm{r})$ is concentrated at $AA$ stacking positions within a small size $b$ as shown in Fig. \ref{TB41Vhopping}(a). Therefore, $\widetilde{t}_{ij}$ in Eq. (\ref{eqn:ttilde}) are dominated by $\mathbf{r}$ near $AA$ stacking positions, are illustrated in the Fig. \ref{TB41Vhopping}(b)-(d). In particular, when $\mathbf{r}$ is near an $AA$ stacking, the gradient operator $\nabla$ in Eq. (\ref{eqn:ttilde}) is of order $1/b$ ($b$ is the diameter at $AA$ stacking within which the charge density is concentrated, see Fig. \ref{TB41Vhopping}(a)), while the term $e(\bm{r}-\bm{R}_j)\times \bm{B}$ is of order $D/\ell_B^2$, where $D$ is the edge length of the hexagonal Moir\'e unit cell (see Fig. \ref{TB41Vhopping}(a)), and $\ell_B$ is the magnetic length. Provided $b\ll \ell_B^2/D$, the term $e(\bm{r}-\bm{R}_j)\times \bm{B}$ will be much smaller than $\nabla$, and thus can be ignored in Eq. (\ref{eqn:ttilde}). The hopping then becomes
\begin{equation}\label{tij}
\widetilde{t}_{ij} \simeq \int d^d\bm{r} \exp\left({ie\int_{c_{ij,\bm{r}}} d\bm{r}'\cdot \bm{A}(\bm{r}')}\right) W^*_{\bm{R}_i}(\bm{r})\hat{H}W_{\bm{R}_j(\bm{r})}\,.
\end{equation}
For $\theta$ near the magic angle, $D\sim 7$nm, the size $b\sim2$nm, so such an ignorance of $e(\bm{r}-\bm{R}_j)\times \bm{B}$ requires $\ell_B\gg \sqrt{bD}\sim4$nm, or $B\ll 40$T. The physics we propose happens around 10T to 15T, so our approximation is accurate. We note that the standard derivation of Peierls substitution \cite{Luttinger1951} also ignores the term $e(\bm{r}-\bm{R}_j)\times \bm{B}$.


Now we consider the nearest, 2nd-nearest, and 3rd-nearest hopping integrals in TB4-1V model, as shown in the Fig. \ref{TB41Vhopping}(b)-(d):

1) The nearest hopping $t$, which is mainly contributed by the overlaps of $W_{\bm{R}_i}$ and $W_{\bm{R}_j}$ localized at two positions $AA_1$ and $AA_2$ as shown in Fig. \ref{TB41Vhopping}(b). When there is no magnetic field, the hopping is given by
\begin{equation}
t \simeq \int_{AA_1}d^2\bm{r}\, W^*_{\bm{R}_i} H W_{\bm{R}_j} + \int_{AA_2} d^2\bm{r}\,W^*_{\bm{R}_i} H W_{\bm{R}_j}\ .
\end{equation}
The two parts contributed by region $AA_1$ and region $AA_2$ are equal, because of $C_{2z}T$ symmetry.
In a magnetic field, the hopping is contributed by two integrals at regions $AA_1$ and $AA_2$ with different Peierls phase factors:
\begin{equation}
	\widetilde t \approx \int_{AA_1}d^2\bm{r}\, e^{ie\int_{c_1} d\bm{r}'\cdot \bm{A}(\bm{r}')}W^*_{\bm{R}_i} H W_{\bm{R}_j}+ \int_{AA_2}d^2\bm{r}\, e^{ie\int_{c_2} d\bm{r}'\cdot \bm{A}(\bm{r}')}W^*_{\bm{R}_i} H W_{\bm{R}_j}=\frac{t}{2}\left(e^{ie\int_{c_1} d\bm{r}\cdot \bm{A}(\bm{r})} + e^{ie\int_{c_2} d\bm{r}\cdot \bm{A}(\bm{r})}\right)\ ,
\end{equation}
where $c_1$ ($c_2$) is the contour of straight lines from $\bm{R_i}$ to position $AA_1$ ($AA_2$) and then to $\bm{R_j}$ (see Fig. \ref{TB41Vhopping}(b)). The two integrals $e^{ie\int_{c_1} d\bm{r}\cdot \bm{A}(\bm{r})}$ and $e^{ie\int_{c_2} d\bm{r}\cdot \bm{A}(\bm{r})}$ differ by a phase factor; this phase is given by the magnetic flux enclosed between $c_1$ and $c_2$.

2) The 2nd hopping $\lambda$ and 3rd nearest hopping $t'$. As shown in Fig. \ref{TB41Vhopping}(c) and (d), respectively, The Wannier functions only overlap at one $AA$ stacking location. From Eq. (\ref{tij}), one obtain
\begin{equation}
	\widetilde{\lambda} = \lambda e^{ie\int_{c_3} d\bm{r}\cdot \bm{A}(\bm{r})}\ ,\qquad\qquad \widetilde{t}' = t' e^{ie\int_{c_4} d\bm{r}\cdot \bm{A}(\bm{r})}\ ,
\end{equation}
where $c_3$ and $c_4$ are the contours along the arrowed (broken) straight lines shown in Fig. \ref{TB41Vhopping}(c) and (d), respectively.

Before we perform the LL calculation, we first review the band structure and topology of the TB4-1V model. From \cite{songz2018}, the energy at high symmetry points $\Gamma$, $M$ and $K$ of the model are given by
\begin{align}
E(\Gamma_1)=\Delta\pm 3(t+t')\,,&~~~~E(\Gamma_2) = -\Delta \pm 3(t+t')\ ,\nonumber\\
E(M_1) = \Delta\pm (t-3t')\,,&~~~~E(M_2) = -\Delta \pm (t-3t')\ ,\nonumber\\
E(K_2K_3) &= \pm\sqrt{\Delta^2 + 27\lambda^2}\ ,
\end{align}
where $\Delta$, $t$, $t'$ and $\lambda$ are the parameters in the Hamiltonian in Eq. (\ref{H-TB41V}). Here $\Gamma_1$, $\Gamma_2$, $K_2K_3$, etc. stand for the irreducible representations formed by the bands, which can be read from the Bilbao server \cite{bradlyn2017}. The energies at $K_M$ and $K_M'$ points are two-fold degenerate, giving rise to protected Dirac cones at $K_M$, $K_M'$.

Here we shall assume $\Delta>0$. The model is in a trivial phase when
$$
\Delta \geq 3|t+t'|\,,~~~\Delta \geq |t-3t'|\,.
$$
in which case the lower two bands have irreducible representations $2\Gamma_2, 2M_2, K_2K_3$. On the contrary, the model becomes topologically nontrivial if
\[
\Delta \leq 3|t+t'|\,,\Delta \leq |t-3t'|\ ,
\]
since the lower two bands have irreducible representations $\Gamma_1 + \Gamma_2, M_1 + M_2, K_2K_3$. They are the same as the irreducible representations of the two TBG Moir\'e bands at one graphene valley. If we also want to make the lower two bands to be flat, we can assume
$$
t'=-\frac{t}{3}\,,~~~\lambda = \frac{2t}{\sqrt{27}}\,.
$$
The band structures of the topological trivial and nontrivial phases are shown in Fig. \ref{TB41Vbandbutterfly} (a) and (c), respectively.

\begin{figure}[tbp]
\centering
\includegraphics[width=6.5in]{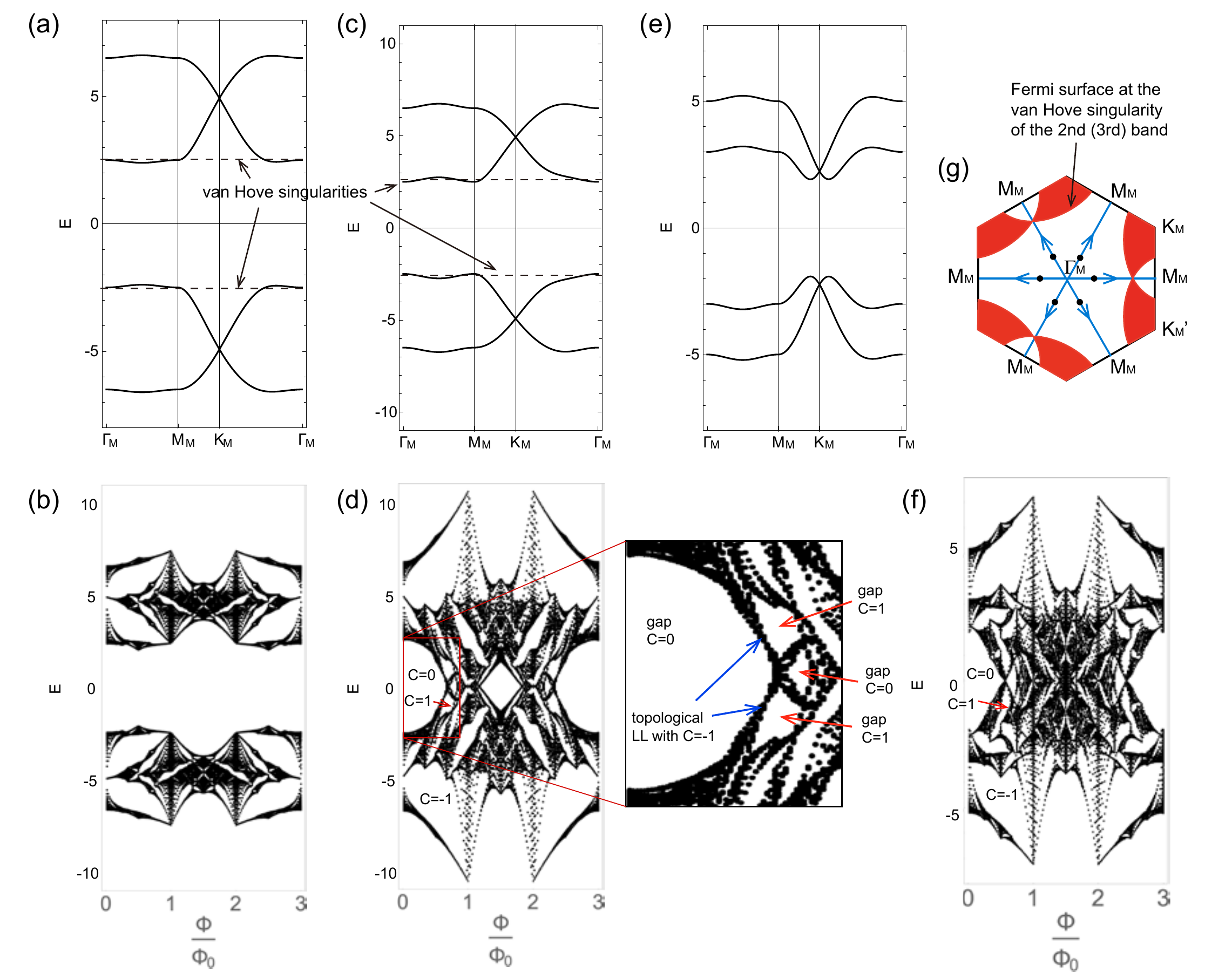}
\caption{(a) The band structure of the topological trivial phase. We choose the parameters as follows: $t = 1$, $\Delta = 4.5$, $t' = -t/3$ and $\lambda = 2t/\sqrt{27}$. (b) The Hofstadter butterfly of the trivial phase in panel (a). (c) Band structure of topological nontrivial phase. The parameters are $t = 9/4$, $\Delta = 2$, $t' = -t/3$ and $\lambda = 2t/\sqrt{27}$. (d) The Hofstadter butterfly of topological nontrivial phase in panel (c). Panels (a)-(d) are also shown in the main text Fig. 2. (e) We change the parameters to $t=2$, $t'=-t/3$, $\lambda = t/\sqrt{27}$ and $\Delta = 1$, which is still in the topological phase.
(f) The Hofstadter butterfly for parameters in (e), which is still entangled between the lower two and the higher two bands. The Zoomed-in inset shows the topological LLs with $C=-1$. (g) The phase transition from trivial to topological phases, which involve six Dirac points (black dots) between the 2nd and the 3rd bands moving along $\Gamma M$ lines from $\Gamma$ to $M$.}
\label{TB41Vbandbutterfly}
\end{figure}

\begin{figure}[htbp]
\centering
\includegraphics[width=7in]{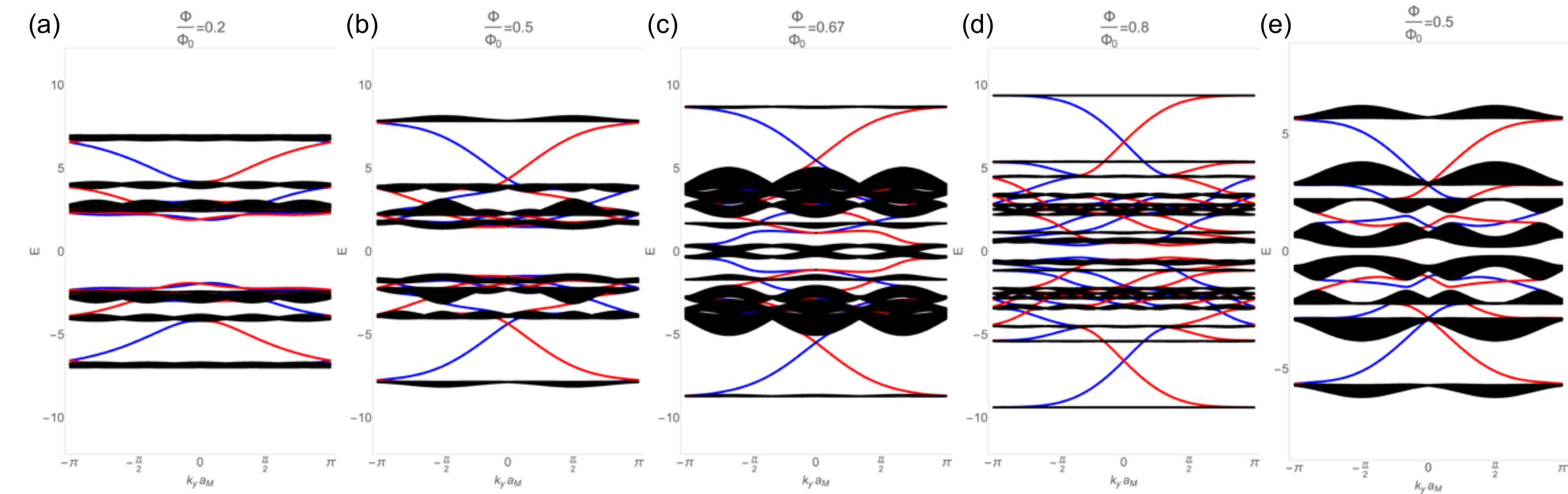}
\caption{(a)-(d) The edge states of the topological phase with different magnetic field. The tight binding parameters are chosen to be the same as in Fig. \ref{TB41Vbandbutterfly} (c) and (d). In this figure the red and green lines stand for edge states on two different edges, and black lines are bulk states. By counting the edge states, we can obtain the in gap Chern numbers in Fig. \ref{TB41Vbandbutterfly} (d). (e) The edge states of TB4-1V with the same parameter choice as Fig. \ref{TB41Vbandbutterfly} (e) at $\Phi/\Phi_0 = 1/2$.}
\label{TB41Vedge}
\end{figure}

The phase transition from the trivial phase to the topological nontrivial phase involves an intermediate metallic phase. During the metallic phase, $6$ Dirac points arise between the 2nd and 3rd bands from $\Gamma_M$ point (the gap between the 2nd and 3rd bands closes at $\Gamma_M$), then move along the six $\Gamma_M M_M$ directions, and finally annihilate each other at three $M$ points (gap between the 2nd and 3rd bands reopens), as shown in Fig. \ref{TB41Vbandbutterfly}(g) (the 6 black dots on the $\Gamma_M M_M$ lines denote the 6 Dirac points between the 2nd and 3rd bands). Such a process flips the relative helicities of Dirac points between the lower two bands (1st and 2nd) at $K_M$ and $K_M'$, and changes the Wilson loop winding number of the lower two bands from $0$ to $1$. Note that the $C_{2x}$ and $C_{3z}$ symmetry requires the 6 Dirac points to be symmetric about each $\Gamma_MM_M$ line.

In addition, the van Hove singularities of the 2nd band and the 3rd band touch each other at certain instant during the process. This can be understood as follows (here the magnetic field $B=0$): the red shaded area in Fig. \ref{TB41Vbandbutterfly}(g) illustrates the region occupied by electrons in the MBZ when the Fermi level is at the van Hove singularity of the 2nd band. At this energy, the two Fermi surfaces (the boundary of the red shaded region) enclosing the two Dirac points at $K_M$ and $K_M'$ (Dirac points between the 1st and 2nd bands) touch each other on the three $\Gamma_MM_M$ lines. Essentially, the van Hove singularity is by definition the energy where the two Fermi surfaces around $K_M$ and $K_M'$ (in the 2nd band) touch each other, and by $C_{2x}$ symmetry they have to touch on the $\Gamma_MM_M$ lines. When the model parameter changes, the shape of this Fermi surface at the 2nd band's van Hove singularity (the boundary of the red shaded region) may change, but the Fermi surface touching points will remain robust on $\Gamma_MM_M$ lines. Since the model is particle-hole symmetric about the 2nd and the 3rd band, the Fermi surface at the 3rd band's van Hove singularity has exactly the same shape.
During the intermediate metallic phase, the 6 Dirac points between the 2nd and 3rd bands move along the $\Gamma_M M_M$ lines. Therefore, there has to be an instant where 3 of the 6 Dirac points (3 because of $C_{3z}$ symmetry) coincide with the touching points of the Fermi surface of the 2nd (and 3rd) band's van Hove singularity on the $\Gamma_MM_M$ lines. At this instant, the 3 Dirac points, the van Hove singularity energy of the 2nd band, and the van Hove singularity energy of the 3rd band are at the same energy; in other words, the van Hove singularities of the 2nd and the 3rd bands touch each other.

We calculate the Hofstadter butterfly by applying a uniform magnetic field to the tight binding model using the Landau gauge $\bm{A} = (0,Bx)$. The magnetic flux through each hexagonal Moir\'{e} unit cell is $\Phi =\frac{3\sqrt{3}}{2}D^2B$, where $D$ is the Moir\'e unit cell edge length (distance between nearest $AB$ and $BA$ stackings) defined in Fig. \ref{TB41Vhopping}(a). Under this gauge, the $y$ direction translation is preserved (see Fig. \ref{TB41Vhopping} for the definition of $x$ and $y$ direction). Under a translation of lattice vector $\bm{a}_2$ (defined in Fig. \ref{TB41Vhopping}(e)), the three hoppings $\widetilde{t},\widetilde{\lambda}$ and $\widetilde{t}'$ in Fig. \ref{TB41Vhopping}(b),(c),(d) will pick up a phase $\widetilde{t}\rightarrow \widetilde{t}e^{i\frac{e}{\hbar}\frac{\Phi}{2}}$, $\widetilde{\lambda}\rightarrow \widetilde{\lambda}e^{i\frac{e}{\hbar}\Phi}$ and $\widetilde{t}'\rightarrow \widetilde{t}'e^{i\frac{e}{\hbar}\Phi}$, respectively. Therefore, when
$$
\frac{\Phi}{2\Phi_0}=\frac{p}{q}\,,~~~p,q\in \mathbb{Z}\,,~~~\mathrm{gcd}(p,q) = 1\ ,
$$
where $\Phi_0=2\pi\hbar/e$ is the flux quanta, all the hoppings recover after a translation $q \bm{a}_2$, namely, the magnetic unit cell is $q$ times of the original unit cell. For twist angles $\theta \sim 1^\circ$, one has $\Phi/\Phi_0=1$ when $B \sim 25\, \rm T$. We then solve the eigenenergies of the TB4-1V tight binding Hamiltonian under magnetic field. The results for topologically trivial and nontrivial phases are shown in Fig. \ref{TB41Vbandbutterfly} (b) and (d), respectively. One can clearly see that the topologically trivial phase has two Hofstadter butterflies bounded within the lower two bands and the higher two bands, respectively. In contrast, the Hofstadter butterflies from the lower two bands and the upper two bands topological non-trivial phase are entangled at large magnetic fields, thus not bounded by their own band widths.
The Hofstadter butterfly is periodic for $\Phi/\Phi_0 \rightarrow \Phi/\Phi_0 + 3$, because the minimal magnetic flux an electron picks up when hopping along a loop connecting different sites is $\Phi/3$ under the modified Peierls substitution we used here. More explicitly, the smallest hopping loop (the triangle around a site $\bm{R}_i$ and two of its nearest neighbours) has a total hopping amplitude (multiplication of hoppings at each step) $t^2\lambda\cos^2(\frac{e}{\hbar}\frac{\sqrt{3}}{4}D^2B)e^{i\frac{e}{\hbar}\frac{\sqrt{3}}{2}D^2B} =t^2\lambda\cos^2(\frac{\pi\Phi}{3\Phi_0})e^{\frac{2\pi i\Phi}{3\Phi_0}}$, which is periodic for $\Phi/\Phi_0 \rightarrow \Phi/\Phi_0 + 3$.



To determine the Chern number in the Hofstadter gaps, we set the open boundary condition along $x$ direction, and solve the energy spectrum and edge states. The Chern number of a gap can be obtained by counting the number of chiral edge states in the gap. We calculate the edge modes of TB4-1V with tight binding parameters in Fig. \ref{TB41Vbandbutterfly} (c) for $\Phi/\Phi_0 = 1/5, 1/2,2/3$ and $4/5$. The results are shown in Fig. \ref{TB41Vedge} (a)-(d), where red and green lines are the edge states on the left and right edges, respectively. In the butterfly in Fig. \ref{TB41Vbandbutterfly} (d) for the topological phase, one sees the Hofstadter butterfly of the lower two bands is heavily connected with that of the higher two bands. In particular, there is
a Hofstadter gap carrying Chern number $C=1$ breaking above the rest butterfly of the lower two bands (see the inset of Fig. \ref{TB41Vbandbutterfly} (d)). Since this model is strongly particle-hole asymmetric (although characterizes the topology of TBG flat bands correctly), its feature is not quantitatively the same as that of the TBG continuum model. However, they are qualitatively the same.


To show this feature is topologically robust and independent of band dispersions, we plot the band dispersions and Hofstadter butterfly of the TB4-1V model with another set of parameters in Fig. \ref{TB41Vbandbutterfly} (e), (f) (see captions therein for the hopping parameters, which are still in the topological phase), and its edge modes at $\Phi/\Phi_0 = 1/2$ in Fig. \ref{TB41Vedge} (e). We can see that the connectness of the Hofstadter butterfly is robust.

\subsection{The TB4-1V model in the magnetic field using the conventional Peierls substitution}

We have demonstrated in Fig. \ref{TB41Vhopping} that the conventional Peierls substitution does not apply to the TB4-1V model in magnetic fields. However, we now show that the qualitative aspects of our topological LL is not sensitive to the type of Peierls substitution we consider. As a comparison, in this subsection we calculate the Hofstadter butterfly of the TB4-1V model using the conventional Peierls substitution.

\begin{figure}[htbp]
\centering
\includegraphics[width = 7in]{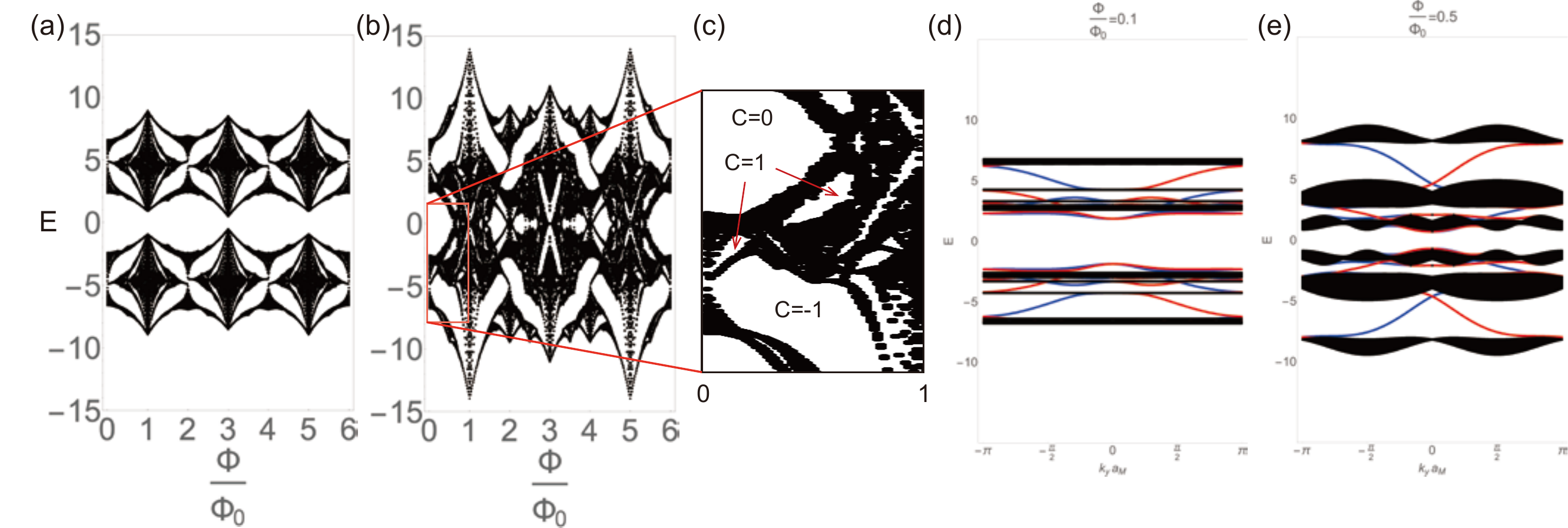}
\caption{The Hofstadter butterfly of the TB4-1V model calculated using the conventional Peierls substitution for (a) the trivial phase with parameters of Fig. \ref{TB41Vbandbutterfly}(a); and (b) the topological phase with parameters of Fig. \ref{TB41Vbandbutterfly}(c). (c) The zoom in of panel (b), where some in-gap Chern numbers are labeled. (d)-(e) The open boundary edge state calculations for the topological phase at $\Phi/\Phi_0=0.1$ and $0.5$, respectively.}
\label{TB41V2}
\end{figure}

Fig. \ref{TB41V2} (a) and (b) show the Hofstadter butterfly of the TB4-1V model calculated using the conventional Peierls substitution, where (a) is in the trivial phase, and (b) is in the topological phase. The period of $\Phi/\Phi_0$ changes to $6$ under this kind of Peierls substitution. This is because in the conventional Peierls substitution, the smallest hopping loop, the triangle connecting site $\bm{R}_i$ with two of its nearest neighbours (which has area $\frac{\sqrt{3}}{4}D^2$, $D$ defined in Fig. \ref{TB41Vhopping}(a)), picks up a gauge phase integrated directly along the triangle edges $\frac{e}{\hbar}\int \bm{A}\cdot d\bm{l}=\frac{\sqrt{3}e}{4\hbar}D^2B=\Phi/6\Phi_0$, which is periodic when $\Phi/\Phi_0$ changes by $6$.
As one can see, the Hofstadter butterfly of the lower two bands is still bounded by the band widths for trivial phase, and unbounded by the band widths and connected with the other bands for the topological phase. Fig. \ref{TB41V2} (d) and (e) give the open boundary edge state calculations in the topological phase. The Hofstadter butterflies under the two kinds of Peierls substitutions we used (in Figs. \ref{TB41Vbandbutterfly} and \ref{TB41V2} respectively) have similar shapes. In particular, for $\Phi/\Phi_0\lesssim1$, one can still see the $C=1$ Hofstadter gap breaking above the rest Hofstadter butterfly of the lower two bands (see the in-gap Chern numbers in the zoom in Fig. \ref{TB41V2} (c)), which qualitatively agrees with the result not using conventional Peierls substitution (Fig. \ref{TB41Vbandbutterfly}(d)).

\subsection{The ten-band effective model in the magnetic field}
\begin{figure}[htbp]
\centering
\includegraphics[width = 7in]{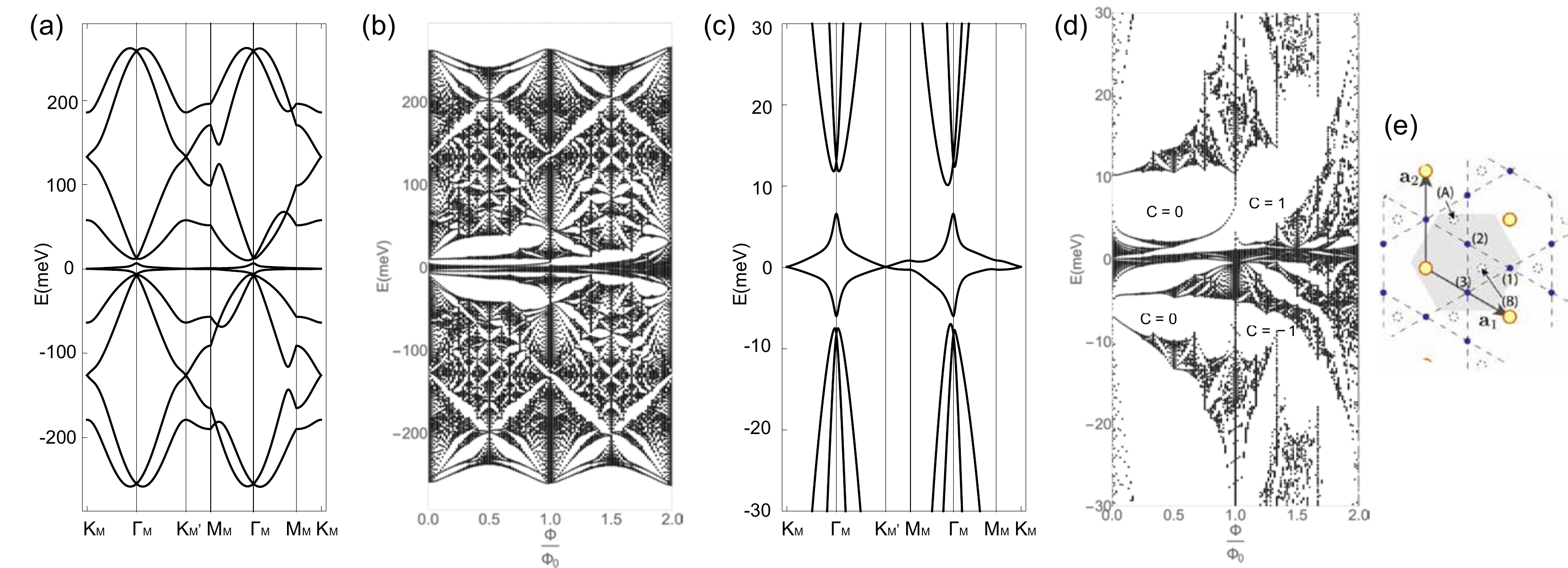}
\caption{(a) The energy band structure of the 10-band tight binding model in Ref.\cite{po2018b}. (b) The Hofstadter butterfly of this model for the band structure in panel (a). (c) Zoom-in of the energy band in panel (a). (d) Zoom-in of the Hofstadter butterfly in panel (d). The Chern numbers in gap are labeled in the figure, which are determined by the open boundary edge state calculations in Fig. \ref{TenBandedge}. (e) The orbital positions in the 10-band model from Ref.\cite{po2018b}.}
\label{TenBandButterfly}
\end{figure}

\begin{figure}[htbp]
\centering
\includegraphics[width=6in]{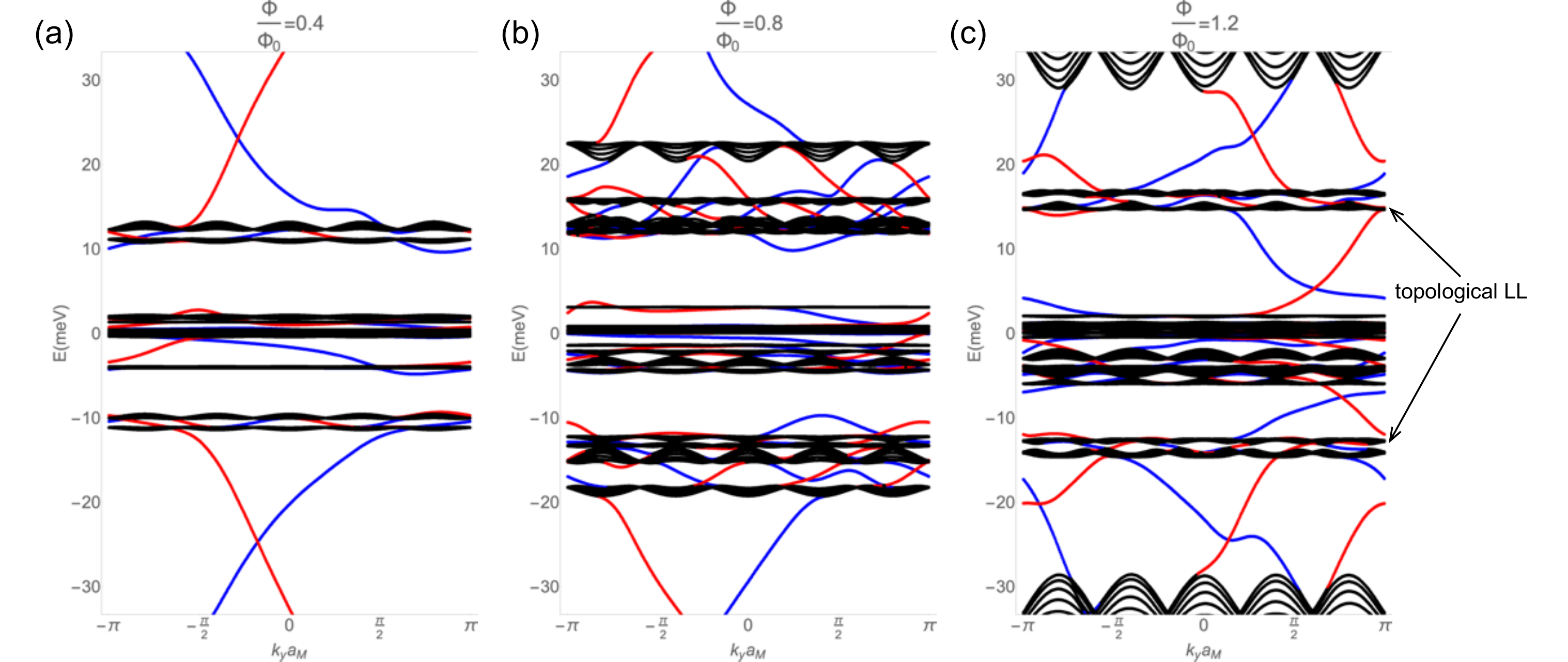}
\caption{(a) The edge states of ten-band model with magnetic field $\Phi/\Phi_0 = 1/10$. (b) The edge states of ten-band model with magnetic field $\Phi/\Phi_0 = 1/5$. (c) The edge states of ten-band model with magnetic field $\Phi/\Phi_0 = 3/10$.}
\label{TenBandedge}
\end{figure}

In this section we calculate the Hofstadter butterfly of the ten-band tight binding model in Ref.\cite{po2018b}, which is given in Eq. (3) of the paper. There are three $p_{z,\pm}$ orbitals on each plaquette center (big yellow hollow circles in Fig. \ref{TenBandButterfly}(e), taken from Ref. \cite{po2018b}), one $s$ orbitals on each Kagome site (blue solid dots labeled by (1),(2),(3) in Fig. \ref{TenBandButterfly}(e)), and two $p_{\pm}$ orbitals on each honeycomb sites (dashed circles labeled by (A),(B) in Fig. \ref{TenBandButterfly}(e)). This ten-band model reproduces the low energy band dispersion of TBG and the topology. There are several orbitals sitting at various positions; we will not consider the shape of the Wannier wave functions, and simply use the conventional Peierls substitution
$$
\widetilde{t}_{ij} = t_{ij} e^{i\int_{c}d\bm{r}'\cdot \bm{A}(\bm{r}')}\ ,
$$
in which the integration is along the straight line $c$ from $\bm{R}_i$ to $\bm{R}_j$. We have enlarged the parameter $\delta$ in the ten-band Hamiltonian $H(t,\delta)$ of Ref. \cite{po2018b} (see their Eq. (3)) to $1.84$, and keep all the other parameters unchanged. In Ref. \cite{po2018b}, they set $\delta=1$. Such a change in $\delta$ does not close any gap, so the band topology remains unchanged. The reason we enlarge $\delta$ is to make the gap between the lowest conduction (valence) band and the second conduction (valence) band larger, so that the topological features of the Hofstadter butterfly of the ten band model can be seen clearer (the same topological feature can still be seen for $\delta=1$, as we have checked).
The band structure and the Hofstadter butterfly are shown in Fig. \ref{TenBandButterfly}. In Fig. \ref{TenBandButterfly}(d) As one can see, around magnetic field $\frac{\Phi}{\Phi_0}=1$, the Hofstadter butterfly of the lowest two bands are connected with higher bands: one Hofstadter band moves above (on the electron side) and one moves below (on the hole side) the energy range of the flat bands, and connects to the higher bands, leading to a $C=+1$ gap and a $C=-1$ gap extending to the energies of higher bands on the electron and hole sides, respectively. This agrees with our expectation from main text Eq. (3). Due to the particle-hole asymmetry of this 10-band tight binding model, the Hofstadter band energies of the positive and negative LLs are not identical; however, we can still identify the topological features on both sides.

To verify that the two Hofstadter gaps have Chern number $\pm1$, we calculate the edge states with the open boundary condition. When the magnetic field is small, as can be seen in Fig. \ref{TenBandedge} (a) and (b) (where red and green are edge states on the left and right open boundaries, respectively), the gap between the flat bands and the higher bands has no edge state; thus, has Chern number $C=0$ (the gaps labeled by $C=0$ in Fig. \ref{TenBandButterfly} (d)). In contrast, when the  magnetic field is large ($\frac{\Phi}{\Phi_0}>1$), there is one edge state in the gap between the butterfly of the lowest conduction (valence) band and the topological LL (Fig. \ref{TenBandedge} (c)), which indicates Chern numbers $C=\pm1$ on the electron and hole sides, respectively (the gaps denoted by $C=+1$ and $C=-1$ in Fig. \ref{TenBandButterfly} (d)).

\subsection{A stable topology example: the Hofstadter butterfly of a two-band Chern insulator}

In the above, we have calculated the Hofstadter butterfly of the $C_{2z}T$ fragile topological models, and have shown that the Hofstadter butterfly of the topological bands are connected with that of the other bands. As we claimed in the main text, the same is true for models with stable topology. Here we show the Hofstadter butterfly of a two-band Chern insulator as an example.

The Chern insulator model we consider is defined on a square lattice, and has a Hamiltonian in the quasimomentum $\mathbf{k}$ space
\begin{equation}\label{Seq-HCI}
H=\sigma_z(M-\cos k_x-\cos k_y)+A(\sigma_x\sin k_x+\sigma_y \cos k_y)\ ,
\end{equation}
where $\sigma_i$ $(i=x,y,z)$ are Pauli matrices, while $M$ and $A$ are constants. We shall set $A=2$. When $|M|>2$, both bands have Chern number $0$ and are trivial. When $|M|<2$, the valence band carries Chern number $1$, and the conduction band carries Chern number $-1$, so both of the two bands are topologically nontrivial.

\begin{figure}[htbp]
\centering
\includegraphics[width = 7in]{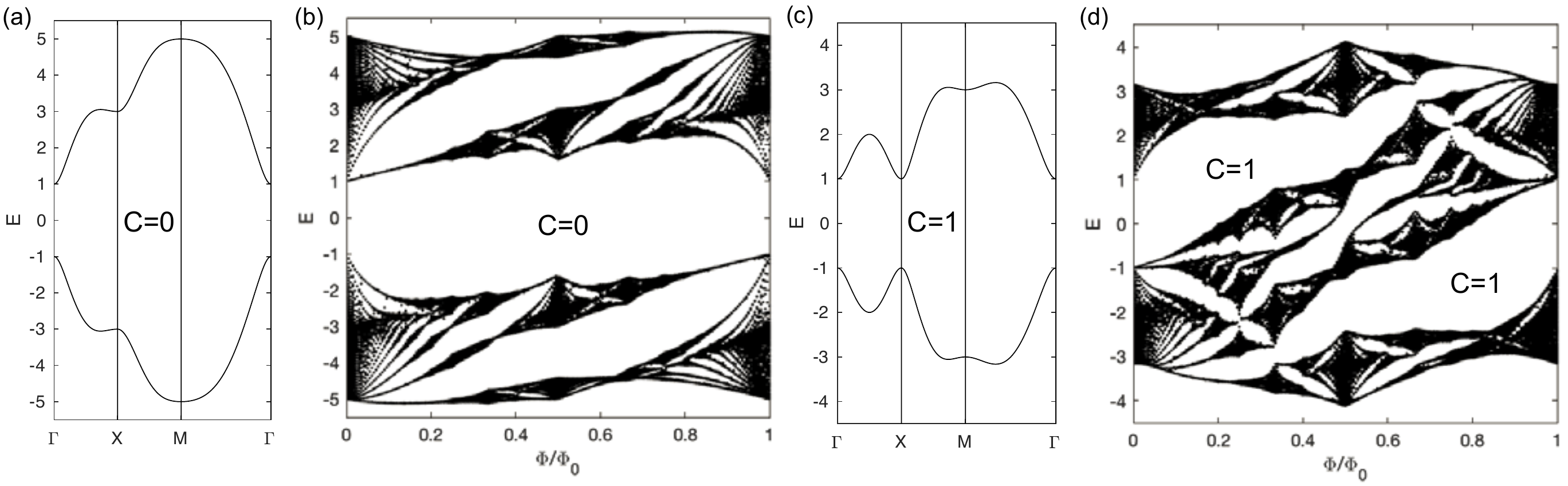}
\caption{The band structure ((a) and (c)) and Hofstadter butterfly ((b) and (d)) of model (\ref{Seq-HCI}), where $\Phi/\Phi_0$ is the flux quanta per plaquette. (a)-(b) illustrate the trivial phase with $M=3$ and $A=2$, where both bands have Chern number $0$. The Hofstadter butterfly of each band is roughly bounded within its own band width. (c)-(d) shows the Chern insulator phase where the two bands have Chern number $\pm1$, respectively. The Hofstadter butterflies of the two bands are clearly connected with each other.}
\label{CIHof}
\end{figure}

As shown in Fig. \ref{CIHof} (a)-(b), the trivial phase simply has the Hofstadter butterfly of each band roughly bounded within its own bandwidth. In contrast, in the nontrivial Chern insulator phase shown in Fig. \ref{CIHof} (c)-(d), the Hofstadter butterfly is connected between the two Chern bands, closing the Chern number $C=1$ gap between the two bands at $\Phi/\Phi_0=1$. The Hofstadter butterfly does not further extend beyond the total energy range of the two bands, because the two bands together are topologically trivial, i.e., have total Chern number $0$.

In general, for a two-band model with a valence band with Chern number $C$ and a conduction band with Chern number $-C$, the band gap between the two bands will close at flux $\Phi/\Phi_0=1/|C|$, where the Hofstadter butterflies of the two bands are connected. This can be proved as follows.

Assume the electron filling fraction (number of electrons per zero-magnetic-field unit cell, not the magnetic unit cell) is $n_C$ in the band gap with Chern number $C$. With only two bands at zero magnetic field, the number of electrons per unit cell $n_C$ has to be bounded by
\begin{equation}\label{Seq-nCbound}
0\le n_C\le 2\ .
\end{equation}
At zero magnetic flux $\Phi=0$, it is easy to see that $n_C=1$. For nonzero magnetic flux, according to the Streda formula \cite{streda1982}, the filling $n_C$ of the Chern number $C$ band gap is given by
\begin{equation}
n_C=1+C\frac{\Phi}{\Phi_0}\ .
\end{equation}
When $\Phi/\Phi_0>1/|C|$, if the Chern number $C$ gap remains robust, we would find $n_C<0$ or $n_C>2$, exceeding the physical bound of Eq. (\ref{Seq-nCbound}). Therefore, the Chern number $C$ band gap is forced to at $\Phi/\Phi_0=1/|C|$, at which the Hofstadter butterflies of the two bands are connected.

\subsection{Heuristic understanding of connected Hofstadter butterfly of topological bands}

The connected Hofstadter butterfly of topological bands can be heuristically understood from the fact that topological bands have no local symmetric Wannier functions. This is understood as follows.

If a set of bands are trivial, one is able to write down local symmetric Wannier functions for the set of bands, based on which a tight-binding model with short-range hoppings can be written down. Assume the tight-binding model is written as
\begin{equation}\label{Seq-GerH}
H=\sum_{i,j,\alpha,\beta} t_{i,\alpha;j,\beta}c^\dag_{i,\alpha}c_{j,\beta}\ ,
\end{equation}
where $c_{i,\alpha}$ is the fermion annihilation operator on site $i$ and orbital $\alpha$, and $t_{i,\alpha;j,\beta}$ is the hopping from site $j$ orbital $\beta$ to site $i$ orbital $\alpha$. In particular, $t_{i,\alpha;i,\alpha}$ stands for the on-site energy of orbital $\alpha$ on site $i$. By the Gershgorin circle theorem, any energy eigenvalue $E$ of the Hamiltonian $H$ is bounded by the inequality
\begin{equation}\label{Seq-Ger}
|E-t_{i,\alpha;i,\alpha}|\le\sum_{j,\beta}|t_{i,\alpha;j,\beta}|\ ,
\end{equation}
where $i$ and $\alpha$ can be any site and orbital. For trivial bands, the hoppings $t_{i,\alpha;j,\beta}$ are short range, so the right-hand side of Eq. (\ref{Seq-Ger}) is finite, and bounds the energies $E$.

When a magnetic field is introduced, the hoppings $t_{i,\alpha;j,\beta}$ in Hamiltonian (\ref{Seq-GerH}) will be replaced by $t_{i,\alpha;j,\beta}e^{i\phi_{i\alpha,j\beta}}$ under the standard Peierls substitution, where $\phi_{i\alpha,j\beta}$ are the gauge phases. Note that this does not change the right-hand-side of Eq. (\ref{Seq-Ger}). Therefore, one concludes that the energies of the Hofstadter butterfly of a set of trivial bands is upper bounded by Eq. (\ref{Seq-Ger}).

For topological bands, there is no local symmetric Wannier functions and thus no short-range tight-binding models. Instead, one can write down nonlocal Wannier functions and hopping models with long range hoppings $t_{i,\alpha;j,\beta}$. One can still apply the Gershgorin circle theorem to obtain Eq. (\ref{Seq-Ger}), but the right-hand-side of Eq. (\ref{Seq-Ger}) is generically divergent due to long range hoppings. Therefore, although one may fine-tune the hoppings so that the energy spectrum at zero magnetic field is within some energy range, one would expect the Hofstadter butterfly at finite magnetic fields to be unbounded, as the phases of the hoppings may be changed dramatically by the Peierls substitution.

When the Hofstadter butterfly of a set of topological bands meets that of another set of bands which trivializes the total band topology of the two sets of bands, one would expect the Hofstadter butterfly to stop growing in energies. This is because the two sets of bands together are trivial, which would allow short-range tight-binding models.

\section{The Continuum Model Hamiltonian under magnetic field}\label{SecII}

When an out of plane magnetic field $B$ is added, the momentum $\mathbf{k}$ is replaced by $\mathbf{k}-e\mathbf{A}/\hbar$, where the gauge potential $\mathbf{A}=(By,-Bx)/2$. Define the magnetic length $\ell_B=\sqrt{\hbar/eB}$. We can then define the LL raising and lowering operators as
\begin{equation}
a=\frac{\ell_B}{\sqrt{2}}\left[(k_x-k_{0x})+i(k_y-k_{0y})+\frac{y}{2\ell_B^2}-i\frac{x}{2\ell_B^2}\right]\ ,\qquad a^\dag=\frac{\ell_B}{\sqrt{2}}\left[(k_x-k_{0x})-i(k_y-k_{0y})+\frac{y}{2\ell_B^2}+i\frac{x}{2\ell_B^2}\right]\ ,
\end{equation}
which satisfies commutation relation $[a,a^\dag]=1$, and $\mathbf{k}_0=(k_{0x},k_{0y})$ are free parameters. The continuum model Hamiltonian of TBG in the magnetic field can then be obtained by the substitution
\begin{equation}\label{kplus}
k_x+ik_y\rightarrow k_{0x}+ik_{0y}+\sqrt{2}\ell_B^{-1}a\ ,\qquad k_x-ik_y\rightarrow k_{0x}-ik_{0y}+\sqrt{2}\ell_B^{-1}a^\dag
\end{equation}
in Eq. (\ref{SeqH}). The basis of the Hamiltonian is the tensor product of the original orbital basis of Eq. (\ref{SeqH}) with the eigen-basis $|l\rangle$ of $a^\dag a$ (which satisfies $a^\dag a|l\rangle=l|l\rangle$). The free parameter $\mathbf{k}_0=(k_{0x},k_{0y})$ is the center momentum of the wave function of each eigenstate $|l\rangle$ in the momentum space ($l\ge0$). For instance, the zero-th Landau level $|0\rangle$ has a momentum space wave function $\psi_0(\mathbf{k})=\langle\mathbf{k}|0\rangle=e^{-\ell_B^2[(k_x-k_{x0})^2+(k_y-k_{y0})^2]}$, which is rotationally symmetric about $\mathbf{k}_0$. Note that the Hamiltonian (\ref{SeqH}) only contains terms linear in momentum $\mathbf{k}$, so after the substitution one get a Hamiltonian no higher than linear order of $a$ and $a^\dag$. By applying a momentum space area cutoff $\mathcal{A}_k=N_{BZ}\Omega_{BZ}$ in the Hamiltonian (\ref{SeqH}) (see Fig. \ref{Mhop}), and a cutoff in the LL number $l\le N$ after the substitution (\ref{kplus}), we can solve for the LLs numerically.

In addition, a Zeeman energy
\[H_{Z}^{\pm}=\pm\mu_B B\ ,\]
where $\mu_B=5.79\times 10^{-5}$ eV$\cdot$T$^{-1}$ is the Bohr magneton, $\pm$ are for spin up and down, respectively, can be easily added.

\section{Edge states and Spectral flow in LL calculations}\label{SecFlow}

By diagonalizing the Hamiltonian in magnetic field $B$ as described in Sec. \ref{SecII} above, one can obtain a LL spectrum which captures the outline of the Hofstadter butterfly at large $B$ fields, as one can see in main text Fig. 3 and the supplementary Figs. \ref{largeB} and \ref{largeB2}. Meanwhile, there are flowing levels in the Hofstadter gaps known as spectral flows, which we will show are due to edge states in the open momentum space, since we do not have periodic boundary conditions usually imposed in conventional Hofstadter butterfly calculations.
In this section, we develop a method to read out the number of edge states in a LL or Hofstadter butterfly gap from the spectral flows in our numerical calculations, from which we can determine the Chern number of the gap.

Before we begin, we comment on the choice of center momentum $\mathbf{k}_0$ in Eq. (\ref{kplus}) in our calculations.
For a finite LL cutoff $N$, the resulting LL spectrum will have a slight dependence on the choice of center momentum $\mathbf{k}_0$ in Eq. (\ref{kplus}), which becomes negligible at large $B$. This is because large $B$ corresponds to small magnetic length $\ell_B$, and the effect of shifting the center momentum is proportional to the dimensionless parameter $k_0\ell_B$, which tends to zero at large $B$. In principle, the Hofstadter bands in the $N\rightarrow\infty$ limit should not depend on the center momentum $\mathbf{k}_0$. This is because a shift from center momentum $\mathbf{k}_0$ to $\mathbf{k}_0+\mathbf{p}$ only changes $(a,a^\dag)\rightarrow (a',a'^\dag)=(a-p_+,a^\dag-p_-)$, where $p_\pm=p_x\pm i p_y$. For $N=\infty$, this only induces a transformation of the LL basis from
\[|l\rangle=\frac{(a^\dag)^{l}}{\sqrt{l!}}|0\rangle\qquad \text{to}\qquad |l'\rangle=\frac{(a^\dag-p_-)^{l'}}{\sqrt{l'!}}e^{p_+a^\dag-\frac{\mathbf{p}^2}{2}}|0\rangle = e^{-\mathbf{p}^2/2}\sum_{l_1=0}^{l'}\sum_{l_2=0}^{\infty}\frac{\sqrt{(l_1+l_2)!}}{l_2!\sqrt{l'!}} \left(\begin{array}{c}l'\\l\end{array}\right) (-p_-)^{l'-l_1}p_+^{l_2}|l_1+l_2\rangle\ .
\]
Since both the old basis $|l\rangle$ and the new basis $|l'\rangle$ are orthonormal and complete, this transformation is unitary, and does not change the energy spectrum.

\subsection{Spectral flow in open real space LL calculations}

In general, in a 2D system with an out-of-plane magnetic field $B$, there will be in-gap chiral edge states on the real space edges due to the Chern numbers of LLs (Hofstadter bands). The standard calculation of the Hofstadter butterfly where the periodic boundary condition is carefully treated \cite{hofstadter1976} is done on a closed manifold without edges. Therefore, only the bulk Hofstadter bands occur, and no in-gap edge states can be obtained. In contrast, if one calculates the energy levels in an out-of-plane magnetic field $B$ on a manifold with edges, one will obtain both the bulk Hofstadter bands and the edge states in the bulk gaps. For instance, if one calculates a model in an out-of plane magnetic field $B$ on a cylindrical strip as shown in Fig. \ref{Sflow}(a), which has an open boundary condition in the $x$ direction and is periodic in the $y$ direction, one expects to see in-gap edge states as a function of $k_y$ (which is a good quantum number in the Landau gauge $(A_x,A_y)=(0,Bx)$) for a fixed magnetic field $B$. Fig. \ref{Sflow}(b) is a schematic plot of the energy levels $E_j(k_y,B_0)$ ($j$ labels different levels) at a fixed magnetic field $B_0$ for a generic model, where only the two Hofstadter bands above and below the gap (blue shaded area) are plotted; between these two bands there exists one chiral edge state coming from the top edge (blue solid line) and one from the bottom edge (red dashed line), respectively. This implies a total Chern number $C=1$ in the gap (of all the bands below the gap).

In Ref. \cite{asboth2017}, it is shown that the edge states can also be read out by plotting the energy levels $E_j(k_y,B)$ as a function of magnetic field $B$ (in $z$ direction) for a fixed $k_y$ (on the cylindrical strip in Fig. \ref{Sflow}(a), and $(A_x,A_y)=(0,Bx)$). For the discussion here, we set $x=0$ ($x=N_x$, lattice constant is set to $1$) at the top edge (bottom edge) labeled in Fig. \ref{Sflow}(a). For a square lattice (with a lattice constant $1$), if a Hofstadter gap (which changes continuously as a function of $B$ in the Hofstadter butterfly) carries Chern number $C$, the authors of Ref. \cite{asboth2017} show that there will be $C$ edge state levels in $E_j(k_y,B)$ (with a fixed $k_y$) flowing across the gap when the flux $\phi_{N_x}=N_xB$ in $N_x\times1$ plaquettes (see Fig. \ref{Sflow}(a)) changes by $2\pi$, which is called the spectral flow.

\begin{figure}[htbp]
\begin{center}
\includegraphics[width=7in]{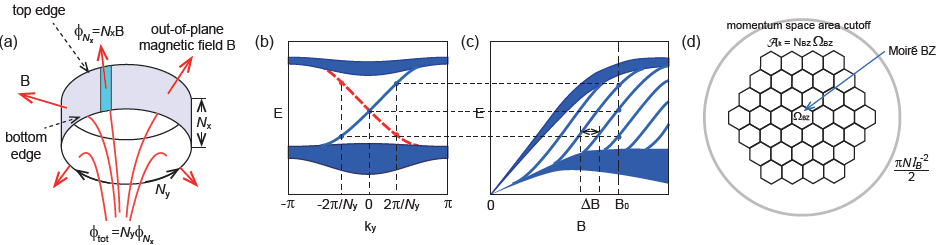}
\end{center}
\caption{(a) The open boundary cylindrical manifold on which Hofstadter bands are calculated. (b) The Hofstadter bands and edge states between them at a fixed magnetic field $B_0$. (c) Illustration of the spectral flow in the Hofstadter gaps for calculation done on manifold of panel (a). If one cut the spectral flows at $B=B_0$, one should obtain the spectrum in panel (b) at all the discrete momentums $k_y^{(m)}=2\pi m/N_y$. (d) The hexagon region of honeycomb lattice illustrates the effective momentum space open boundary manifold our LL calculation is done on, which has an area $N_{BZ}\Omega_{BZ}$, with $\Omega_{BZ}$ being the area of the MBZ. The big circle illustrates the range of the total flux $N\Phi_0$ inserted in our calculation ($\Phi_0=2\pi\hbar/e$), where $N$ is the LL number cutoff ($|a^\dag a|\le N$).}
\label{Sflow}
\end{figure}

Here we generalize this result to the LL spectrum on a generic 2D open manifold via the following observation. Assume the cylindrical strip in Fig. \ref{Sflow}(a) has period $N_y$ (number of sites) in the $y$ direction. $k_y$ can then only take $N_y$ discrete values $k_y^{(m)}=2\pi m/N_y$ ($0\le m\le N_y-1$). If one plots all the energy levels $E_j(k_y^{(m)},B)$ of all $k_y^{(m)}$ versus the magnetic field $B$, one will see a spectrum flow in a Hofstadter gap with nonzero Chern number as shown in Fig. \ref{Sflow}(c). The in-gap levels at a fixed $B=B_0$ in Fig. \ref{Sflow}(c) (blue dots) are simply the edge states at the momentums $k_y^{(m)}=2\pi m/N_y$ in the edge state dispersion at $B=B_0$ shown in Fig. \ref{Sflow}(b). In the example of Fig. \ref{Sflow}(b) where the Chern number in the gap is $C=1$, if we fix a constant Fermi energy in this gap (which can be chosen arbitrarily), as $\phi_{N_x}$ increases by $2\pi$, each momentum $k_y^{(m)}$ will have one level crossing the Fermi energy in the gap; thus in total one would expect $N_y$ levels of all the momentums $k_y^{(m)}$ to flow across this Fermi energy in the gap. Equivalently, this means when $\phi_{N_x}$ increases by $2\pi/N_y$, or the total flux through the manifold $\phi_{tot}=N_y\phi_{N_x}=N_xN_yB$ increases by $2\pi$ (the magnetic field increases by $\Delta B=2\pi/(N_xN_y)$), there will be one level flowing across the Fermi energy in the gap. Therefore, the magnetic field distance $\Delta B$ of two neighbouring flowing levels along a constant energy line in the gap with Chern number $C=1$ (see Fig. \ref{Sflow}(c)) is given by $\Delta B=2\pi/(N_xN_y)$.

This result can also be understood from a flux threading argument. In increasing the out-of-plane $B$ field, we can fix the gauge field to be $(A_x,A_y)=(0,0)$ on the top edge ($x=0$, labeled in Fig. \ref{Sflow}(a)), while at point $(x,y)$ the gauge field is $(A_x,A_y)=(0,Bx)$. This correspond to a 3D magnetic field configuration as shown in Fig. \ref{Sflow}(a), where the $B$ field is thread from the loop of the bottom edge ($x=N_x$) and then come out of the cylinder surface, keeping the flux through the top edge $0$. Accordingly, the total gauge flux threaded in the loop of the bottom edge is equal to the total out-of-plane flux through the cylinder surface $\phi_{tot}=N_y\phi_{N_x}=N_xN_yB$. If we view the bottom edge as a 1D system with $|C|$ chiral fermions, a change of the threaded flux $\phi_{tot}$ by $2\pi$ will lead to a level shift for each chiral fermion. Therefore, there will be in total $|C|$ levels shifted on the bottom edge when $\phi_{tot}$ changes by $2\pi$. Meanwhile, no level shifts occur on the top edge, as the flux threaded in the loop of the top edge is fixed at $0$.

For Landau level (Hofstadter band) calculations of a lattice on a generic 2D open manifold with a boundary, the above principle can be generalized as follows. Assume the total area of the manifold is $\mathcal{A}_{tot}$. Then if one plots all the energy levels versus magnetic field $B$, in a Hofstadter gap with Chern number $C$, one would expect to see $|C|$ flowing levels when the total flux $\phi_{tot}=B\mathcal{A}_{tot}$ changes by a flux quanta $\Phi_0=2\pi\hbar/e$ (here we recover the units $\hbar$ and $e$), and the sign of $C$ determines the direction of flow. In fact, this can be understood from the celebrated Streda formula for a 2D system in a magnetic field $B$ \cite{streda1982}:
\begin{equation}\label{streda}
e\frac{dn}{dB}=\sigma_{xy}\ ,
\end{equation}
where $e$ is the electron charge, $n$ is the occupied electron number density in the bulk, and $\sigma_{xy}$ is the Hall conductance. If the system is situated on an open manifold with total area $\mathcal{A}_{tot}$, the electron number density $n=N_{occ}/\mathcal{A}_{tot}$, where $N_{occ}$ is the number of occupied electron states below the Fermi energy $\epsilon_F$. When the Fermi energy $\epsilon_F$ is in a Hofstadter butterfly gap (well defined in a certain range of $B$) with Chern number $C$, the Hall conductance $\sigma_{xy}=C\frac{e^2}{2\pi\hbar}$ is invariant versus $B$. Therefore, for fixed total area $\mathcal{A}_{tot}$, when the magnetic field changes by $\Delta B$, the Streda formula (\ref{streda}) tells us the change $\Delta N_{occ}$ in number of occupied states $N_{occ}$ is
\begin{equation}\label{flow}
\frac{1}{\mathcal{A}_{tot}}\frac{\Delta N_{occ}}{\Delta B}=\frac{dn}{dB}=\frac{\sigma_{xy}}{e} =\frac{C}{\Phi_0}\ ,
\end{equation}
where $\Phi_0=2\pi\hbar/e$. Therefore, when the total magnetic flux changes by $\Delta \phi_{tot}=\mathcal{A}_{tot}\Delta B=\Phi_0$, one has $\Delta N_{occ}=C$, indicating there are $C$ levels flowing across the Fermi level $\epsilon_F$, and thus across the gap since $\epsilon_F$ can be chosen anywhere in the gap. In particular, because the Hofstadter gap does not collapse as a function of $B$, the $C$ levels flowing across the gap have to come from (1) the physical edge states for a manifold with boundary, or (2) edge states made by a change in flux (the flux is threaded through a clear surface area).

\subsection{Spectral flow in open momentum space LL calculations}

The above principle in Eq. (\ref{flow}) can be applied to our case by duality to momentum space. Recall that in our numerical calculation of LLs, we first take a momentum space area cutoff $\mathcal{A}_k=N_{BZ}\Omega_{BZ}$ (see Fig. \ref{Mhop} and also Fig. \ref{Sflow}(d)) in the continuum model Hamiltonian Eq. (\ref{SeqH}) before we make the substitution in Eq. (\ref{kplus}) of $k_\pm$ into $a,a^\dag$. Here $\Omega_{BZ}$ is the area of the Moir\'e BZ. Therefore, we are in fact doing the LL calculation on a momentum space open manifold of area $\mathcal{A}_k$ (with a momentum space tight-binding Hamiltonian (\ref{SeqH}), instead of a real space open manifold). On the other hand, it is known that the momentum space and real space are dual to each other in the LL physics (not only for the zeroth LL), by which we can write down a formula similar to the Streda formula (\ref{flow}) for the momentum space. This can be most easily seen in the Landau gauge $(A_x,A_y)=(0,Bx)$, under which $k_y$ is a good quantum number for Landau levels. the translation operator in $y$ direction is therefore $P_y=k_y-(eB/\hbar)x$. Equivalently, one can view $P_{k_x}=-(\hbar/eB)P_y=x-(\hbar/eB)k_y$ as the translation operator in the momentum space $k_x$ direction, and $x=i\partial_{k_x}$ is the good quantum number (guiding center) in the momentum space representation. If one redefines
\[\widetilde{B}=\hbar^2/(e^2B)\ ,\]
one can rewrite $P_{k_x}$ as $P_{k_x}=x-(e\widetilde{B}/\hbar)k_y$, and the problem is mapped into the momentum space. Therefore, one could view the magnetic field $B$ in the real space as a ``magnetic field" $\widetilde{B}=\hbar^2/(e^2B)$ in momentum space.


Since our calculation is done on a momentum space manifold with area $\mathcal{A}_k=N_{BZ}\Omega_{BZ}$, we can apply the spectral flow principle in Eq. (\ref{flow}) to momentum space: in a Hofstadter gap, the number of spectral flow levels $\Delta N_{occ}$ in a ``magnetic field" interval $\Delta \widetilde{B}$ satisfy
\begin{equation}\label{ntilde}
\frac{1}{\mathcal{A}_{k}}\frac{\Delta N_{occ}}{\Delta \widetilde{B}}=\frac{d\widetilde{n}}{d\widetilde{B}} =\frac{C_K}{\Phi_0}\ ,
\end{equation}
where $\widetilde{n}=N_{occ}/N_{BZ}\Omega_{BZ}$ is the electron ``number density" in the momentum space, and $C_K\in\mathbb{Z}$ can be understood as a momentum space dual Chern number of the Hofstadter gap (we will explain its relation with the actural Chern number $C$ in the below). Equivalently, one can rewrite the above formula as
\begin{equation}\label{CK}
C_K=\frac{1}{N_{BZ}}\frac{dN_{occ}}{d(\Phi_0/\Phi)}\ ,
\end{equation}
where $\Phi=B \Omega=4\pi^2B/\Omega_{BZ}$ is the magnetic flux per real space unit cell area $\Omega$. Or explicitly, this gives
\begin{equation}
\Delta N_{occ} = \frac{ C_KN_{BZ}\Omega_{BZ}}{\Phi_0}\Delta \widetilde{B}=-\frac{ C_KN_{BZ}\Omega_{BZ}}{\Phi_0}\frac{\hbar^2}{e^2B^2}\Delta B=-\frac{ C_KN_{BZ}\Omega_{BZ}\Phi_0}{4\pi^2B^2}\Delta B\ .
\end{equation}
Therefore, one expects to see $\Delta N_{occ}=1$ spectral flow level every real magnetic field interval
\begin{equation}\label{DB}
\Delta B=-\frac{1}{C_K}\frac{2\pi eB^2}{\hbar N_{BZ}\Omega_{BZ}}\ .
\end{equation}
This formula allows us to determine $C_K$ in the Hofstadter gap. This spectral flow is due to the edge states on the boundary of the finite momentum space of area $\mathcal{A}_k$.

The dual Chern number $C_K$ is not yet the real Chern number $C$ which is related with the Hall conductance $\sigma_{xy}$. This is because the momentum space density $\widetilde{n}=N_{occ}/N_{BZ}\Omega_{BZ}$ in Eq. (\ref{ntilde}) is not the real space electron density $n$. In this method, note that each energy level is a Landau level which has real space size $2\pi\ell_B^2$, we conclude the real space electron density away from charge neutral point is
\begin{equation}
n=\frac{N_{occ}}{2\pi\ell_B^2 N_{BZ}}=\frac{\tilde{n}\Omega_{BZ}}{2\pi\ell_B^2} =\frac{\tilde{n}\Omega_{BZ}B}{\Phi_0}\ ,
\end{equation}
where $N_{occ}$ is the number of occupied states counted from the charge neutral point. The reason to have $N_{BZ}$ in the denominator is because, by having a momentum space of $N_{BZ}$ Brillouin zones, we have duplicated the number of physical states by a factor $N_{BZ}$, namely, $N_{occ}/N_{BZ}$ is the true number of physical states. Then according to the real space streda formula (\ref{flow}), the actual Chern number $C$ giving Hall conductance is given by
\begin{equation}
C=\Phi_0\frac{dn}{dB}=\Omega_{BZ}\left(\widetilde{n}+B\frac{d\widetilde{n}}{dB}\right) =\frac{N_{occ}}{N_{BZ}}-\frac{\Phi_0}{\Phi}C_K\ .
\end{equation}
Therefore, the Chern number $C$ in a Hofstadter gap can be determined in this method by counting the number of occupied states $N_{occ}$ and extracting out the dual Chern number $C_K$ from the spectral flow.


As we have shown, the spectral flow distance $\Delta B$ in Eq. (\ref{DB}) only depends on $N_{BZ}$, so one may wonder whether the LL number cutoff $N$ matters. In fact, the cutoff $N$ is the total number of flux quanta one applies in the entire momentum space. The big circle in Fig. \ref{Sflow}(d) illustrates the area in which the $N\Phi_0$ magnetic flux is uniformly distributed, which has an area $2\pi N\ell_B^{-2}$, and is centered at the center momentum $\mathbf{k}_0$ we choose in our LL numerical calculations (Sec. \ref{SecII}). In the derivation of Eq. (\ref{DB}), the spectral flow is induced by the change of the total flux within the momentum space area $\mathcal{A}_k$ (the lattice area illustrated in Fig. \ref{Sflow}(d)). If the area inside the big circle in Fig. \ref{Sflow}(d) is bigger than $\mathcal{A}_k$, the momentum space ``magnetic field" $\widetilde{B}$ is applied uniformly in the momentum space area $\mathcal{A}_k$. Therefore, the total flux in the momentum space open manifold $\mathcal{A}_k$ increases as $\widetilde{B}$ increases, leading to Eq. (\ref{DB}). On the contrary, if the big circle in Fig. \ref{Sflow}(d) is smaller than and inside the area $\mathcal{A}_k$, the total flux in the area $\mathcal{A}_k$ will be constantly $N\Phi_0$, independent of $\widetilde{B}$, and one would not expect the spectral flows of Eq. (\ref{DB}). Therefore, to see the spectral flow given by Eq. (\ref{DB}), the big circle within which the total flux is distributed has to be greater than the open manifold area $\mathcal{A}_k$ in momentum space, namely, $N\Phi_0/\widetilde{B}>\mathcal{A}_k$. This gives a requirement that the flux per unit cell
\begin{equation}
\frac{\Phi}{\Phi_0}=\frac{2\pi e B}{\Omega_{BZ}\hbar}>\frac{N_{BZ}}{N}\ , 
\end{equation}
for the spectral flow picture to be valid.

A rigorous theoretical derivation of the above method will be given in a separate paper \cite{LLmethod}.

\subsection{Numerical results of the Hofstadter spectrum of TBG}

\begin{figure}[htbp]
\begin{center}
\includegraphics[width=7in]{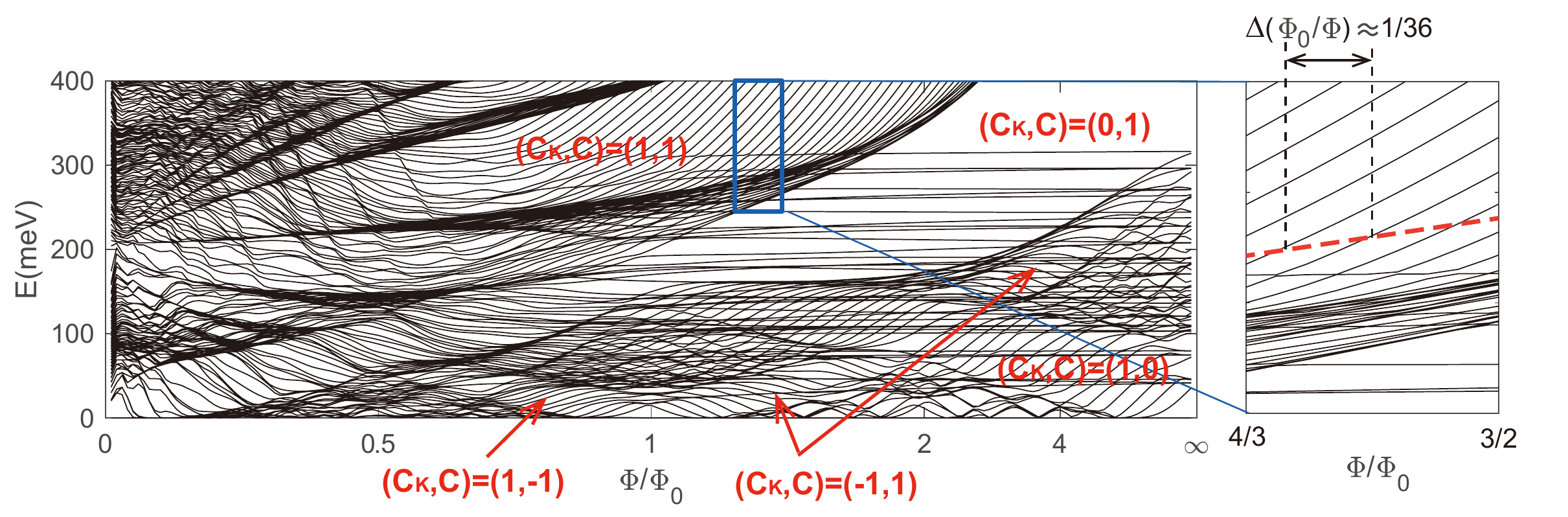}
\end{center}
\caption{Extraction of the momentum space dual Chern number $C_K$ from the spectral flow, and determination of the actual Chern number $C$. The $x$ axis labels $\Phi/\Phi_0$, which is plotted linear in $\Phi/\Phi_0$ from $0$ to $1$, and mapped to $x=2-\Phi_0/\Phi$ for $\Phi/\Phi_0>1$ (so that infinite flux is mapped to a finite value on $x$ axis). The zoom-in panel on the right illustrate the determination of $C_K$ from spectral flow, where $\Delta(\Phi_0/\Phi)\approx1/36=1/N_{BZ}$, yielding a dual Chern number $C_K=1$. }
\label{Nflow}
\end{figure}

Fig. \ref{Nflow} illustrates how we can apply the spectral flow method to determine the dual Chern number $C_K$ and the actual Chern number $C$ in a Hofstadter gap. In the numerical Landau level calculation, we have set the number of (Moir\'e) Brillouin zones to $N_{BZ}=36$, and the Landau level cutoff $N_L=60$, so that the spectral flow behaves as expected when $\Phi/\Phi_0\gtrsim N_{BZ}/N_L=0.6$. Besides, we have used the zero angle approximation, so that the spectrum is strictly particle-hole symmetric, and we only plotted the positive energy spectrum. In order to cover the entire range of $\Phi/\Phi_0$ from 0 to infinity, we define the $x$ axis coordinate as $x=\Phi/\Phi_0$ when $0\le\Phi/\Phi_0\le1$, and $x=2-\Phi_0/\Phi$ for $\Phi/\Phi_0>1$, so that infinite flux is mapped to $x=2$ (we still label it as $\Phi/\Phi_0=\infty$ for convenience).

The zoom-in panel on the right illustrate the determination of $C_K$ from spectral flow: the red thick dashed line parallel to the Hofstadter band edge (the denser region) below it cuts a number of flowing levels, and the distance between neighbouring flowing levels along the cut gives the inverse flux change $\Delta(\Phi_0/\Phi)$ for $\Delta N_{occ}=1$ level to flow out of the Hofstadter band below the gap. In the gap of the zoom-in panel, one finds $\Delta(\Phi_0/\Phi)\approx1/36=1/N_{BZ}$, which gives a dual Chern number $C_K=1$. Similarly, one can obtain the dual Chern number $C_K$ of other gaps, some of which are labeled in Fig. \ref{Nflow}.

\begin{figure}[htbp]
\begin{center}
\includegraphics[width=6.5in]{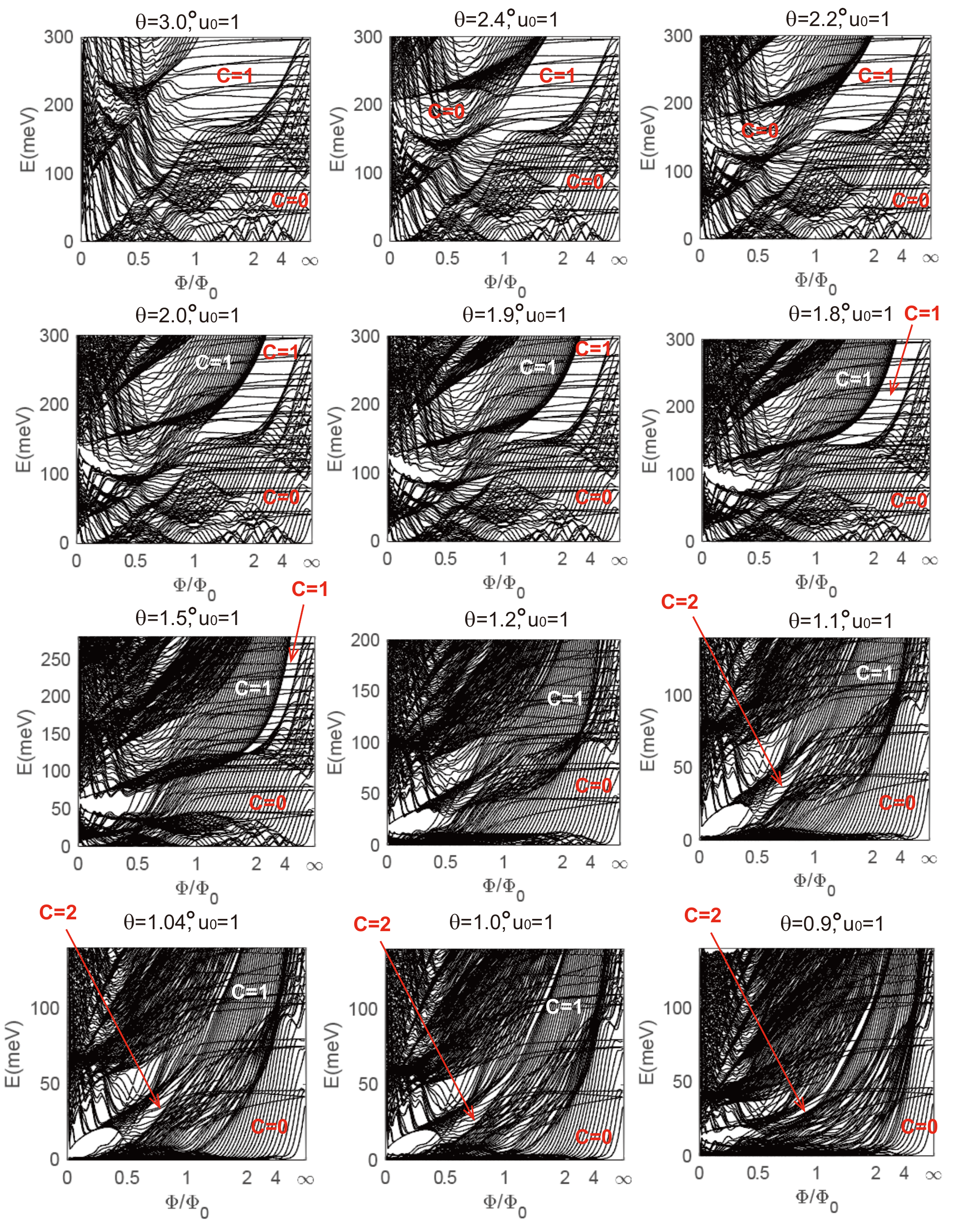}
\end{center}
\caption{Numerical Hofstadter butterfly and spectral flows for $u_0=1$ (no corrugation) at various angles. The angle $\theta$ decreases from $3^\circ$ to $0.9^\circ$. Some of the in-gap Chern numbers $C$ determined from our spectral flow method are labeled. As $\theta$ decreases, one can clearly see a transition around $1.9^\circ$, where the extended $C=1$ gap breaks into two, giving way to the connected $C=0$ gap. Another transition happens around $1.1^\circ$, where the Hofstadter is reconnected at $\Phi/\Phi_0=1/2$ and $1$. The horizontal lines at large $\Phi/\Phi_0$ are spurious modes which should be ignored (see Eq. (\ref{spurious0})).}
\label{largeB}
\end{figure}

\begin{figure}[htbp]
\begin{center}
\includegraphics[width=6.5in]{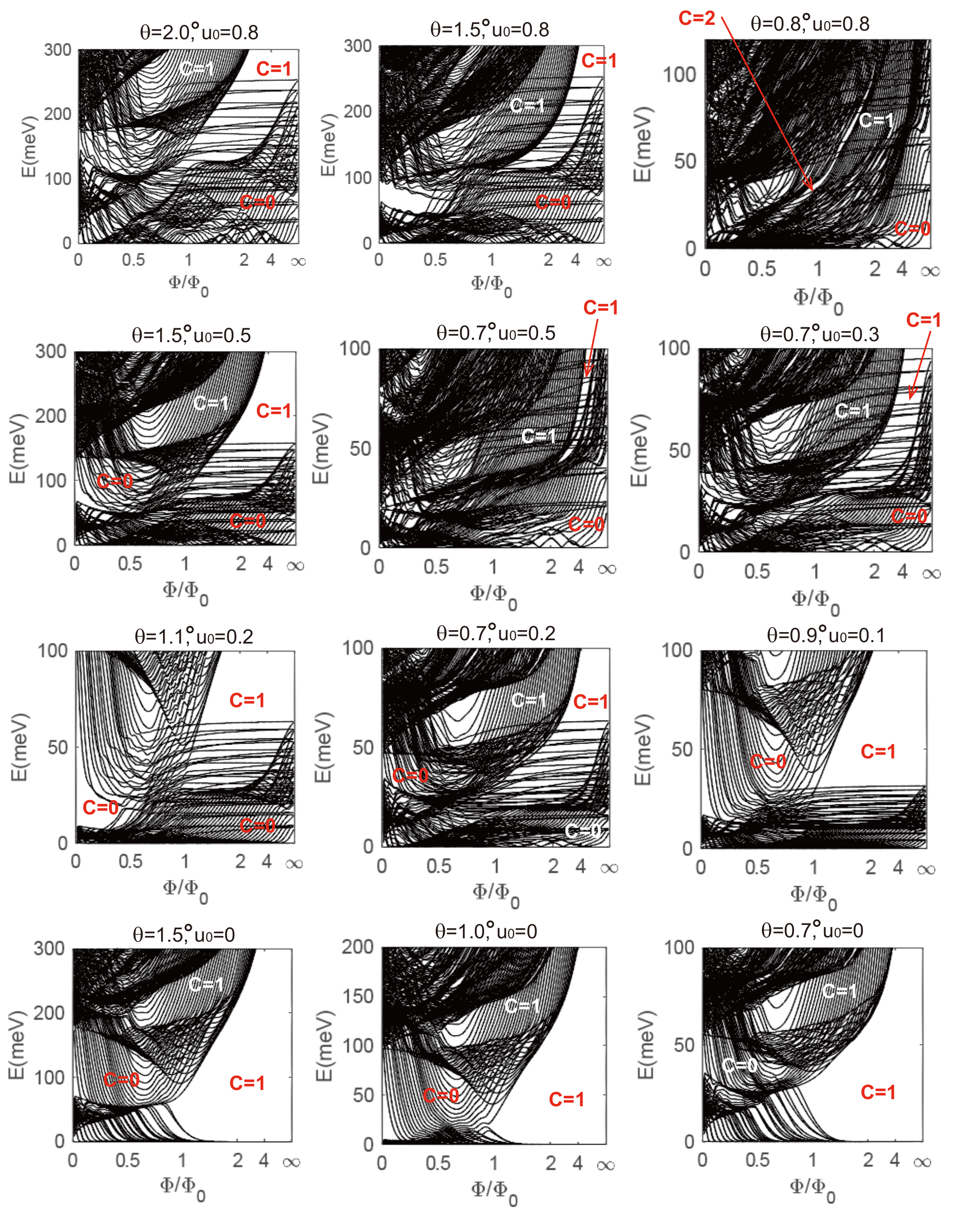}
\end{center}
\caption{Examples of numerical Hofstadter butterfly and spectral flows for $0\le u_0<1$ at various angles. Some of the in-gap Chern numbers $C$ determined from our spectral flow method are labeled. One can identify each Hofstadter spectrum with the three phases $\Lambda_1$, $\Lambda_2$ and $\Lambda_3$ in the main text Fig. 3(m). The horizontal lines at large $\Phi/\Phi_0$ (dropping down from small $\Phi/\Phi_0$) are spurious modes which should be ignored (see Eq. (\ref{spurious0})).}
\label{largeB2}
\end{figure}

\begin{figure}[htbp]
\begin{center}
\includegraphics[width=2.8in]{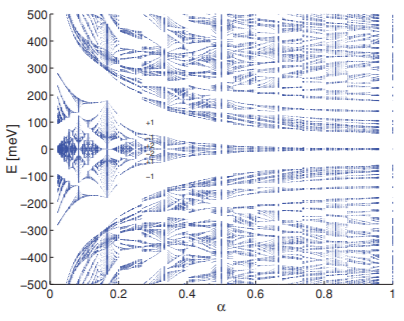}
\end{center}
\caption{Hofstadter spectrum of TBG calculated in \cite{bistritzer2011a} Fig. 2 (taken from their paper) for $\theta=2^\circ$ (system parameters such as Fermi velocity therein may slightly differ from ours), where the $x$ axis is $\Phi_0/6\Phi$ in our convention. One can clearly see the $C=\pm1$ gap (the two most obvious gaps) extending to higher bands and infinite $\Phi/\Phi_0$, which is in phase $\Lambda_1$ we identified.}
\label{MacHof}
\end{figure}

Next, to figure out the actual Chern number $C$ of a Hofstadter gap, one needs to count the number of levels $N_{occ}$ between the charge neutral point and the Hofstadter gap at a particular magnetic flux $\Phi/\Phi_0$. There is, however, a subtlety in the counting. As one can see from Fig. \ref{Nflow}, at large $\Phi/\Phi_0$, there are a number of horizontal energy levels which do not disperse with respect to $\Phi/\Phi_0$. These levels are spurious levels due to Landau level number cutoff $N$, and should not be counted in $N_{occ}$. More explicitly, consider a Dirac fermion in a magnetic field:
\begin{equation}\label{spuriousH}
H=\frac{\sqrt{2}v}{\ell_B}\left(\begin{array}{cc}
0&a-p_+\ell_B/\sqrt{2}\\
a^\dag-p_-\ell_B/\sqrt{2} &0
\end{array}
\right)\ ,
\end{equation}
where $a$ and $a^\dag$ are lowering and raising operators of Landau levels, and $p_\pm=p_x\pm ip_y$ denotes the momentum position of the Dirac point. Assume one takes a Landau level cutoff $N$. In the large $B$ limit, $p_\pm\ell_B\rightarrow 0$, and one finds there are two zero modes under the Landau level cutoff $N$:
\begin{equation}\label{spurious0}
\psi_A=\left(\begin{array}{c}0\\ |0\rangle\end{array}\right)\ ,\qquad\qquad \psi_B=\left(\begin{array}{c} |N\rangle\\ 0\end{array}\right)\ .
\end{equation}
The first zero mode $\psi_A$ is the physical zero mode of Dirac fermion we know, while the second zero mode $\psi_B$ is a spurious zero mode due to the cutoff. In the TBG continuum model, there are $2N_{BZ}$ Dirac fermions coupled via momentum space hopping $w$. Therefore, one will get $2N_{BZ}$ spurious zero modes at large $B$, which couple only among themselves via the momentum space hopping $w$. Therefore, these modes do not disperse with respect to $B$, and should be ignored since they are unphysical. More detailed explanation will be given in a separate theoretical paper \cite{LLmethod}.

By ignoring the nondispersive unphysical modes, one is able to count the number of states $N_{occ}$ between a gap and the charge neutral point. For instance, for the gap of the zoom-in panel in Fig. \ref{Nflow}, one finds $N_{occ}\approx 2N_{BZ}$ at flux $\Phi/\Phi_0=1$, and therefore the actual Chern number of the gap is $C=(N_{occ}/N_{BZ})-C_K\Phi_0/\Phi=1$. One can check that the result doesn't depend on the flux $\Phi/\Phi_0$ one looks at. Similarly, we can obtain the actual Chern number of other gaps, as shown in Fig. \ref{Nflow}.

Fig. \ref{largeB} shows the Hofstadter butterfly of the TBG continuum model at various angles for $\Phi/\Phi_0$ from $0$ to $\infty$, where the corrugation parameter is $u_0=1$ (no corrugation or relaxation). In the calculations, we have used the zero angle approximation, so that the Hofstadter butterfly is particle-hole symmetric, and only the positive energy spectrum is drawn. There are three phases which we describe below.

1) At large angles $\theta\gtrsim1.9^\circ$, one can see there is a $C=1$ Hofstadter gap extending from the charge neutral point in the lowest two bands all the way to higher bands as $B$ increases, so that the Hofstadter butterfly of the lowest two fragile topological bands are connected with that of the higher bands. This is phase $\Lambda_1$ in the main text Fig. 3(m). This agrees with our expectation from main text Eq. (3) as well as the Hofstadter butterflies of the fragile topological tight-binding models in Refs. \cite{songz2018,po2018b}. Namely, fragile topology leads to a Hofstadter butterfly connected with other bands, closing the gap between the fragile topological bands and some other bands. We note that this extended $C=1$ gap in the TBG Hofstadter butterfly can also be seen in the large angle Hofstadter calculated using periodic method in Ref. \cite{bistritzer2011a} (Fig. 2 therein, captured as Fig. \ref{MacHof} here).

2) As the angle decreases to intermediate angles $1.1^\circ<\theta<1.9^\circ$, the $C=1$ gap closes at $\Phi/\Phi_0=1$ and breaks into two disconnected gaps, giving way to the trivial $C=0$ gap between the lowest conduction band and the next band, which is phase $\Lambda_2$ in the main text Fig. 3(m). The Hofstadter butterfly of the lowest two bands is no longer connected with higher bands at any finite $\Phi/\Phi_0$.

However, one should also consider the Hofstadter spectrum at infinite magnetic field $\Phi/\Phi_0=\infty$. This is because the Hofstadter butterfly of the continuum model is not periodic within any finite magnetic fields. In order to have a periodic Hofstadter butterfly like that of a tight-binding model, one needs to include
the $\Phi/\Phi_0=\infty$ point, for which one identifies $\Phi/\Phi_0=+\infty$ with $\Phi/\Phi_0=-\infty$. In this way, the $\Phi/\Phi_0=\infty$ point can be viewed as the half period point of the Hofstadter butterfly.
This is because the continuum model can be viewed as a tight binding model with infinite orbitals per unit cell, which therefore has an infinite large Hofstadter period.

In this sense, the Hofstadter butterfly of phase $\Lambda_2$ is in fact still connected, but is connected at point $\Phi/\Phi_0=\infty$. This behavior can be seen from the numerical calculations in Figs. \ref{largeB} and \ref{largeB2}, where the Hofstadter spectrum above the $C=0$ gap approaches zero energy near $\Phi/\Phi_0=\infty$, tending to connect with the Hofstadter butterfly of the lowest two bands at zero energy and close the $C=0$ gap.
This connection of spectrum can also be seen analytically from the continuum model. In the $B\rightarrow\infty$ limit, as we have explained in Eq. (\ref{spurious0}), only the zero mode
\[\psi_A=\left(\begin{array}{c}0\\ |0\rangle\end{array}\right)\]
remains at low energy (the spurious zero mode $\psi_B$ is unphysical and should be ignored), which is a zero energy eigenstate of the Dirac Hamiltonian (\ref{spuriousH}) when $B\rightarrow\infty$. Since each momentum site $\mathbf{Q}_m$ of the continuum model (see Fig. \ref{Mhop}) has such a Dirac Hamiltonian (the diagonal $2\times2$ blocks in Eq. (\ref{SeqH})), there will be a zero mode $\psi_A=(0,|0\rangle)^T$ on each momentum site at low energies as $B\rightarrow\infty$.

The continuum model at $B\rightarrow\infty$ at low energies therefore becomes a tight binding model in the momentum space honeycomb lattice (Fig. \ref{Mhop}): each momentum site has an orbital $\psi_A=(0,|0\rangle)^T$ with zero on-site energy, while neighbouring sites along a bond $\mathbf{q}_j$ has a hopping given by $\psi_A^\dag (wT_j)\psi_A=u_0w$. This is simply the tight binding model of monolayer graphene, so one immediately learns that the spectrum is gapless at zero energy, and spans an energy range $[-3u_0w,3u_0w]$. This agrees with our numerical calculations in Figs. \ref{largeB} and \ref{largeB2}, and proves that the Hofstadter butterfly is gapless and connected at zero energy at $\Phi/\Phi_0=\infty$.

We note that the infinite magnetic field $B=\infty$ is time reversal invariant ($B=+\infty$ and $B=-\infty$ are identified). Therefore, the above momentum space tight binding model at $B=\infty$ has $C_{2z}T$ symmetry, which protects the gaplessness of the spectrum. Since the fragile topology of TBG at zero field is protected by $C_{2z}T$, this shows that the connectivity of Hofstadter butterfly is exactly due to the fragile topology.

3) When the angle is below $1.1^\circ$, the Hofstadter spectrum enters phase $\Lambda_3$, and becomes reconnected between the lowest two bands and higher bands at $\Phi_0/\Phi=1/2$ and $1$. A clear feature one can see in the last row of Fig. \ref{largeB} is the emergence of a $C=2$ gap extending from the lowest conduction band to higher bands. In addition, the fact that the Hofstadter butterfly also reconnects at $\Phi_0/\Phi=1$ indicates the reconnection of the $C=1$ gap in the $\Lambda_1$ phase, although this gap is not quite clear in Fig. \ref{largeB}. Therefore, phase $\Lambda_3$ has both a $C=2$ gap and a $C=1$ gap from the lowest conduction band extending to higher bands.

We then explore the Hofstadter spectrum of more angles and corrugations $0\le u_0\le 1$, some of which are shown in Fig. \ref{largeB2}. Based on our calculations, we find a phase diagram of the three phases $\Lambda_{1,2,3}$ with respect to $\theta$ and $u_0$ as shown in main text Fig. 3(m). In particular, phases $\Lambda_1$ and $\Lambda_3$ which has connected Hofstadter butterfly between the lowest two bands and higher bands at finite $\Phi/\Phi_0$ dominate the parameter space, while phase $\Lambda_2$ has the Hofstadter butterfly connected between the lowest two bands and higher bands at $\Phi/\Phi_0=\infty$.

\section{Numerical calculation of small magnetic field Landau levels}\label{SecNLL}

We now use our numerical method to calculate the LLs at small magnetic fields $B$. Throughout the calculations in this section, we assume no corrugation, namely, $u_0=1$. For small $B$, one needs a larger LL number cutoff $N$ to obtain the full Hofstadter spectrum, since the largest LL orbital in momentum space has a size $\sqrt{N}\ell_B^-1= \sqrt{NeB/\hbar}$, which needs to be larger than the MBZ size $k_\theta$ for the spectrum obtained to be complete.

However, in the case $\sqrt{N}\ell_B^-1<k_\theta$, one can still obtain the LLs contributed by the band edge at some momentum $\mathbf{p}$ by choosing the center momentum $\mathbf{k}_0$ at point $\mathbf{p}$. The calculation in this small $B$ limit simply reduces to that of the $k\cdot p$ model near point $\mathbf{p}$.

In the following LL calculation for different angles $\theta$, we set the cutoffs to $N_{BZ}=36$, and $N=90$. Besides, in the calculation, we can easily apply (or not) the zero twist angle approximation, which gives PHS (PHNS) LL spectrum. The calculation without the zero twist angle approximation is more realistic, but in general the LLs in these two cases have little differences (except the charge neutral point is shifted in energy by about $-2$meV, which does not change the physics).

\subsection{LLs at exactly the magic angle with quadratic band touching}
Fig. \ref{quadratic2} shows the numerical results of low energy LLs for exactly the magic angle $\theta_m=1.0^\circ$ with $\alpha=0.605$. In the calculation, we did not use the zero angle approximation, so the spectrum is PHNS. In particular, the $K_M,K_M'$ Dirac points are no longer at zero energy, but at about $-2.3$meV, which is the charge neutral point of the TBG. The magic angle LL spectrum in larger energy and magnetic field intervals can be seen in Fig. \ref{largeB}(c).

At the magic angle, the Dirac points at $K_M,K_M'$ become quadratic band touching (Eq. (\ref{qLL})), as shown in Fig. \ref{quadratic2}(d). A quadratic band touching has a LL spectrum $E_{\pm l}\propto\pm\sqrt{l(l-1)}B$ linear in $B$, where $l\ge0$. There are two zero mode LLs ($l=0$ and $l=1$).
Fig. \ref{quadratic2}(a) and (b) show the LL spectrum with zero Zeeman field, which are calculated with center momentum $\mathbf{k}_0$ at $K_M$ point and $M_M$ point, respectively. The two results for different $\mathbf{k}_0$ are slightly different, but one can see the main features remain the same, including the positions of LLs, and the rough shapes of Hofstadter butterfly at large $B$ ($\gtrsim2$T). For $B$ within $10$T, the LLs/Hofstadter butterfly of the lowest two Moir\'e bands are mostly concentrated within an energy range $2$meV (as shown in Fig. \ref{quadratic2}(a) and (b)).

\begin{figure}[htbp]
\begin{center}
\includegraphics[width=7in]{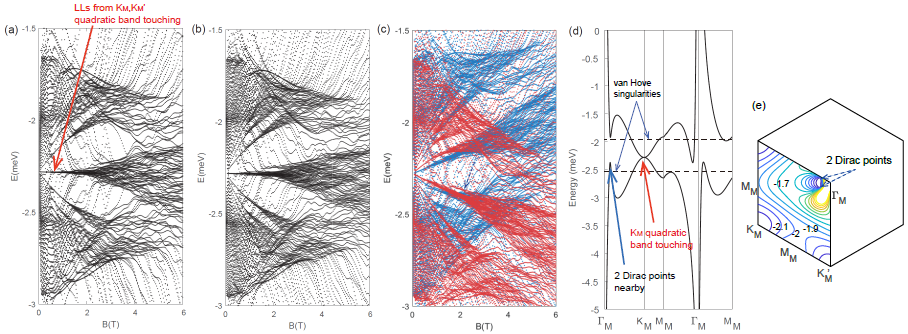}
\end{center}
\caption{Low energy LLs at exactly the magic angle $\alpha=0.605$ ($\theta=1.0^\circ$) where the Dirac fermions at $K_M$ and $K_M'$ are quadratic. (a) LLs calculated with center momentum $\mathbf{k}_0$ at $K_M$ point. (b) LLs calculated with center momentum $\mathbf{k}_0$ at $M_M$ point. (c) LLs with Zeeman splitting, where the center momentum $\mathbf{k}_0$ is at $M_M$. Red and blue are for spin up and down, respectively. (d) The corresponding band dispersion at the magic angle (PH asymmetric). (e) The energy contour plot of the lowest conduction band. The number near the contours labels the energy (meV).}
\label{quadratic2}
\end{figure}

At small magnetic fields, the LLs around energy $-2.3$ meV are contributed by the quadratic band touching at $K_M$ and $K_M'$, which have the expected linear dispersion in $B$.
One may also notice the LL spectrum are denser around energies $-2.0$ meV and $-2.5$ meV, and these come from two van Hove singularities (saddle points) in the Moire bands (The LLs merge into van Hove singularities). The LLs above $-2.0$meV and below $-3$meV are from the band maximum and minimum at the $\Gamma_M$ point (see Fig. \ref{quadratic2}(e), note that for $\theta=1.0^\circ$ there is no band maximum/minimum near $M_M$, instead there is only a saddle point near $M_M$). The band structure also contains 6 Dirac fermions (of the same helicities) very close to the $\Gamma_M$ point, but their LLs cannot be seen in Fig. \ref{quadratic2}.
This is because these $6$ Dirac fermions are too close to each other (the blue dashed arrow in the energy contour plot Fig. \ref{quadratic2}(e) points out the rough positions of two of them in $1/3$ of the MBZ), and have too large Fermi velocities; the large Fermi velocities make the energies of their non-zero-mode LLs (proportional to Fermi velocity) quickly increase and merge into the van Hove singularities readily at very small $B$ ($\lesssim0.04$T). Besides, the extremely close distances among them yield large quantum hoppings among their zero mode LLs (see Sec. \ref{SecZM} and Sec. \ref{SecShift}A for discussion of such hoppings), so their zero mode LLs already strongly deviate from the energy of the six Dirac points at very small $B$ ($\lesssim0.04$T), which cannot be resolved at the scale of Fig. \ref{quadratic2}(a) and (b). In Fig. \ref{104}(f) we show an example: the numerical calculation of the LLs of these 6 Dirac points near $\Gamma_M$ for $\theta=1.04^\circ$ (the $6$ Dirac points are robust for $|\theta-\theta_m|\lesssim 0.1^\circ$), where we find these LLs quickly merge into the van Hove singularities at $B\sim0.04$T. This is far below the scale of magnetic fields we will discuss, so we can safely ignore these LLs for $B\gtrsim0.1$T.

At large magnetic fields, each LL expands into one or several Hofstadter bands with a finite energy span (Fig. \ref{quadratic2}). Since the method we used has a momentum space cutoff $N_{BZ}$ in the continuum model and a LL number cutoff $N$, we cannot properly take into account the magnetic translation symmetry, so we cannot obtain a clean Hofstadter butterfly; we also observe in-gap edge states called spectral flows \cite{asboth2017} (see Sec. \ref{SecFlow} for detailed explanations), which can be seen clearer for larger $B$. These in-gap spectral flows slightly depend on the choice of center momentum $\mathbf{k}_0$, while in contrast, the shape of the Hofstadter bands are independent of the choice of center momentum $\mathbf{k}_0$. We develop a method to link the in-gap spectral flows \cite{asboth2017} with the in-gap Chern numbers in our system, which is given in Sec. \ref{SecFlow}.


The Zeeman field is important for $\theta$ around the magic angle, where the bandwidth of the lowest two Moir\'e bands is small. Fig. \ref{quadratic2}(c) shows the LLs with Zeeman splitting taken into account, where red and blue are for spin up and down, respectively. The theoretical Zeeman splitting energy is comparable and even larger than the orbital LL energy spacing in the lowest conduction/valence bands. Therefore, the spin degeneracy is significantly broken. We plot the schematic LLs at the magic angle in Fig. \ref{FigShopping} based on the numerical result in Fig. \ref{quadratic2}.


\subsection{Near the magic angle}
When the twist angle $\theta$ is near the magic angle, the $K_M$ and $K_M'$ points will host Dirac fermions with nonzero Fermi velocity $v^*$, and there are several other Dirac points in the Moir\'e BZ not at high symmetry points, but on high symmetry lines (in the PHS approximation) \cite{songz2018}. When $\theta>1.12^\circ$, there are no other Dirac points except for the 2 Dirac points at $K_M$ and $K_M'$. When $\theta$ decreases to $1.12^\circ$ (using the PHS band structure), there are 12 Dirac points pairwisely created on the 6 $\Gamma_MK_M$ lines. As $\theta$ further decreases towards the magic angle $\theta_m=1.0^\circ$, 6 of the Dirac points stay on the $\Gamma_MK_M$ lines close to the $\Gamma_M$ point; while the other $6$ Dirac points first move along $\Gamma_MK_M$ to $\Gamma_M$ point, then move along $\Gamma_MM_M$ lines to $M_M$, and lastly move along $M_MK_M$ lines to $K_M$ ($K_M'$), leading to quadratic band touchings at $K_M$ ($K_M'$) when $\theta$ reaches $\theta_m$. (If one does not impose the PHS approximation, the Dirac points on $\Gamma_MK_M$ lines will slightly deviate from the $\Gamma_MK_M$ lines.) These additional Dirac points will also contribute to the LLs. The $l$-th LL of a linear Dirac fermion has an energy proportional to $\pm\sqrt{lB}$. In particular, each Dirac fermion has a zero mode.

\begin{figure}[htbp]
\begin{center}
\includegraphics[width=6.8in]{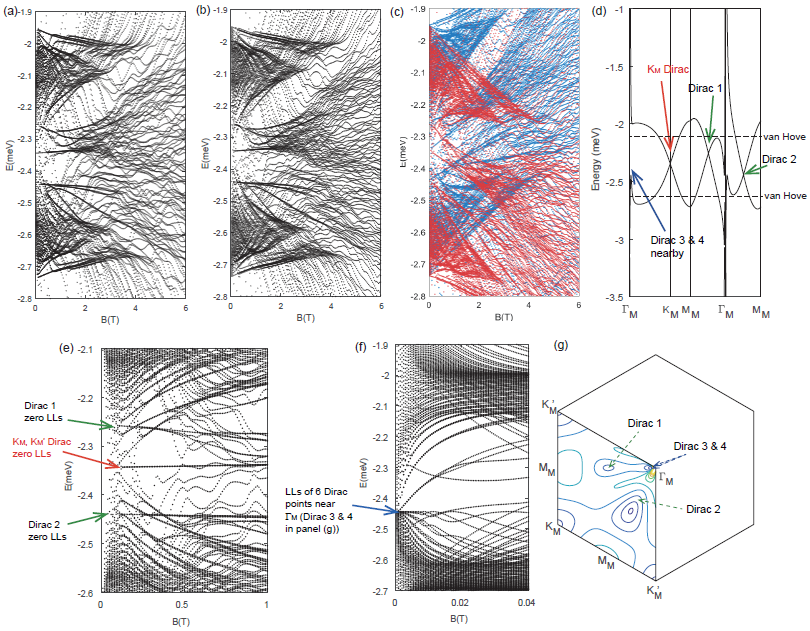}
\end{center}
\caption{Low energy LLs for $\alpha=0.583$ ($\theta=1.04^\circ$). (a) LLs calculated with center momentum $\mathbf{k}_0$ at $K_M$ point. (b) LLs calculated with center momentum $\mathbf{k}_0$ at $M_M$ point. (c) LLs with Zeeman splitting, where the center momentum $\mathbf{k}_0$ is at $M_M$. Red and blue are for spin up and down, respectively. (d) The corresponding band dispersion at this angle, $\alpha=0.583$ (PH asymmetric). (e) Panel (b) zoomed in for $0$T $<B<1$T (with the center momentum $\mathbf{k}_0$ at $M_M$). (f) The LLs for $0$T $<B<0.04$T calculated with center momentum $\mathbf{k}_0$ chosen at one of the 6 Dirac points near $\Gamma_M$ (Dirac $3$ or $4$ in panel (g)), where one can see the LLs from these Dirac points near $\Gamma_M$. (g) The energy contour plot of the lowest conduction band (PHNS). The number near the contours labels the energy (meV).}
\label{104}
\end{figure}

\begin{figure}[htbp]
\begin{center}
\includegraphics[width=6.5in]{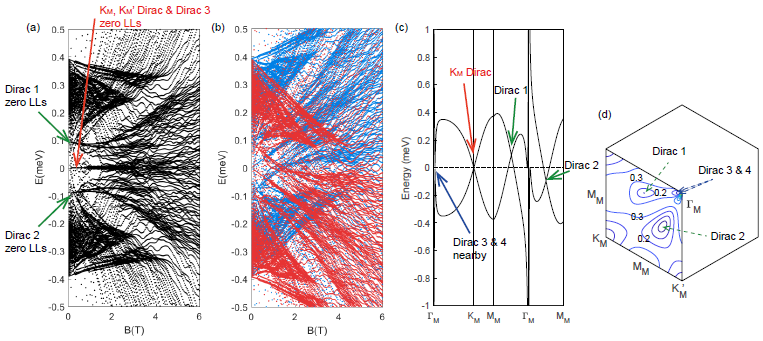}
\end{center}
\caption{Low energy LLs for $\alpha=0.583$ ($\theta=1.04^\circ$) with zero twist angle approximation used. (a) LLs calculated with center momentum $\mathbf{k}_0$ at $M_M$ point. (b) LLs with Zeeman splitting, where the center momentum $\mathbf{k}_0$ is at $M_M$. Red and blue are for spin up and down, respectively. (c) The corresponding band dispersion at this angle, $\alpha=0.583$ (PHS). (d) The energy contour plot of the lowest conduction band (PHS). The number near the contours labels the energy (meV).}
\label{104PHS}
\end{figure}

\begin{figure}[htbp]
\begin{center}
\includegraphics[width=7in]{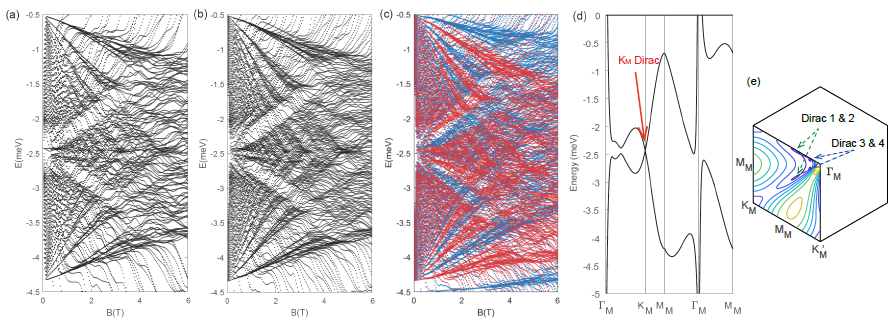}
\end{center}
\caption{LLs near charge neutral point for $\alpha=0.554$ ($\theta=1.1^\circ$). (a) LLs calculated with center momentum $\mathbf{k}_0$ at $K_M$ point. (b) LLs calculated with center momentum $\mathbf{k}_0$ at $M_M$ point. (c) LLs with Zeeman splitting, where the center momentum $\mathbf{k}_0$ is at $M_M$. Red and blue are for spin up and down, respectively. (d) The corresponding band dispersion at this angle, $\alpha=0.554$ (PH asymmetric). (e) The energy contour plot of the lowest conduction band.}
\label{110}
\end{figure}

Fig. \ref{104} shows the LLs (without the zero twist angle approximation, therefore PHNS) for $\theta=1.04^\circ$ ($\alpha=0.583$). For the case, there is one Dirac fermion at each $K_M$ and $K_M'$, respectively; furthermore, there are $6$ Dirac fermions on $\Gamma_M$-$M_M$ lines ($3$ have an energy above the charge neutral point, the other $3$ have an energy below the charge neutral point, labeled by ``Dirac 1" and ``Dirac 2" in Fig. \ref{104}(d) and (g), respectively), and $6$ Dirac fermions close to $\Gamma_M$ point near the labeled arrows ``Dirac 3 \& 4" in Fig. \ref{104}(d) and (g).

At small $B$ field ($\lesssim0.5$T), in Fig. \ref{104}(a), (b) and (e) (without Zeeman field, (a) and (b) are calculated with the center momentum $\mathbf{k}_0$ at $K_M$ and $\Gamma_M$, respectively, while (e) is the zoom in of (b)), one can see the zero modes from the Dirac fermions at $K_M$ and $K_M'$ at energy $-2.34$ meV (the charge neutral point, see red arrows in Fig. \ref{104}(d) and (e)). In addition, there are $3$ zero modes at $-2.25$ meV and another $3$ at $-2.43$ meV, which are from the $3$ Dirac fermions above charge neutral point and the $3$ Dirac fermions below charge neutral point on $\Gamma_M$-$M_M$ lines, respectively (see green arrows labeled by ``Dirac 1 zero LLs" and ``Dirac 2 zero LLs" in Fig. \ref{104}(e)). The LLs of the $6$ Dirac fermions close to $\Gamma_M$ point, however, cannot be seen at the scale of Fig. \ref{104}(a), (b) and (e). As we explained in subsection \ref{SecNLL}A, this is because of their large Fermi velocities and small momentum separations, which make their LLs (including the zero mode LLs) quickly deviate from the energy of these 6 Dirac points as $B$ increases, and grow to energies comparable to the van Hove singularity. To resolve these LLs, we calculate the small $B$ field ($0$T to $0.04$T) LL spectrum by choosing the center momentum $\mathbf{k}_0$ at one of the 6 Dirac points near $\Gamma_M$, and the result is plotted in Fig. \ref{104}(f). As Fig. \ref{104}(f) shows, the LLs around $-2.45$meV are those of the 6 Dirac points near $\Gamma_M$, and $-2.45$meV is the energy of the $6$ Dirac points. In particular, one notes that these LLs quickly become nondegenerate and deviate from the energy of the $6$ Dirac points; these LLs readily merge into the van Hove singularities (around $-2.1$meV and $-2.6$meV) when $B\sim0.04$T. We note that the LLs of the other Dirac points (at $K_M,K_M'$ or on $\Gamma_MM_M$ lines) are not seen in Fig. \ref{104}(f). This is because the LL number cutoff $N$ makes us only be able to calculate the LLs within a momentum space range $\sqrt{N}l_B^{-1}\propto\sqrt{NB}$ around the center momentum $\mathbf{k}_0$. For $B=0.04$T and $N=90$ we take, the range $\sqrt{N}l_B^{-1}\sim0.25k_\theta$ is much smaller than the size of MBZ $k_\theta$. Therefore, with the center momentum $\mathbf{k}_0$ near $\Gamma_M$ point, we cannot see the LLs from the other Dirac points far away from $\Gamma_M$ at $B\leq0.04$T. In this paper, we shall only discuss magnetic fields $B\gg0.1$T, in which range there will be no low energy LLs from the $6$ Dirac points near $\Gamma_M$.

When $B\lesssim0.5$T, the 3 zero modes (of 3 Dirac points on $\Gamma_M$-$M_M$ lines) around $-2.25$ meV (and the $3$ around $-2.43$ meV) are approximately degenerate, due to the degeneracy among the 3 Dirac points. As magnetic field becomes larger ($B\gtrsim 0.5$T), one sees the 3 zero modes (of Dirac points on $\Gamma_M$-$M_M$ lines) around $-2.25$ meV ($-2.43$ meV) are no longer degenerate, which is due to mutual hopping between them, and they can no longer be viewed as isolated Dirac fermions. Meanwhile, the zero modes from $K_M$ and $K_M'$ remain rather degenerate for $B\lesssim3$T (see Fig. \ref{104}(a) and (b)) before developing into broad Hofstadter bands (in fact, even for $B>3T$, the zero modes from $K_M$ and $K_M'$ become broad bands, but are still not well separated, as shown in Fig. \ref{104}(a)). Therefore, at small $0.5$T $\lesssim B\lesssim3$T with zero Zeeman, there are two kinds of low energy LLs: those from $K_M$ and $K_M'$ which have spin, graphene valley and Moir\'e valley degeneracies, and those from Dirac points on $\Gamma_MM_M$ lines (not at high symmetry points) with only spin and graphene valley degeneracy when $B\gtrsim 0.5$T.

When Zeeman field splitting and graphene valley degeneracy are taken into account, for $0.5$T $\lesssim B\lesssim3$T, the LLs from $K_M,K_M'$ will be $4$-fold degenerate (with respect to $K_M,K_M'$ and $K,K'$), while those from Dirac points on $\Gamma_MM_M$ will be only $K,K'$ $2$-fold degenerate (although $C_{3z}$ ensures these $\Gamma_MM_M$ Dirac points to be $3$-fold degenerate, this $3$-fold degeneracy of their LLs is broken by mutual hopping for $B\gtrsim0.5$T). As shown in Fig. \ref{104}(d), these two kinds of LLs from $K_M,K_M'$ and from Dirac points on $\Gamma_MM_M$ are close in energy, and they may cross each other so that their order in energies changes as the function of $B$. This would lead to the Landau fan shift which we discuss in Sec. \ref{SecShift}(c).

For comparison, we also calculated the LLs with the zero twist angle approximation (with center momentum $\mathbf{k}_0$ at $M_M$ point) as shown in Fig. \ref{104PHS}, where the LLs are PHS. As one can see, the LLs are not much different from that in Fig. \ref{104}, and one can still the zero mode LLs from $K_M$, $K_M'$ (the red arrow in Fig. \ref{104PHS}(a)) and those from Dirac points on the $\Gamma_MM_M$ lines (green arrows in Fig. \ref{104PHS}(a)).

Furthermore, we present here another near magic angle LL data for $\alpha=0.554$ ($\theta=1.1^\circ$) in Fig. \ref{110} (without zero twist angle approximation), where there are $2$ Dirac fermions at $K_M$ and $K_M'$, 6 Dirac fermions close to $\Gamma_M$ point, and also $6$ Dirac fermions approximately on $\Gamma_M$-$K_M$ and $\Gamma_M$-$K_M'$ lines. The general understanding of LLs should be similar to the above, but in this case all the Dirac points are extremely close to the charge neutral point, making ito difficult to resolve the zero modes of each Dirac fermion.

\subsection{Far away from the magic angle}

In this section, we show the LLs calculated for higher angles away from the magic angle, in which case there are only two Dirac fermions at $K_M$ and $K_M'$ per spin per graphene valley. As the angle increases from the magic angle, the band width becomes larger and larger, and in comparison the Zeeman splitting becomes less and less important, so the spin degeneracy can be treated as unbroken for large enough twist angles. Figs. \ref{130} shows the LLs for $\theta=1.3^\circ$ ($\alpha=0.466$), where panels (a)-(b) are calculated without Zeeman energy, and panel (c) is calculated with Zeeman energy. In this case, the conduction (valence) band only has a single band maximum (minimum) at $\Gamma_M$ point, as shown in Fig. \ref{130}(d) (There is no band maximum or minimum near $M_M$, instead there is only a saddle point on the $\Gamma_MM_M$ line, see the contour plot Fig. \ref{130}(e)).

In Fig. \ref{130}(a), one can clearly see the LLs from Dirac fermions at $K_M$ and $K_M'$ near zero energy(red arrow in Fig. \ref{130}(a)); while at high energies there are LLs from the $\Gamma_M$ band maximum (blue arrow in Fig. \ref{130}(a)). In this case, one expects the Dirac LLs around the charge neutral point to be spin, graphene valley and Moir\'e $8$ fold degenerate, yielding a Landau fan at electron density $n_0=0$ at fillings
\[
\cdots,-20,-12,-4,4,12,20,\cdots\ .
\]
While near the band maximum $n=n_s$ (4 electrons per Moir\'e unit cell), the LLs are contributed by the $\Gamma_M$ band maximum, and are thus only spin and graphene valley $4$-fold degenerate. This gives rise to a Landau fan at $n_0=n_s$ at fillings
\[
0,-4,-8,-12,\cdots\ ,
\]
which occurs on the side of electron density $n<n_s$. This agrees with the observation of TBG at large angles \cite{cao2016}.

\begin{figure}[htbp]
\begin{center}
\includegraphics[width=7in]{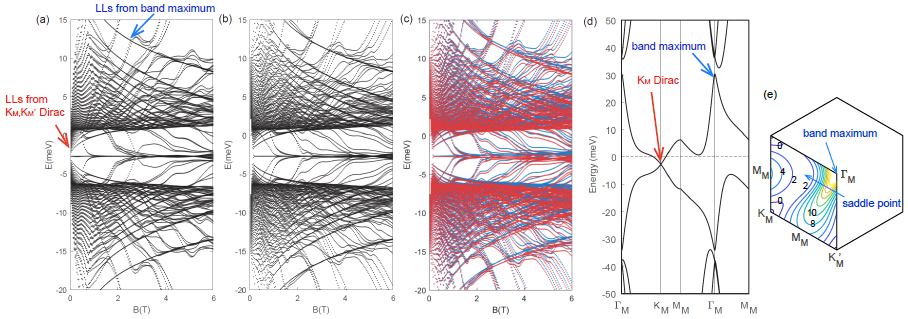}
\end{center}
\caption{LLs near charge neutral point for $\alpha=0.466$ ($\theta=1.3^\circ$). (a) LLs calculated with center momentum $\mathbf{k}_0$ at $K_M$ point. (b) LLs calculated with center momentum $\mathbf{k}_0$ at $M_M$ point. (c) LLs with Zeeman splitting, where the center momentum $\mathbf{k}_0$ is at $M_M$. Red and blue are for spin up and down, respectively. (d) The corresponding band dispersion at this angle, $\alpha=0.466$ (PH asymmetric). (e) The energy contour plot of the lowest conduction band, where the numbers are energies in meV.}
\label{130}
\end{figure}

\section{Landau fans near the magic angle, Landau fan shift}\label{SecShift}

This section is a supplementary material for the Landau fan discussions in the main text (pages 4) for twist angles $\theta$ near but not equal to the magic angle $\theta_m$. In both the main text and this section, we adopt the zero angle approximation so that the TBG bands are PHS.

\subsection{Broken of Landau level degeneracy}

In the 4th paragraph of page 4 of the main text, we mentioned that the $K_M,K_M'$ degeneracy is broken when the magnetic field $B$ becomes large. This is due to the mutual hopping between the Dirac LLs at $K_M$ and $K_M'$. As shown in Fig. \ref{Bperp}(a), the shaded areas at $K_M$ and $K_M$ denote the momentum space LL wave functions of the Dirac fermions at $K_M$ and $K_M$. The sizes of the LL wave functions in the momentum space are proportional to $\ell_B^{-1}\propto\sqrt{B}$. Therefore, when $B$ becomes large, the LL wave functions at $K_M$ and $K_M$ will overlap and hop with each other, so that their degeneracies are broken. The calculation of the second model $H'(\mathbf{k})+H_p(\mathbf{k})$ (with Dirac points of the same helicity) in Sec. \ref{SecZM} is an example showing the degeneracy breaking of the two zero mode LLs at two different Dirac points.

Similarly, the degeneracy of the LLs of other multiple band maxima/minima (e.g., 3 band maxima near 3 $M_M$ points) will also be broken at large $B$, due to the mutual hoppings among them. Therefore, at large $B$, the LLs of TBG are only graphene valley $K,K'$ 2-fold degenerate (spin degeneracy is broken by Zeeman).

From our numerical calculations (with Zeeman energy), we observe that the $K_M,K_M'$ Dirac LLs are $4$-fold degenerate (Moir\'e valley $K_M$,$K_M'$ and graphene valley $K,K'$) for $B\lesssim3$T, and become $2$-fold for $B\gtrsim 3$T. The LLs from the 3 band tops (bottoms) near the $M_M$ point are approximately $6$-fold degenerate (3 $M_M$ points and graphene valley $K,K'$) for $B\lesssim1$T, and break down to $2$-fold for $B\gtrsim 1$T.

\subsection{Landau fans under Zeeman energy}



\begin{figure}[tbp]
\begin{center}
\includegraphics[width=7in]{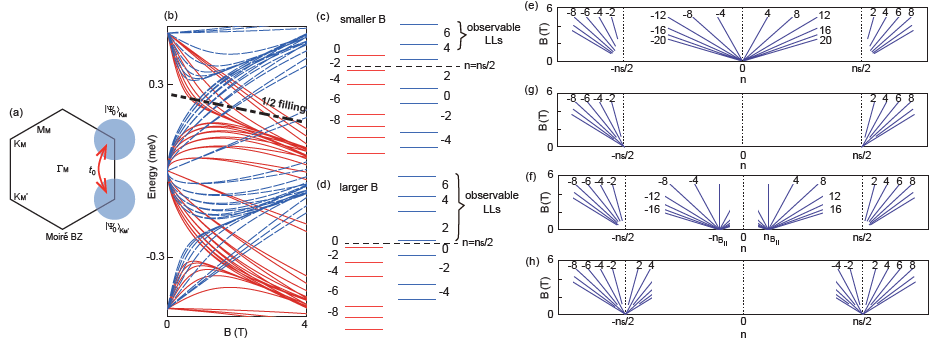}
\end{center}
\caption{(a) Illustration of the quantum tunneling between LLs at $K_M$ and $K_M'$, which leads to a Moir\'e valley degeneracy breaking at large $B$. (b) The LLs with Zeeman splitting with respect to $B$. (c)-(d) Illustration of the LLs near filling $n_0=n_s/2$ at a smaller Zeeman field (panel (c)) and at a larger Zeeman field (panel (d)), where the left (red) levels are the spin $\uparrow$ conduction band top LLs, and the right (blue) levels are the spin $\downarrow$ Dirac LLs at $K_M,K_M'$. The spin $\downarrow$ LLs exceeding the highest spin $\uparrow$ LL are observable in the Landau fan at $n_0=n_s/2$. (e)-(h) Landau fans under out-of-plane magnetic field $B$ and a fixed in-plane magnetic field $B_\parallel$: (e) $B_\parallel=0$; (f) $B_\parallel<B_c=W/2\mu_B$; (g) $B_\parallel=B_c$; and (h) $B_\parallel>B_c$, where $W$ is the conduction (valence) band width.}
\label{Bperp}
\end{figure}


In paragraph 5 of page 4 of the main text, we have mentioned the Dirac LLs from $K_M,K_M'$ with Zeeman splitting are $4$-fold degenerate at small magnetic field $B$, and become $2$-fold degenerate for $B\gtrsim 3$T (from numerical calculation in Fig. \ref{104PHS}) due to the $K_M,K_M'$ degeneracy breaking. Therefore, at $n_0=0$ one has the dominant Landau fan at $\nu_j=4j$ ($j\in\mathbb{Z}$) for small $B$ as shown in Fig. \ref{Bperp}(b), which breaks into a $2$-fold Landau fan at large $B$.


Further, we show in paragraph 4 of main text page 4 that the strong Zeeman splitting can also give rise to Landau fans at half fillings $n=\pm n_s/2$, where $n_s$ is the electron density to fully fill a $4$-fold degenerate Moir\'e band at $B=0$. Here we describe this half-filling Landau fan in more details. As shown in the main text Fig. 4(a), assume the Zeeman splitting $E_Z(B)=2\mu_BB>W$, where $W$ is the conduction (valence) band width. In this case, the $K_M,K_M'$ Dirac points of the spin $\downarrow$ band (which are stable since spin and orbital are decoupled) are higher than the entire spin $\uparrow$ conduction band. Therefore, for electron density $n=n_s/2$, if $E_Z(B)>W$, the entire spin $\uparrow$ conduction band will be occupied, and the Fermi level will be fixed at the energy of the $K_M,K_M'$ Dirac points of the spin $\downarrow$ band. This enables us to see the Landau fan of the spin $\downarrow$ ($\uparrow$) Dirac fermions at $n=n_s/2$ ($n=-n_s/2$). One expects this Landau fan to be $2$-fold degenerate, since it appears at relatively large field $B\gtrsim W/2\mu_BB$.

More explicitly, the fan at the $l$-th spin $\downarrow$ Dirac LL ($l\in\mathbb{Z}$) can be seen above $n_0\approx n_s/2$ when
\begin{equation}\label{LW}
E_l(B)+E_Z(B)\gtrsim W\ ,
\end{equation}
with $E_l(B)\approx v^*\text{sgn}(l)\sqrt{2|l|\hbar eB}$ is the energy of the $l$-th Dirac LL, and $E_Z(B)=2\mu_BB$.
This gives the condition for the $l$-th spin $\downarrow$ Dirac LL to be seen:
\begin{equation}\label{Bbound}
B\gtrsim\left(\sqrt{\frac{v_*^2|l|\hbar e}{8\mu_B^2}+\frac{W}{2\mu_B}}-\text{sgn}(l)\frac{v_*\sqrt{|l|\hbar e}}{2\sqrt{2}\mu_B}\right)^2\ .
\end{equation}
Therefore, if one increase $B$ from zero, the higher (i.e., the more positive) $l$ LLs will be seen earlier. This is illustrated in Fig. \ref{Bperp}(c)-(d). As a result, as shown in Fig. \ref{Bperp}(e), the higher filling lines (representing $\rho_{xx}$ minima) at $n_0=n_s/2$, (i.e., higher LL number $l$) can be observed at smaller $B$ (given by the bound \ref{Bbound} for LL number $l$).
In particular, from Eq. (\ref{LW}), the observation of the $l<0$ spin $\downarrow$ Dirac LLs in the Landau fan at $n_0=n_s/2$ requires much larger $B$ than that of the $l>0$ spin $\downarrow$ LLs, since $E_l(B)$ is negative (positive) for negative (positive) $l$. For instance, for the band structure of $\theta=1.04^\circ$ (see Fig. 4(a) in the main text, or Fig. \ref{band}(d) and (i)), one estimates that the $l=1$ LL can be seen around $2$T at $n_0=n_s/2$, the $l=0$ LL can be seen around $4$T, while the observation of the $l=-1$ LL requires a magnetic field above $6$T. Therefore, the Landau fan is easier to be seen on the $n>n_s/2$ side than the $n<n_s/2$ side. This agrees with the half-filling Landau fans observed in experiments \cite{cao2018,yankowitz2018}.

We now consider the effect of a fixed in-plane magnetic field $B_{\parallel}$ in addition to the out-of-plane field $B_\perp$, which changes the Zeeman energy to $E_Z^\pm(B)=\pm\mu_B(B_\perp^2+B_\parallel^2)^{1/2}$. Only $B_\perp$ contributes to the LL orbital effect. At out-of-plane field $B_\perp=0$, the spin $\downarrow$ and $\uparrow$ (along $B_\parallel$) $K_M,K_M'$ Dirac points are shifted to energies $\pm\mu_BB_\parallel$, respectively. Therefore, the Landau fan at $n_0=0$ will split into two fans at $n_0=\pm n_{B_\parallel}$ (Fig. \ref{Bperp}(f)), where $n_{B_\parallel}\approx \frac{4\mu_B^2B_\parallel^2}{\pi^2\hbar^2 v_*^2}$ for small $B_\parallel$, which is the electron density at Fermi energy $\mu_BB_\parallel$. Each Landau fan is still $K,K'$ and $K_M,K_M'$ $4$-fold ($K,K'$ $2$-fold) degenerate for $B\lesssim3$T ($B\gtrsim3$T).

Meanwhile, the Landau fans at $n_0=\pm n_s/2$ will now be seen when $\sqrt{B_\perp^2+B_\parallel^2}>B_c=W/2\mu_B$, thus require a smaller $B_\perp$. In particular, when $B_\parallel$ reaches $B_c=W/2\mu_B$, the density at Fermi energy $\mu_BB_\parallel$ will reach $n_{B_\parallel}= n_s/2$. Thus the Landau fans at $n_0=\pm n_{B_\parallel}$ will merge with those at $n_0=\pm n_s/2$ (both of which are due to the spin $\downarrow$ ($\uparrow$) Dirac LLs at $K_M,K_M'$), as shown in Fig. \ref{Bperp}(g).
Furthermore, when $B_\parallel>B_c$, the spin $\downarrow$ and $\uparrow$ bands are more and more separated, and the Landau fans at $n_0=\pm n_s/2$ will be seen clearer, as shown in Fig. \ref{Bperp}(h). For $\theta=1.04^\circ$, the band theory of TBG predicts a $B_c$ around $5$T.

\subsection{LL crossing and fan shift}
As can be seen in Fig. \ref{band}, additional non-high symmetry points Dirac crossings may arise between two flat bands when $\theta$ is near the magic angle \cite{songz2018}.
For instance, for $1.02^\circ<\theta<1.07^\circ$ ($0.570<\alpha<0.593$), there are 6 Dirac points along $\Gamma_MM_M$ lines per graphene valley, 3 above and 3 below the charge neutral point, and there are $6$ Dirac points  along $\Gamma_MK_M$ concentrated around $\Gamma_M$ point (see Fig. \ref{band}(d) and (i)),
which will contribute additional low energy LLs.
As we have numerically shown in Fig. \ref{104}, the LLs from these additional Dirac points are only $2$-fold (graphene valley $K,K'$) degenerate for $B\gtrsim0.5$T, due to mutual hopping among these Dirac LLs.
Fig. \ref{Lcrossing}(a) gives a schematic plot of the zero mode LLs of the Dirac points on $\Gamma_MM_M$ lines with Zeeman energy (black dashed lines; for energy $E>0$, the lower three dashed lines with negative slopes have spin $\uparrow$, and the higher three dashed lines with positive slopes have spin $\downarrow$), while Fig. \ref{Lcrossing}(b) is a zoom-in plot. The LL energies in the figure are illustrative instead of accurate (in principle, the LLs begin to expand into Hofstadter bands instead of having a definite energy for $B\gtrsim 1.5$T).

As shown in Fig. \ref{Lcrossing}(b), these additional 2-fold degenerate LLs (black dashed lines) may cross with the Dirac LLs of $K_M,K_M'$ which are approximately 4-fold degenerate (for $B$ not too large).
Such LL crossings will induce a shift of the dominant Landau fan $\nu_j=4j$ by $2$ around $n_0=0$ as shown in Fig. \ref{Lcrossing}(c).
For example, we can look at the crossings of the lowest 2-fold zero mode LL (thick dashed line) with a set of $4$-fold LLs (blue and red solid lines) at magnetic fields $B_j$ ($j\ge1$) as shown in Fig. \ref{Lcrossing}(b). The green arrow points out the crossing at magnetic field $B_1$. For $B>B_1$, above zero energy one first has two 4-fold LLs, then encounters the lowest 2-fold LL, followed by more 4-fold LLs. This yields Landau fan filling factors at $\nu_j=\{4,8,10,14,18,\cdots\}$. While for $B_2<B<B_1$, above zero energy one first has three 4-fold LLs, then encounter the lowest 2-fold LL. Therefore, the Landau fan will be shifted to $\nu_j=\{4,8,12,14,18,\cdots\}$.
In general, whenever the 2-fold LL crosses with the $j$-th 4-fold LL, $\nu_j$ will be shifted from $4j$ to $4j-2$. Similar fan shift may occur on the hole doping side. We note that in Fig. \ref{Lcrossing} we have ignored the particle-hole asymmetry of TBG, but in reality the band structure is particle-hole non-symmetric. At the single particle level, the particle-hole asymmetry is small, and one expects the fan shift to occur on both the electron side and the hole side. Such fan shift phenomena is observed on the hole doping side of TBG under pressure in the recent experiment \cite{yankowitz2018}, but not on the electron side. If the hole side fan shift can be explained by our single particle picture, the absence of the electron side fan shift might be due to the existence of a correlated insulating phase at $n=n_s/4$ on the electron side in their experiment (while no insulating phase at $n=-n_s/4$ on the hole side), whose many-body effects may overwhelm the Landau fan shift at $n>0$.

The latest experimental measurements indicate the Zeeman splitting in TBG is much smaller than we expected here from the TBG band theory \cite{priv}. Therefore, the Landau fans observed in the experiments may not have the same origin as that in our theoretical analysis here.

\begin{figure}[tbp]
\begin{center}
\includegraphics[width=5.5in]{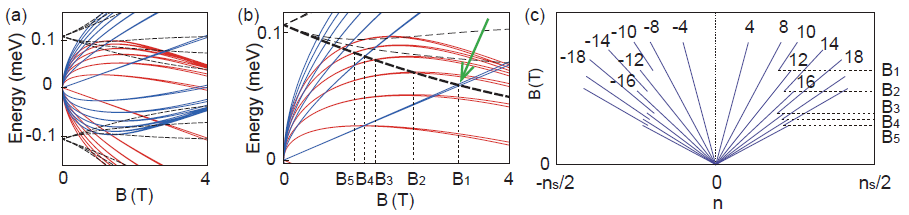}
\end{center}
\caption{(a) Schematic illustration of LLs with Zeeman energy when there are additional Dirac points, where red (blue) solid lines are the spin up (down) $K_M,K_M'$ Dirac LLs, while black dashed lines are zero mode LLs of additional Dirac points on $\Gamma_MM_M$ lines. For energy $E>0$, the lower three dashed lines with negative slopes have spin $\uparrow$, and the higher three dashed lines with positive slopes have spin $\downarrow$. (b) Zoom in of panel (a), where the magnetic fields $B_i$ label the crossings between the lowest 2-fold LL and the various 4-fold LLs. (c) The Landau fan shift at $n_0=0$ due to the crossing between 2-fold and 4-fold LLs in (a).}
\label{Lcrossing}
\end{figure}



\section{Landau fans of exactly magic angle with quadratic band touching}\label{SecQ}

In the main text and supplementary Sec. \ref{SecShift} we have discussed the Landau fans for twist angles $\theta$ near the magic angle $\theta_m$, mostly using $\theta=1.04^\circ$ as an example. In this section, we discuss the Landau fans at exactly the magic angle $\theta_m=1.0^\circ$ ($\alpha=0.605$), where the first conduction and valence bands have a quadratic band touching at $K_M$ and $K_M'$. Additionally, there are 6 Dirac points close to $\Gamma_M$. The main features of the Landau fans remain, except that the Landau fan sequence may change.

At low energies, the quadratic band touching at $K_M$ and $K_M'$ is described by the effective Hamiltonian
\begin{equation}\label{qLL}
H_{\text{eff}}\approx\frac{\hbar^2}{2M_*}\left(\begin{array}{cc}0&(k_x\pm ik_y)^2\\(k_x\mp ik_y)^2&0\end{array}\right)\ ,
\end{equation}
where $M_*$ is the effective mass. By the substitution in Eq. (\ref{kplus}) (with center momentum chosen at $\mathbf{k}_0=(0,0)$), one finds the Landau level spectrum of quadratic Dirac fermion $E_n=\pm\sqrt{l(l-1)}eB\hbar/M_*$. In particular, there are two zero mode LLs for each quadratic Dirac fermion. Therefore, the quadratic fermions at $K_M$ and $K_M'$ together contribute 4 zero mode-LLs per spin per graphene valley. The spectrum is linear in $B$. These are the key differences from the LLs of a linear dispersion Dirac fermion.

\begin{figure}[htbp]
\begin{center}
\includegraphics[width=7in]{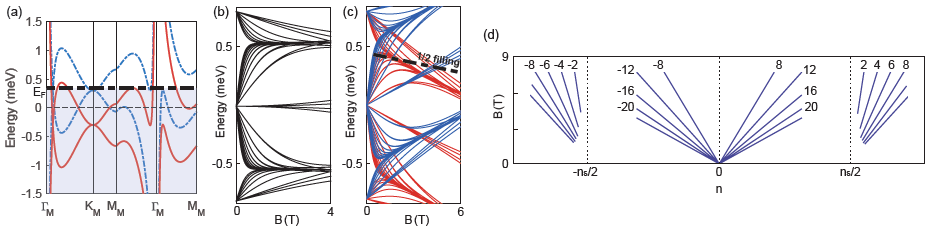}
\end{center}
\caption{(a) The first conduction and valence bands of $\theta=\theta_m=1.0^\circ$ ($\alpha=0.605$) at graphene valley $K$ with a Zeeman energy of $B\approx 6$T, where the red solid (blue dashed) line represents spin up (down). (b) Illustration of the main branches of LLs of the first bands without Zeeman energy. (c) The LLs with Zeeman energy, where the red solid (blue dashed) line represents spin up (down) LLs. (d) The expected Landau fans resulting from LLs in (c).}
\label{FigShopping}
\end{figure}

Fig. \ref{FigShopping}(a) shows the lowest two Moir\'e bands (particle-hole symmetric) with a Zeeman energy of $B\approx6$T at magic angle $\theta_m$ (zero angle approximation used), where one can see the quadratic band touching at $K_M$ point. The LLs without and with Zeeman energy are shown in Fig. \ref{FigShopping}(b) and (c), respectively. They are schematically plotted based on our quantum and semiclassical LL numerical calculations. At small magnetic fields ($0.5$T $\lesssim B\lesssim3$T), the LL spectrum in Fig. \ref{FigShopping}(b) around the charge neutral point is as expected in Eq. (\ref{qLL}). The nonzero mode LLs are 2-fold degenerate per spin per graphene valley, while the zero mode LL is 4-fold degenerate per spin per graphene valley. Therefore, for small $B\lesssim3$T, one would expect to see dominant Landau fans near the charge neutral point at fillings
\[
\nu_j=\cdots,-16,-12,-8,0,8,12,16,\cdots\ ,
\]
as shown in Fig. \ref{FigShopping}(d).

As the magnetic field increases $B\gtrsim 3$T, the Moir\'e valley $K_M,K_M'$ degeneracy is broken due to the quantum hopping between LLs at $K_M$ and $K_M'$ (see Fig. \ref{Bperp}(a)). In particular, the degeneracy of the 4 zero mode LLs at $K_M$ and $K_M'$ (for one spin and one graphene valley) will all be broken. This is because of the following: the two zero mode LLs at $K_M$ ($K_M'$) have different wave functions $|\Psi_0\rangle_{K_M}$ and $|\Psi_1\rangle_{K_M}$ ($|\Psi_0\rangle_{K_M'}$ and $|\Psi_1\rangle_{K_M'}$). Therefore, the hopping amplitude $t_0$ between $|\Psi_0\rangle_{K_M}$ and $|\Psi_0\rangle_{K_M'}$ is different from the hopping $t_1$ between $|\Psi_1\rangle_{K_M}$ and $|\Psi_1\rangle_{K_M'}$. Accordingly, the four zero mode LLs of $K_M$ and $K_M'$ will split into four distinct energies $\pm t_0$ and $\pm t_1$, so their $4$-fold degeneracy will be totally broken. Further, there could also be a nonzero hopping $t_{01}$ between $|\Psi_0\rangle_{K_M}$ and $|\Psi_1\rangle_{K_M'}$ (between$|\Psi_1\rangle_{K_M}$ and $|\Psi_0\rangle_{K_M'}$), which will further split the four zero modes. Therefore, the only remaining LL degeneracy is then the $2$-fold graphene valley $K,K'$ degeneracy (with spin degeneracy broken by Zeeman). Accordingly, for large $B\gtrsim3$T, the Landau fan at the charge neutral point is expected to be seen at all even fillings $\nu_j=2j$ ($j\in\mathbb{Z}$) (which is not shown in Fig. \ref{FigShopping}(b)).

Yet another possibility at large $B>\gtrsim3$T is, if it happens that the hoppings among zero modes satisfy $t_0\approx t_1$ and $t_{01}\approx 0$ (or $t_0\approx t_1\approx0$, $t_{01}\neq0$), the four zero mode LLs (for one spin and one graphene valley) will only be broken to two 2-fold degenerate LLs. In this case, one would expect the dominant Landau fan fillings near the charge neutral point to be
\[
\nu_j=\cdots,-16,-14,-12,-10,-8,-4,0,4,8,10,12,14,16,\cdots\ .
\]
This requires a fine tuning of $t_0$, $t_1$ and $t_{01}$, which is less likely.

At half filling $n\approx\pm n_s/2$, following the same argument we had in the main text (paragraph 4 on page 4) and supplementary Sec. \ref{SecShift}B, one expects Landau fans at even fillings $\nu_j$ arising at finite magnetic fields $B$ as shown in Fig. \ref{FigShopping}(d).



\section{Semiclassical calculation of LLs}\label{SecSemi}
In this section we explain the semiclassical determination of the LL spectrum of the Moir\'e bands. The band edges and band touching points can provide a significant amount of information about LLs at small field \cite{Alexandradinata2017}. It is known that
each LL will occupy an area in the momentum space:
\begin{equation}\label{eqn:area}
	\Delta \Omega_{k} = \frac{2\pi}{\ell_B^2}\ .
\end{equation}
Therefore, semiclassically, the energies of LLs correspond to a set of consecutive equal-energy contours with enclosed momentum space areas spaced by $2\pi/\ell_B^2$. Around a band edge with quadratic dispersion, the LL energy takes the form $E_l \sim (l+\frac{1}{2}) B$. around a Dirac point the energy has another form $E \sim \sqrt{lB}$ with one zero mode. Between the top band edge and bottom band edge, there is always at least one van-Hove singularity (VHS) in the band. The VHS is the energy where the equal-energy contour changes from electron pockets into hole pockets. Equivalently, we can say that the area inside the electron pocket diverges when the Fermi energy reaches the VHS. From Eq. (\ref{eqn:area}) we can claim that the semiclassical LLs from the top and bottom band edges will approach the VHS at large $B$. However, the zero mode LLs of Dirac points or quadratic band touchings will remain at zero energy in the semiclassical picture (which is not true in quantum calculations), as they correspond to zero energy contours which have zero momentum space area.

Our semiclassical picture of LLs is based on the equal energy contour area in momentum space. The constant energy contour plot in the lowest conduction Moir\'e band (PHS) for angles from $1.0^\circ$ to $1.8^\circ$ can be found in Fig. \ref{band}(k)-(o). We can also take part of the quantum effect into consideration, including the correct number of zero modes from Dirac fermions and quadratic band touchings.
The semiclassical LLs with necessary quantum corrections are given by the following rules ($e$ and $\hbar$ set to 1):
\begin{itemize}
	\item around a Dirac point, the LLs have the form $E_l = \pm v\sqrt{2l B}$ with one zero mode, where the coefficient $v$ is the Dirac Fermi velocity;
	\item around a quadratic band touching point, LLs are given by $E_l =M^{-1}\sqrt{l(l-1)}B$ with two zero modes (this is from the quantum results instead of classical), where the coefficient $M$ is the band mass;
	\item around a trivial quadratic band edge, the LLs are $E_l =M^{-1} (l + 1/2)B$, where $M$ is the quadratic band mass;
	\item the LL energies will not increase to infinity (with the exception of a finite number of topological ones), instead they will merge into the VHS;
	\item if a number of Dirac points or quadratic touching points are degenerate, their LL degeneracy will break down when their equal-energy contours merge together.
\end{itemize}
With these rules, we semiclassically plot the LLs of TBG. For simplicity we take the PHS approximation. The steps are as follows: first, we determine the energies of all the band minima $E_{j}^{\rm min}$, maxima $E_{j}^{\rm max}$, Dirac point energy $E_{j}^{\rm D}$, quadratic band touching energy $E_{j}^{\rm Q}$ and van-Hove singularities $E_{j}^{\rm vH}$ from the band structure of the lowest two Moir\'e bands. For each band minimum, maximum or quadratic band touching, we extract out its effective mass $M$ (the dispersion is $\pm k^2/2M$); for each Dirac point, we extract out its Fermi velocity $v$. If the mass or velocity is anisotropic, we take the average mass $M=2(M_1^{-1}+M_2^{-1})^{-1}$ or the average velocity $v=(v_1+v_2)/2$, where $M_1$ and $M_2$ ($v_1$ and $v_2$) are the maximal and minimal mass (velocity) among all directions.

\begin{figure}[htbp]
\centering
\includegraphics[height=7cm]{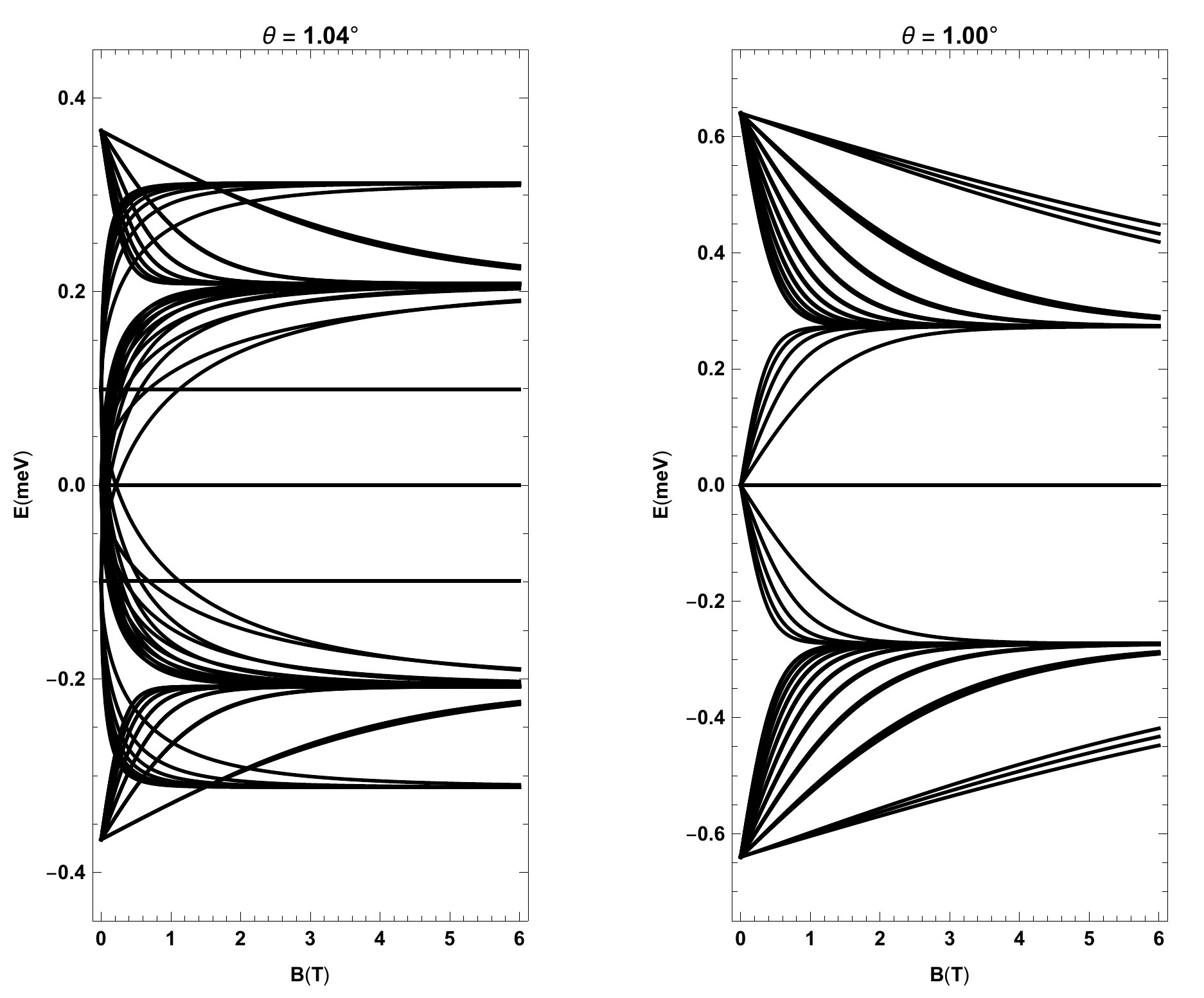}
\caption{The schematic semiclassical LLs without Zeeman field of (a) $\theta = 1.04^\circ$ and (b) $\theta =\theta_m= 1.0^\circ$ (the magic angle).}
\label{semiclassicalLL}
\end{figure}

Second, assume we have identified a quadratic band minimum $E_1^{\rm min}$ with an average effective mass $M$, and the constant energy contour surrounding it is obstructed by van Hove singularity $E_1^{\rm vH}$ as the energy increases. We then take an approximate ansatz $E_{l}(B)=(E_1^{\rm vH}-E_1^{\rm min})\tanh\left[M^{-1}(E_1^{\rm vH}-E_1^{\rm min})^{-1}(l+\frac{1}{2})B\right]$ to plot its semiclassical LLs ($l\ge0$). This ansatz yields the expected small $B$ behavior of LLs $E_l=M^{-1}(l+\frac{1}{2})B$, while at large $B$ the LLs merge into the van Hove singularity, $E_{l}(B\rightarrow\infty)\rightarrow E_1^{\rm vH}$. This is accurate enough for the purpose here (the semiclassical results are used as a supplement to our quantum numerical calculations). Similarly, if we have identified a quadratic band touching $E_1^{\rm Q}$ with an average effective mass $M$, we take the LL ansatz $E_{l}^\pm(B)=\pm(E_\pm^{\rm vH}-E_1^{\rm Q})\tanh\left[M^{-1}(E_\pm^{\rm vH}-E_1^{\rm Q})^{-1}\sqrt{l(l-1)}B\right]$, where $E_\pm^{\rm vH}$ are the van Hove singularities above and below the quadratic band touching $E_1^{\rm Q}$ (by which the energy contours surrounding the quadratic band touching are obstructed). And lastly if we have a Dirac point at energy $E_1^{\rm D}$ with an average velocity $v$, its semiclassical LLs are assumed to be $E_{l}^\pm(B)=\pm(E_\pm^{\rm vH}-E_1^{\rm min})\tanh\left[v(E_\pm^{\rm vH}-E_1^{\rm min})^{-1}\sqrt{2lB}\right]$, where $E_\pm^{\rm vH}$ are the corresponding van Hove singularities above and below the Dirac point.

Lastly, in addition to the band minima, maxima, Dirac points, quadratic band touchings and van-Hove singularities, there are energies $E_{j}^{\rm merge}$ where the energy contours around several degenerate band minima (maxima) merge together. Such degeneracy is usually due to symmetries; for instance, when there is a band maximum at $M_M$ point of the MBZ, the $C_{3z}$ symmetry indicates the band maxima at all the three $M_M$ points are degenerate. When the semiclassical LLs from these degenerate band minima (maxima) reach the contour merging energy, their degeneracy is broken; to simulate that, we simply add a small splitting among them (which is only qualitative).

In Fig. \ref{semiclassicalLL} (a) and (b), we show the semiclassical LL spectrum for $\theta = 1.04^\circ$ and $\theta = 1.00^\circ$ (the magic angle) obtained from the above procedure. They capture well the main sets of LLs obtained from our quantum calculations in Sec. \ref{SecNLL}.

In the $\theta = 1.04^\circ$ case, we take the Dirac points on $K_M$, $K_M'$ and $\Gamma_MM_M$ lines, and the quadratic band tops/bottoms along $\Gamma_MM_M$ into consideration. Since the 6 Dirac points near $\Gamma_M$ are extremely close to each other and have large Fermi velocities, their LLs quickly merge into the van Hove singularities as $B$ increases (they merge at $B\sim0.04$T, see Sec. \ref{SecNLL}B), and can be ignored when the minimal semiclassical LL orbital area (around $2\pi/\ell_B^2$) grows large.

\end{widetext}

\end{document}